\documentclass[letterpaper,showpacs,onecolumn,eqsecnum,superscriptaddress,nofootinbib]{revtex4}
\usepackage[utf8]{inputenc}
\usepackage{amssymb,amsmath,amsthm,amsfonts,epsfig,times,natbib}
\usepackage{graphicx} 

\usepackage[usenames,dvipsnames]{color}
\usepackage{epstopdf}
\usepackage{amsmath,amssymb}
\usepackage{tensor}
\usepackage{mathtools}
\usepackage{amsbsy}
\usepackage{bm,url}
\usepackage[linktocpage,hidelinks]{hyperref}
\usepackage{soul}
\usepackage{commath}
\usepackage[scaled]{beramono}
\usepackage[T1]{fontenc}
\usepackage{graphicx}
\usepackage[caption=false]{subfig}

\usepackage[normalem]{ulem}
\usepackage{soul}

\begin{document}
\title{
Head-on collisions of $\ell$-boson stars}

\author{V\'ictor Jaramillo} 
\affiliation{Instituto de Ciencias Nucleares, Universidad Nacional
  Aut\'onoma de M\'exico, Circuito Exterior C.U., A.P. 70-543,
  M\'exico D.F. 04510, M\'exico}

\author{Nicolas Sanchis-Gual}
\affiliation{Departamento  de  Matem\'atica  da  Universidade  de  Aveiro and  Centre  for  Research  and  Development  in Mathematics  and  Applications  (CIDMA), Campus  de  Santiago,  3810-183  Aveiro,  Portugal}
\affiliation{Departamento de
  Astronom\'{\i}a y Astrof\'{\i}sica, Universitat de Val\`encia,
  Dr. Moliner 50, 46100, Burjassot (Val\`encia), Spain}

\author{Juan Barranco}
\affiliation{Departamento de F\'isica, Divisi\'on de Ciencias e Ingenier\'ias,
Campus Le\'on, Universidad de Guanajuato, Le\'on 37150, M\'exico}

\author{Argelia Bernal}
\affiliation{Departamento de F\'isica, Divisi\'on de Ciencias e Ingenier\'ias,
Campus Le\'on, Universidad de Guanajuato, Le\'on 37150, M\'exico}

\author{Juan Carlos Degollado} 
\affiliation{Instituto de Ciencias F\'isicas, Universidad Nacional Aut\'onoma 
de M\'exico, Apdo. Postal 48-3, 62251, Cuernavaca, Morelos, M\'exico}

 \author{Carlos Herdeiro}
\affiliation{Departamento  de  Matem\'atica  da  Universidade  de  Aveiro and  Centre  for  Research  and  Development  in Mathematics  and  Applications  (CIDMA), Campus  de  Santiago,  3810-183  Aveiro,  Portugal}

\author{Miguel Megevand}
\affiliation{Instituto de F\'isica Enrique Gaviola, CONICET. Ciudad Universitaria, 5000 C\'ordoba, Argentina}

\author{Dar\'{\i}o N\'u\~nez}
\affiliation{Instituto de Ciencias Nucleares, Universidad Nacional
  Aut\'onoma de M\'exico, Circuito Exterior C.U., A.P. 70-543,
  M\'exico D.F. 04510, M\'exico}


\date{February 2022}

\begin{abstract}
Fully non-linear numerical evolutions of the Einstein-(multi)--Klein-Gordon equations are performed  
to study head-on collisions of $\ell$-boson stars.
Despite being spherically symmetric, $\ell$-boson stars have a (hidden) 
frame of reference, used in defining their individual multipolar fields. To assess the impact of their relative orientation, we perform simulations with different angles between the axes of the two colliding stars.
Additionally, two scenarios are considered for the colliding stars: that they are composites of either the same or different scalar fields.
Despite some model-specific behaviours, the simulations generically indicate that: 1) the collision of two sufficiently (and equally) massive  stars leads to black hole formation;  2) below a certain mass threshold the end result of the evolution is a bound state of the composite scalar fields, that neither disperses nor collapses into a black hole within the simulation time;
3)  this end product (generically) deviates from spherical symmetry and the equipartition of the number of bosonic particles between the different scalar fields composing the initial boson stars is lost, albeit not dramatically.
This last observation indicates, albeit without being conclusive, that the end result of these collisions belongs to the previously reported larger family of equilibrium  multi-field boson stars,  generically non-spherical, and of which $\ell$-boson stars are a symmetry enhanced point. 
We also extract and discuss the waveforms from the collisions studied.

\end{abstract}




\maketitle
\section{Introduction}
\label{sec:introduction}
The advent of the gravitational wave (GW) era promises to deliver invaluable information on some of the most prominent challenges in theoretical physics. Amongst these is the nature of the dark Universe.  At the time of writing, the LIGO-Virgo and now Kagra (LVK) collaborations released three public catalogues from the first three science runs O1+O2~\cite{LIGOScientific:2018mvr}, O3a~\cite{LIGOScientific:2020ibl} and O3b~\cite{LIGOScientific:2021djp}, reporting 85 events. These events are providing invaluable information about black hole populations~\cite{LIGOScientific:2018jsj}, constraints on dark energy models~\cite{LIGOScientific:2017vwq,Creminelli:2017sry,Baker:2017hug,Ezquiaga:2017ekz} and even tantalizing  hints about the nature of dark matter~\cite{Bustillo:2020syj}.

LVK searches are performed using “matched filtering”, a data analysis technique to detect characteristic signals in noisy data, which requires a library of theoretical waveform models. In this respect, the issue of degeneracy has been an understated caveat in GW detections. The black hole interpretation seems vindicated within the Kerr black hole paradigm of general relativity. There is, however, a lack of alternative models for which waveforms have been accurately produced, in order to assess whether matched filtering really selects general relativity black holes within a more extensive library of theoretical templates.

Within this empty landscape of alternatives, bosonic stars offer a unique opportunity. First constructed in the late 1960s~\cite{Kaup68,Ruffini69} for massive, complex scalar fields (and more recently for massive complex vector fields~\cite{brito2016proca}) minimally coupled to Einstein's gravity, these are self-gravitating solitonic solutions that are both compact~\cite{Schunck:2003kk,Herdeiro:2017fhv,Herdeiro:2019mbz} and dynamically robust in regions of their parameter space~\cite{Seidel90,Liebling:2012fv,Guzman04,Guzman:2009xre,sanchis2017numerical,sanchis2019nonlinear}, forming from fairly generic initial data and reaching the equilibrium state by "gravitational cooling"~\cite{Seidel:1993zk,di2018dynamical,sanchis2019nonlinear}, possibly complemented by GW emission (in non-spherically symmetric evolutions).
Thus, they offer an (non-black hole) alternative relativistic two body problem which, unlike neutron stars, need not have electromagnetic counterparts. As a matter of critical importance, the well-posedness/hyperbolicity of the evolutions is under control, and  currently available computational infrastructures can be used with fairly minor adaptations, such as the Einstein Toolkit~\cite{EinsteinToolkit:web,loffler2012f,ZilhaoWitekCanudaRepository}. This state of affairs contrasts with modified gravity, wherein even promising models may face fundamental issues, such as the breakdown of hyperbolicity, see $e.g$~\cite{Ripley:2019aqj}. Thus, one can use bosonic star binaries to produce waveform banks that share features, but also present differences, with respect to those generated by the vanilla black hole binaries of general relativity. In fact, one such recent analysis raised the intriguing possibility that a real GW event~\cite{LIGOScientific:2020iuh} could be interpreted as a collision of vector bosonic stars~\cite{Bustillo:2020syj}, showing, at least as a proof of concept, how such interpretation would lead to the discovery of a fundamental, ultralight dark matter particle.

The simplest bosonic stars arise in single (complex) field models. A realization of the last few years, however, is that there is a wider landscape of bosonic star families when allowing multi-field models. 
Among these different possibilities, one of the most interesting configurations are the $\ell$-boson stars \cite{Alcubierre:2018ahf}, as it has been proved that they are the only stable configuration \cite{Sanchis-Gual:2021edp}, bestowing them with a central role in the possible scalar field configurations. $\ell$-boson stars are spherically symmetric solutions arising in a model with $2 \ell+1$ complex scalar fields, wherein the individual fields carry a multipolar structure but the composite object is spherical and static. Note that $\ell$-boson stars reduce to standard boson stars in the particular case where $\ell=0$. Subsequent studies  showed these solutions are dynamically robust in regions of the parameter space~\cite{Alcubierre:2019qnh,Alcubierre:2021mvs} and also unveiled they can be regarded as part of a wider family of multi-field, multi-frequency bosonic stars~\cite{Jaramillo:2020rsv,Sanchis-Gual:2021edp,Alcubierre:2021psa}. 
$\ell$-boson stars in contrast to $\ell=0$ regular boson stars have greater compactness and they are hollow in the central region. This empty space becomes larger as $\ell$ increases as  well as their mass and reaches the maximum compactness as $\ell \to \infty$ \cite{Alcubierre:2021psa}. The maximum compactness for the $\ell$-boson stars almost doubles the maximum compactness pf $\ell=0$ boson stars. Following this line of thoughts
it becomes an interesting problem to evolve $\ell$-boson stars in binaries, both to further test their dynamical robustness and to obtain new waveform templates, that enlarge the effort of constructing a vaster library of non-Kerr waveforms. These are precisely the goals of this paper.

In this work we shall study head-on collisions of $\ell$-boson stars, as the simplest ``binaries" of these objects. Even in this simplest scenario, the multi-field nature of $\ell$-boson stars allows more possibilities than when a single field is considered. Firstly, despite being described by a spherically symmetric metric and total energy-momentum tensor, the composite nature of $\ell$-boson stars endows the individual fields with a (hidden) 
frame of reference with respect to which the multipolar structure of the individual fields is defined. Thus, even for head-on collisions starting from rest, there is the additional degree of freedom of misaligning the hidden spin axes of the individual stars. Secondly, since one is entertaining the possibility of many scalar fields, it fits such rationale to allow the $\ell$-boson stars to be composed  by the same, or by different, scalar fields. We shall dub the former (latter) scenarios as tackling coherent (incoherent) states. As we shall see, both  choices, $i.e.$ alignment and coherence,  lead to important changes in the dynamics, although some generic trends can also be inferred from the sample of simulations performed.

We have focused on binary $\ell$-boson stars with $\ell=1$, using several initial configurations and evolving them while analysing the spatial distribution of the scalar fields and computing the GW emission. 
When the individual stars are massive and compact enough, it is found that the remnant is a black hole. However, if the sum of the masses of both stars does not greatly exceed the maximum mass of the corresponding family of solutions, (see Fig.  \ref{fig:mass_vs_frequency} below), the collision forms a gravitationally bound scalar field configuration.
Whereas in the merger of two standard $\ell=0$ boson stars some of the final configurations clearly tend towards another $\ell=0$ boson star \cite{Palenzuela:2006wp,Palenzuela:2007dm,bezares2017final,bezares2018gravitational,sanchis2019head}, our results indicate that  
the merger of two $\ell$-boson stars, albeit remaining in a bound state, do not necessarily lead to an $\ell$-boson star.

The waveforms generated from the head-on collision of the $\ell$-boson stars, like in the usual case of boson stars with $\ell=0$, present very peculiar features, which make them significantly different from the waveform of a black hole collision, even {\it nudus oculus}.  As we show in the present work, the waveform produced by the head-on collision of $\ell$-boson stars has a richer structure depending on the parameters of the initial configuration (such as the relative alignment), which makes a stronger case for the generation of catalogues to be included in the LVK libraries.

This paper is organized as follows. In the next section, we present the main ideas needed to construct $\ell$-boson stars, in particular explaining the two different scenarios that shall be considered: when taking two spatially separated lumps composed by the same fields, called coherent states, or when taking two spatially separated lumps composed of different fields, called incoherent states. Next, we describe the numerical implementation of these configurations and in section \ref{sec:head-on} we define the quantities that will be analyzed, together with the GWs during the evolution of the system. We present our results in section \ref{sec:results} and conclude with some final remarks. Throughout the paper we use natural units, $c=G=\hbar=1$.

\section{Models for \texorpdfstring{$\ell$}{l}-boson star binaries}
\label{sec:model}

A single $\ell$-boson star is described by an odd number $\mathcal{N}$ 
of complex scalar fields, each with a harmonic time dependence, of the form
\begin{equation} \label{eq:scalar_anzats}
    \Phi_{\ell m}(t,r,\vartheta,\varphi) = e^{-i\omega t} \phi_{\ell}(r) Y_{\ell, m}(\vartheta, \varphi) \ ,
\end{equation}
where $Y_{\ell, m}(\vartheta, \varphi)$  are the standard spherical harmonics.
Notice that the angular momentum number $\ell$ for a given solution is fixed and a single star has $\mathcal{N}=2\ell+1$ fields, corresponding to each possible value of $m$ within the range $-\ell,\dots 0,\dots,+\ell$.
A key ingredient to get a spherically symmetric solution of the Einstein-Klein-Gordon (EKG) system is 
that the field amplitude $\phi_{\ell}(r)$ is precisely the same for all $m$.

We shall be considering models without self-interactions amongst the different scalar fields. These fields, therefore, only see each other via gravity. Still, when considering two stars we may choose that the composing fields of the stars are equal or are different. In this work we shall consider both possibilities. This is reminiscent 
of the description of coherent and incoherent states in quantum mechanics; for the former case, a macroscopic number of quanta all pile into the same momentum state, being used to describe lasers and superfluids \cite{Lancaster:2014pza}.

Our approach is as follows.
The binary system we consider is governed  by the EKG theory and the field equations for the metric $g_{\mu\nu}$ are  
\begin{eqnarray} \label{eq:einstein}
R_{\mu\nu}- \frac{1}{2}g_{\mu\nu}R = 8\pi \left( T_{\mu\nu}^{(1)} + T_{\mu\nu}^{(2)}\right) 
:= 8\pi T_{\mu\nu} \ ,
\end{eqnarray}
where $R_{\mu\nu}$ is the Ricci tensor and $R =g^{\mu\nu}R_{\mu\nu}$.
The matter content is given by either one or two sets of $2\ell+1$ complex scalar fields $\Phi_{\ell m}$, each with a stress-energy tensor of the form
\begin{equation}\label{eq:EMT}
T_{\mu\nu}^{(i)} = \frac{1}{2}\sum_{m = -\ell}^\ell\left[\nabla_\mu\bar\Phi_{\ell m}^{(i)}\nabla_\nu\Phi_{\ell m}^{(i)} + \nabla_\mu\Phi_{\ell m}^{(i)}\nabla_\nu\bar\Phi_{\ell m}^{(i)}
- g_{\mu\nu}\left( \nabla_\alpha\bar\Phi_{\ell m}^{(i)}\nabla^\alpha\Phi_{\ell m}^{(i)} + \mu^2\bar\Phi_{\ell m}^{(i)}\Phi_{\ell m}^{(i)} \right)\right],
\end{equation}
where $i=1, 2$;
$\bar \Phi_{\ell m}$ denotes the complex conjugate of $\Phi_{\ell m}$ and $\mu$ is the mass of the scalar field particle,  which we assume is the same for all fields. This assumption amounts to consider that all different scalar fields belong to a larger multiplet.

As expected, each complex scalar field satisfies the Klein-Gordon equation:
\begin{eqnarray}\label{eq:kleingordon}
g^{\mu\nu}\nabla_{\mu} \nabla_{\nu}\Phi_{\ell m}^{(i)} -\mu^2\Phi_{\ell m}^{(i)}=0 \ .
\end{eqnarray}

\subsection{Coherent and incoherent states}
As mentioned above, in this work we consider two possible systems.
We shall refer to the first system as coherent state and to the second as incoherent state. 
The functional description of both states is as follows.

i) Coherent states. 
For this case, both $\ell$-boson stars are made up of  the same set of scalar fields. 
Such a scenario is modeled with a single set of fields, initially accumulated at two (essentially) disjoint spatial regions, that is:
\begin{equation}
\Phi_{\ell m}^{(1)} \neq 0 \ , \qquad {\rm and} \qquad 
\Phi_{\ell m}^{(2)} = 0 \ , \qquad
\forall \quad m \ . 
\end{equation}
For this system there are $2\ell+1$  independent fields to describe the binary.
In this scenario 
a single set of fields fills up spacetime, which can nonetheless clump at two different locations, forming two $\ell$-boson stars centered at different positions.

ii) Incoherent states.
Here, each star is  composed by a set of 
$2\ell+1$ fields, being different for each star. This requires to turn on both sets of fields discussed
\begin{equation}
\Phi_{\ell m}^{(1)} \neq 0 \ , \qquad {\rm and} \qquad  \Phi_{\ell m}^{(2)}\neq 0 \ , \qquad \forall \quad m \ .
\end{equation}
Consequently, there are $2(2\ell+1)$  independent fields to describe the binary. Notice, however, that in both systems the interaction between any of the fields is only through gravity.

\section{Numerical implementation}
\label{sec:numerical}

In order to describe the dynamics of the binary it becomes necessary to evolve $2\ell +1$ complex fields for the coherent system and 
$2(2\ell +1)$ complex fields for the incoherent system.
For concreteness we focus in this work on the simplest non-trivial $\ell=1$ case yielding 3 fields for coherent states and 6 for incoherent states.

\subsection{Initial data}
\label{sec:initial_data}

Initial data for the binary system are obtained using a superposition of two isolated $\ell$-boson stars. 
The construction of single isolated $\ell$-boson stars
is described in detail in 
Ref.~\cite{Alcubierre:2018ahf}; here we outline a brief description of the procedure.

The starting point is to consider a static and spherically symmetric spacetime of the form
\begin{equation}\label{eq:3metric_esph}
ds^2 =  -\alpha^2dt^2+
\gamma_{jk}dx^jdx^k = -\alpha^2dt^2+
a^2dr^2+r^2\,d\Omega^2 \ ,   
\end{equation}
where $\alpha$ and $a$ are functions of $r$, $d\Omega^2$ is the line element on the unit 2-sphere 
and the scalar fields that compose the $\ell$-boson stars have a harmonic time dependence
given by~\eqref{eq:scalar_anzats}. According to this assumptions it has been shown in~\cite{Alcubierre:2018ahf} that
even though the scalar field oscillates in time the stress-energy tensor is time independent and
the EKG equations yield static  solutions that are described by the following set of ordinary differential equations:
\begin{equation}
\label{eq:ekgsph_p}
\phi_{\ell}'' = - \phi_{\ell}' \left( \frac{2}{r} + \frac{ \alpha'}{\alpha}
- \frac{a'}{2a} \right)
+ a \phi_{\ell} \left( \mu^2 + \frac{\ell(\ell+1)}{r^2} - \frac{\omega^2}{\alpha^2} \right) \: ,
\end{equation}
\begin{equation}\label{eq:ekgsph_a}
\frac{a'}{a} =  \frac{(1-a)}{r} + 4 \pi r a \left[ \frac{(\phi_{\ell}')^2}{a} 
+ \phi^2_{\ell} \left( \mu^2 + \frac{\ell(\ell+1)}{r^2}
+ \frac{\omega^2}{\alpha^2} \right) \right]  \: , 
\end{equation}
\begin{equation}\label{eq:ekgsph_alpha}
\frac{\alpha'}{\alpha}=  \frac{(a-1)}{r} + \frac{a'}{2a}
-  4 \pi r a {\phi^2_{\ell}} \left( \mu^2 + \frac{\ell(\ell+1)}{r^2} \right) \: ,
\end{equation}
where a prime denotes derivative with respect to r.
By studying the Klein-Gordon equation close the origin $r=0$ one finds that the scalar field behaves as $\phi_{\ell} \sim \phi_0 r^{\ell}$. For a given value of $\phi_0$, and 
demanding that the scalar field has an exponential decay and the metric is Minkowski at infinity, the EKG system  becomes a non-linear eigenvalue problem for the frequency $\omega$.

\begin{figure}
\includegraphics[width=0.45\textwidth]{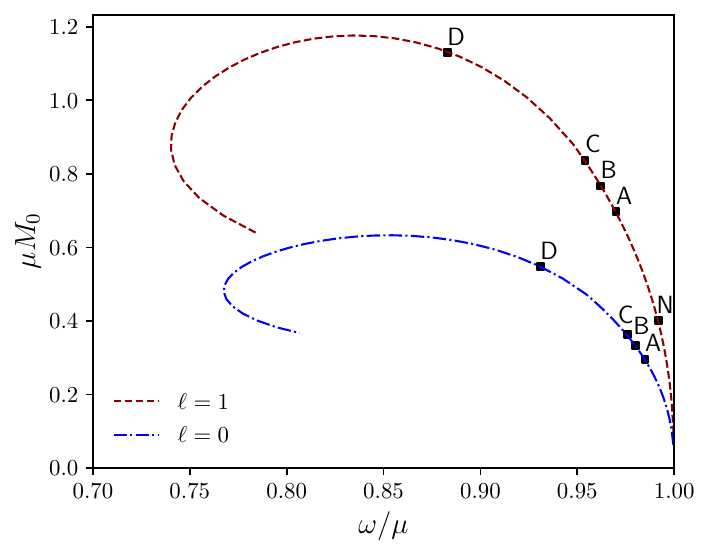}
\includegraphics[width=0.45\textwidth]{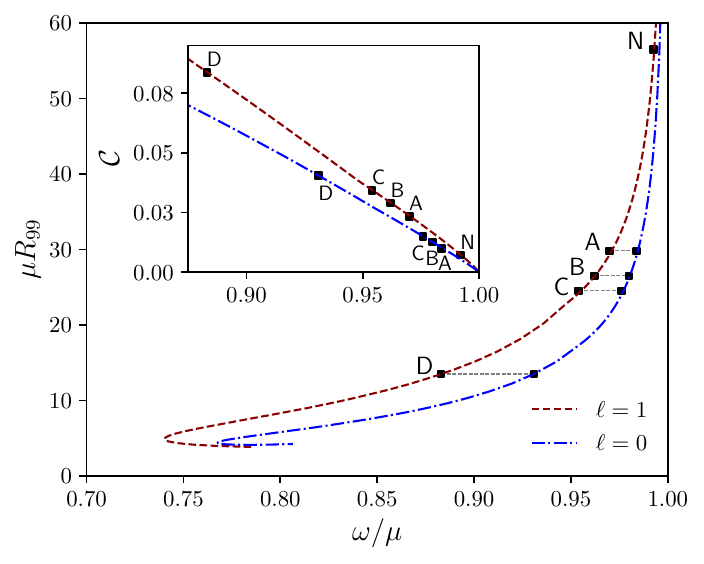}
\caption{ADM mass (left) and radius (right) $vs.$ frequency for static $\ell$-boson stars with $\ell=0,1$. The maximal value of the boson star mass for $\ell=0$ and $\ell=1$ is $\mu M_0=0.63$ and $\mu M_0=1.18$ respectively. The black squares correspond to the solutions used in the head-on collision analysis.
In the inset of the right plot the compactness is shown as a function of the frequency.
} 
\label{fig:mass_vs_frequency}
\end{figure}

The equilibrium configurations are found by integrating numerically Eqs.~(\ref{eq:ekgsph_p})-(\ref{eq:ekgsph_alpha}), considering appropriate boundary conditions, by means of a shooting method using the frequency $\omega$ as the shooting parameter.
The solutions are identified by the value of $\omega$, although for some ranges of $\omega$ there may be more than one solution, defining different branches --see Fig.~\ref{fig:mass_vs_frequency} (left).

$\ell$-boson stars share many features with the single field $\ell=0$ boson stars.  Both exist only for a limited range of frequencies and achieve a maximum 
Arnowitt-Deser-Misner (ADM) mass. Fig.~\ref{fig:mass_vs_frequency} (left)
 displays the mass of $\ell$-boson stars versus $\omega$.
 The maximum mass solution separates stable from unstable configurations as described in \cite{Alcubierre:2021mvs}. In this work we shall only consider configurations in the stable branch. More concretely, the stars we shall use as initial data for the evolution below are marked with a box on the existence curve in Fig.~\ref{fig:mass_vs_frequency}.
 We define the boundary of the star as the radius of the spherical surface that encloses $99\%$ of the mass; this radius is referred to as $R_{99}$ and it is displayed for the solutions in Fig.~\ref{fig:mass_vs_frequency} (right).
 We define the compactness of the stars as $\mathcal{C}:=M/R_{99}$ and it is displayed as an inset in the same panel.
In order to use the infrastructure provided by the Einstein Toolkit we transform the solutions to the usual Cartesian coordinates, $x^{\mu}= (t,r,\vartheta,\varphi)\rightarrow x^{\mu}= (t,x,y,z)$ as
\begin{eqnarray}\label{eq:sp_coordinates}
x = r\cos \varphi \sin\vartheta \ , \qquad
y = r\sin \varphi \sin\vartheta \ ,  \qquad
z = r\cos\vartheta \ .
\end{eqnarray}
In our present investigation we also include configurations
that involve an intermediate rotation in the angles to describe a relative misalignment between the stars, as already discussed above and described in detail below.
The initial data for the binaries used in this work are obtained by a linear superposition of the isolated solutions of two stars given by the spatial metric and a set of scalar fields $\{ \gamma_{jk}^{(-)}, \Phi^{(-)}_{\ell m}\}$ and $\{ \gamma_{jk}^{(+)}, \Phi^{(+)}_{\ell m}\}$ for star 1 and star 2, respectively. 
We shall consider the stars are 
centered at 
$(x_c,0,0)$ and  $(-x_c,0,0)$. We also consider the stars to be initially at rest (at $t=0$).  

In this work we restrict our attention to ground state solutions, for which the amplitudes $\Phi_{\ell m}$ have no nodes. These are the solutions exhibited in Fig.~\ref{fig:mass_vs_frequency}.

We take as initial data for the spatial metric the superposition~\cite{Palenzuela:2006wp,Palenzuela:2007dm,bezares2017final,bezares2018gravitational,sanchis2019head,bezares2022gravitational}: %
\begin{equation}
\gamma_{jk}=\gamma_{jk}^{(+)}(x-x_c,y,z)+\gamma_{jk}^{(-)}(x+x_c,y,z)-\hat\gamma_{jk}(x,y,z).
\end{equation}
where $\hat\gamma_{\mu\nu}$ is the flat spatial metric.
For the scalar fields we need to distinguish between the two systems presented above.
For coherent states we construct the field as
\begin{equation}
\Phi^{(1)}_{m}(t=0,x,y,z)=\Phi^{(+)}_{m}(x-x_c,y,z)+\Phi^{(-)}_{m}(x+x_c,y,z)    \ ,
\end{equation}
and $\Phi^{(2)}_{m}(t,x,y,z)=0$. 
Whereas for incoherent states, the field is constructed as
\begin{eqnarray}
\Phi^{(1)}_{m}(t=0,x,y,z)=\Phi^{(+)}_{m}(x-x_c,y,z) \  , \nonumber\\ 
\Phi^{(2)}_{m}(t=0,x,y,z)=\Phi^{(-)}_{m}(x+x_c,y,z)
\ .
\end{eqnarray}

Due to the nonlinearity of Einstein equations these initial data introduce constraint violations. How this effect has been tracked and controlled throughout the simulations is discussed in Appendix \ref{app:codevalid}. 

By means of the transformation \eqref{eq:sp_coordinates},
the stress-energy tensor of each star is defined with respect to a Cartesian frame $(x,y,z)$. For the initial data, instead of taking the spherical harmonics of each star defined with respect to the same Cartesian frame,  
we may consider a relative misalignment of the two corresponding Cartesian frames, by performing a rigid rotation. This allows a more general scenario, in which the stars have arbitrary initial orientations. As we shall see, this has an interesting impact on the dynamics during the merger and in the resulting configuration.

\begin{figure}
\begin{center} 
\rule{0.3in}{0in}$\mathbb{I}$\rule{1.2in}{0in} $R_z(\pi)$ \rule{1.0in}{0in} $R_y(\pi)$
\\
\rotatebox{90}{\quad$|Y_{1,-1}(\vartheta,\varphi)|$}
\includegraphics[width=3.3cm,height=2.6cm]{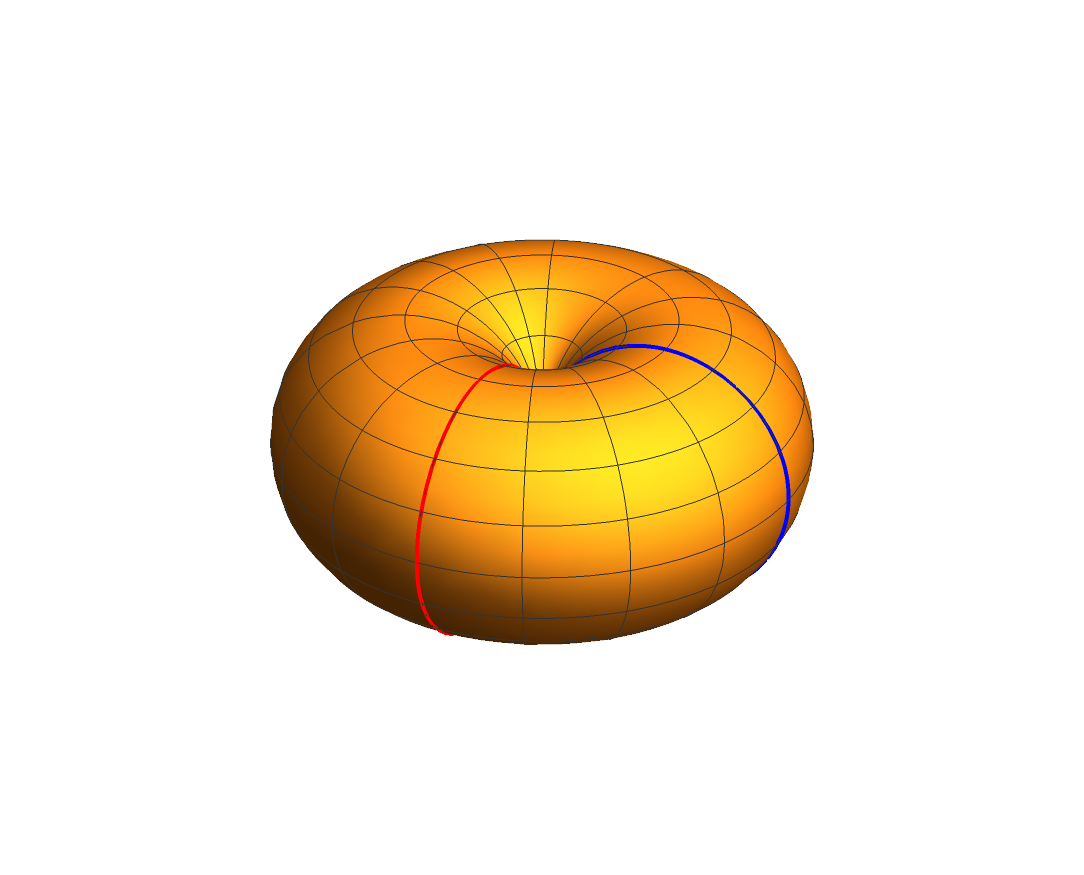}
\includegraphics[width=3.3cm,height=2.6cm]{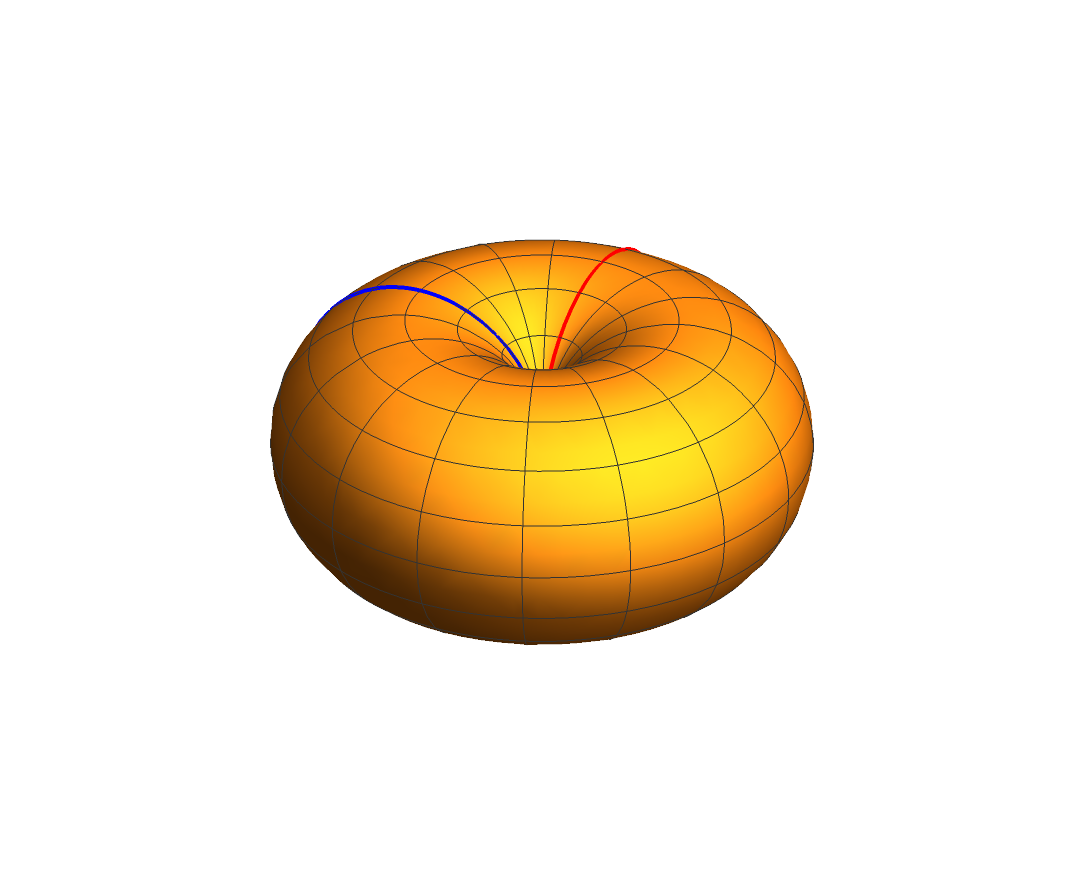}
\includegraphics[width=3.3cm,height=2.6cm]{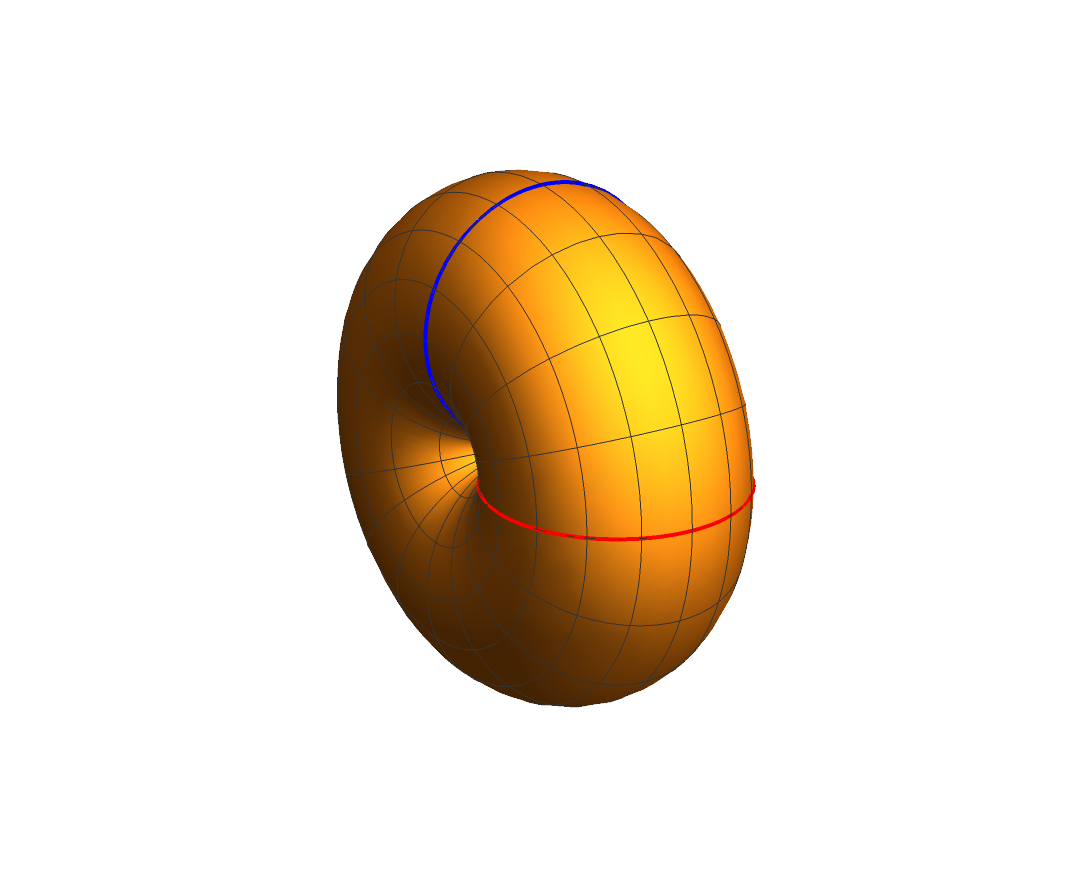}
\\
\rotatebox{90}{\;\quad$|Y_{1,0}(\vartheta,\varphi)|$}\;\;
\includegraphics[width=3.2cm,height=2.6cm]{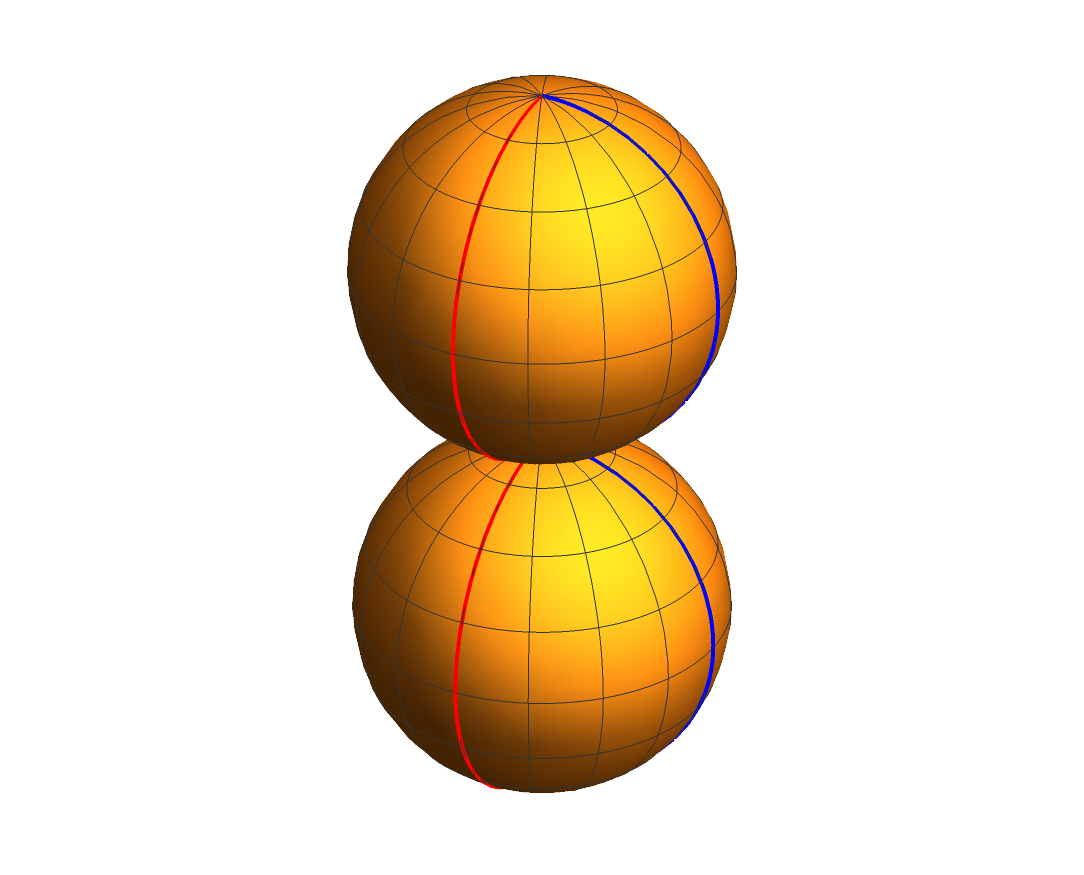} 
\includegraphics[width=3.2cm,height=2.6cm]{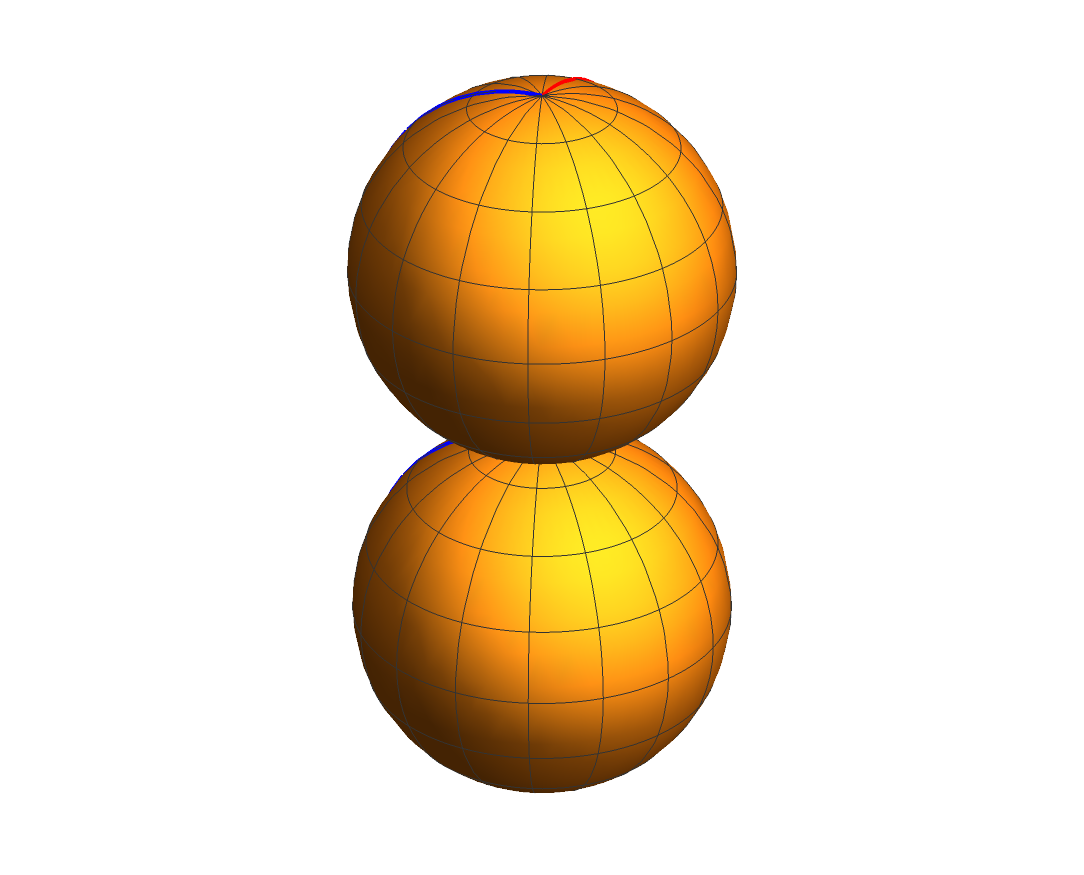}
\includegraphics[width=3.2cm,height=2.6cm]{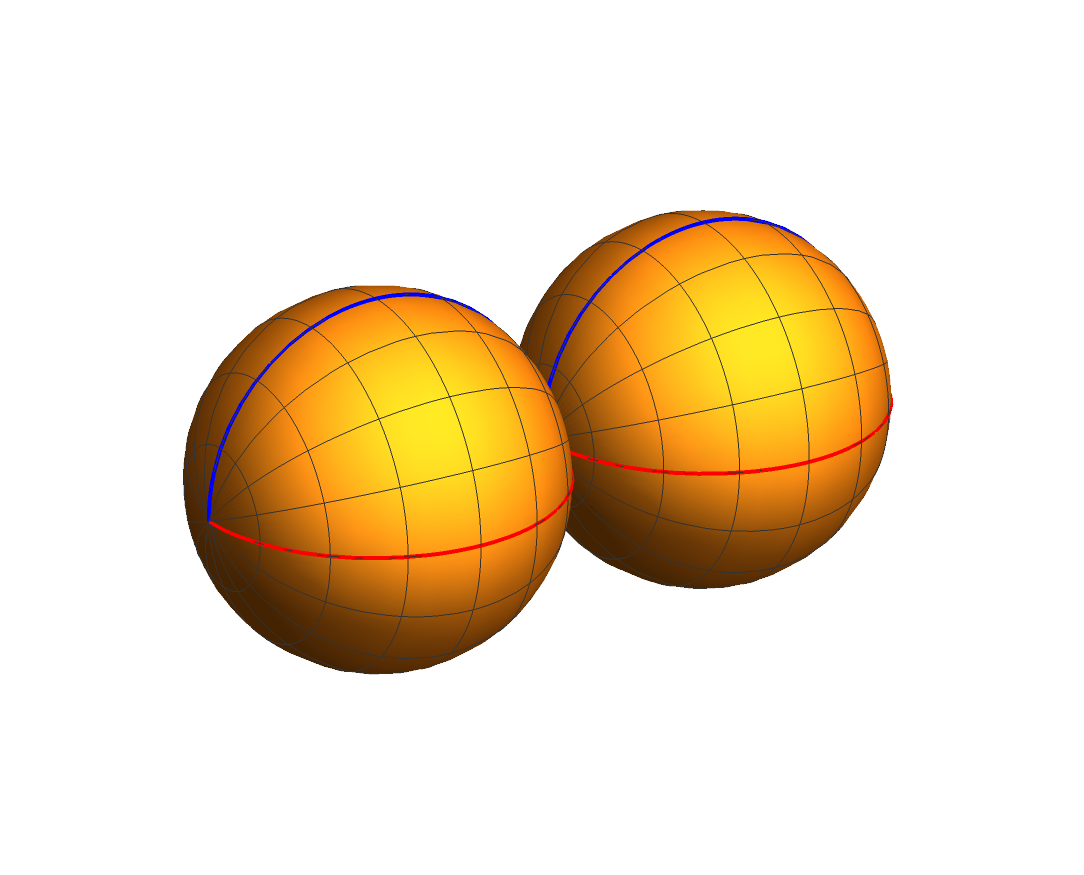}
\\
\rotatebox{90}{\;\;\quad$|Y_{1,1}(\vartheta,\varphi)|$}
\includegraphics[width=3.3cm,height=2.6cm]{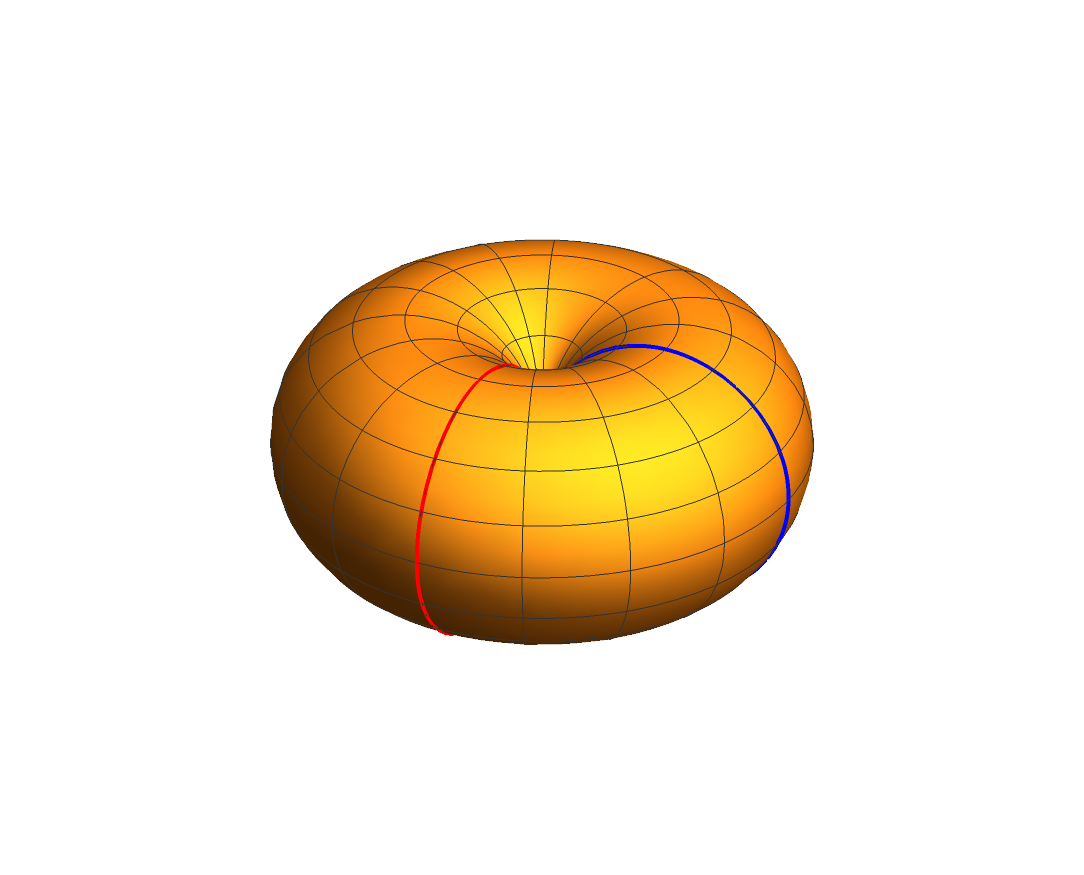}
\includegraphics[width=3.3cm,height=2.6cm]{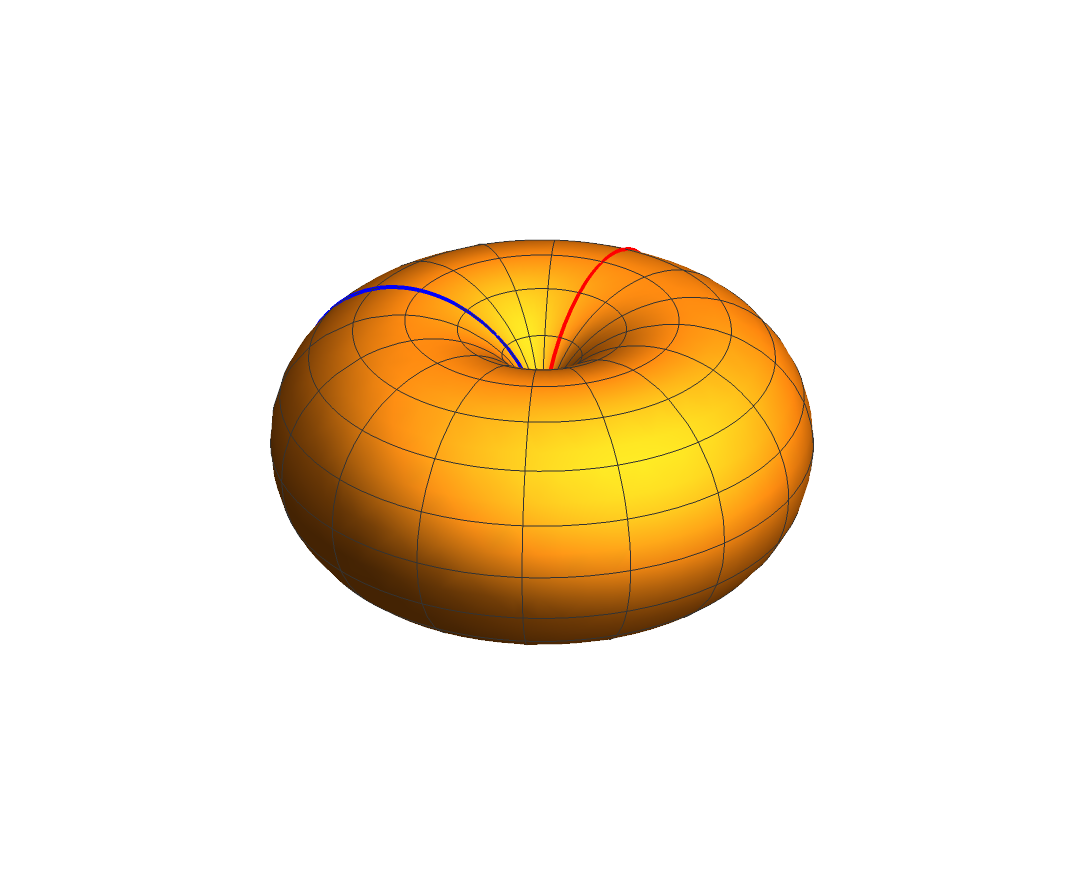}
\includegraphics[width=3.3cm,height=2.6cm]{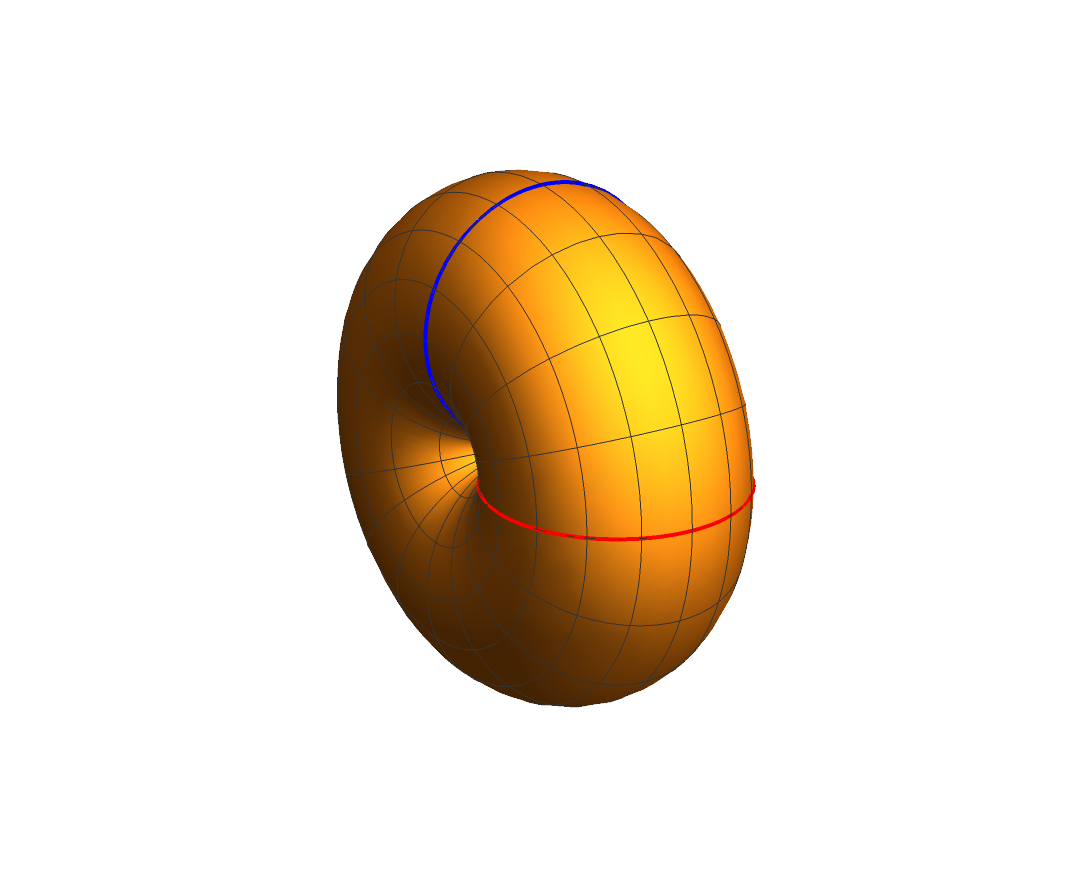}
\\
 \caption{Amplitude of spherical harmonics for the modes $m=-1$, $m=0$, $m=1$, that compose a boson star with $\ell=1$.
The first column is taken as a reference for the usual alignment; the second column shows the effect of a rotation $R_z(\pi)$; and the third column represents a rotation $R_y(\pi)$. As a reference, we have drawn a a red and a blue line on the surfaces. This is particularly useful when comparing the first two columns, which would otherwise look identical. 
}
 \label{fig:harmonic}
\end{center}
 \end{figure}

In order to model such non aligned stars we define an intermediate set of coordinates  
$x'=r\cos\varphi\sin\vartheta,\ y'=r\sin\varphi\sin\vartheta,\ z'=r\cos\vartheta$, 
and perform a transformation of the form
$\mathbf{x}=R_i(\delta)\mathbf{x}'$, 
where $i=x,y,z$ are the Einstein Toolkit Cartesian coordinates and $R_i(\delta)$ is the rotation matrix for an angle $\delta$ around the $i$-axis.
The effect of the rotations $R_z(\pi)$, $R_z(\pi/2)$ and $R_y(\pi)$ can be visualized in Fig. \ref{fig:harmonic}.

\subsection{Evolution Scheme}
\label{sec:evolution}

Our numerical simulations are performed using the open source Einstein Toolkit infrastructure \cite{EinsteinToolkit:web}.
The Einstein equations are integrated in time
using the Baumgarte-Shapiro-Shibata-Nakamura (BSSN) formulation \cite{Baumgarte:1998te}. We use the Method of Lines of the \textsc{MoL} thorn to solve the equations, via
a fourth order Runge-Kutta scheme provided by the \textsc{McLachlan} thorn~\cite{brown2009turduckening,reisswig2011gravitational}. 
The scalar field evolution code has been recently employed to study the stability of $\ell$-boson stars \cite{Jaramillo:2020rsv} and it is described in more detail in that reference.

All the evolutions were made using the 1+log time slicing condition for the lapse $\alpha$, 
and the {\it Gamma-driver} condition for the shift $\beta^{i}$ \cite{Alcubierre08a}.
We have used the thorn \textsc{AHFinder}~\cite{diener2003new} to follow the formation of an apparent horizon (AH) during the evolution. We have also monitored the Hamiltonian and momentum
constraints~\cite{Gourgoulhon2012} to check the convergence during the evolution.

The numerical grid 
consists of nested cubes with six refinement levels and it is handled using the fixed mesh refinement grid hierarchy included in the \textsc{Carpet} arrangements. 
 
In all simulations the stars have their centers placed at $\mu x_c=25$.
The computational domain consists of a box with  $\mu x_{\text{max}}=500=\mu y_{\text{max}}=\mu z_{\text{max}}$, with a grid structure of six refinement levels.
The waveform extraction is made at $\mu r=100, 120$. 
The spatial domain of the refinement levels is $\{500,50,50,25,25,10\}$. We set the resolution on the finest
level to $\mu\{\Delta x, \Delta y, \Delta z\}= 0.3125$. This corresponds to at least 86 points across the initial stars' diameter (defined as $2R_{99}$). The choice of a fixed mesh refinement is mainly motivated by simplicity. The last refinement level that do not contain the stars is included in order to have enough resolution in case a black hole forms. Refinement levels of the grid can produce reflections. However, we do not see any significant impact of the crossing of the mesh boundaries on the dynamics and on the determination of the different quantities (mass, particle number, Hamiltonian constraint). The Cartesian grid can induce a $m=4$ perturbation (see, for instance,~\cite{espino2019dynamical,di2020dynamical}).

\section{Head-on Collision Dynamics}
\label{sec:head-on}
Once we have constructed the initial data, we allow the systems to evolve freely while monitoring the Hamiltonian and momentum constraints to check the accuracy of the results. 
As described above, both the evolution and the number of scalar fields are different depending on the system. For the coherent states we evolve 3 complex fields whereas for the incoherent states we evolve 6 fields. 
We have found that the main differences between the two systems occur near the plunge and this is reflected in the emitted GW signal and in the posterior outcome of the plunge.
We quantify the differences using some analysis quantities, such as the mass of the final configuration, the number of particles, and the GW signal.

\subsection{Analysis Quantities}

Starting from the static superposition initial data, 
the stars approach each other and eventually collide. 
In some of the scenarios we have established that the final object after the merger is a black hole.
To diagnose its appearance in the evolution we use the \textsc{AHFinder} thorn
and then compute the mass of the black hole through the apparent horizon area $A$, using the relation  $M_{BH}=\sqrt{A}/16\pi$, which is valid for a Schwarzschild black hole. The use of the Schwarzschild relation relies on no-hair theorems for static spherically symmetric black holes, ruling out final equilibrium configurations with scalar hair, even with a harmonic time dependence~\cite{Pena:1997cy}. Such no-hair theorems are, however, circumvented for spinning black holes~\cite{Herdeiro:2014goa}. 

We focus primarily on the configurations for which the end state has no horizon and the rest mass density is non zero. 
For the final time reached in our evolution, the final remnant is a localized, perturbed distribution of the different complex scalar fields. However, it is not possible to determine whether that object corresponds to an $\ell$-boson star. We describe some of its properties in the next sections. 

For the total gravitational mass of localized solutions we use the
Komar integral.

\begin{equation}\label{eq:KM}
M=-\int_{\Sigma} \sqrt{\gamma}d^3x \,\alpha\,(2 T^t{}_t-T^{\mu}{}_\mu) \ ,
\end{equation}
where  $\Sigma$  is  a  spacelike  slice  extending  up  to  spatial  infinity, $\gamma$ is the determinant of the 3-metric induced on that slice and $\alpha$ is the lapse function.
To describe the end state of the collision we compute the Noether charge associated with the total bosonic number of particles $N$, which is defined as
\begin{equation}\label{eq:nump_tot}
N= \int_{\Sigma} \sqrt{\gamma} d^3 x\, \alpha \,g^{0\mu}j_\mu^{(i)}   \ ,
\end{equation}
where $j_\mu^{(i)} = \sum_{m}\frac{i}{2} ({\bar\Phi^{(i)}}_m\nabla_\mu\Phi_m^{(i)}-\Phi_m^{(i)}\nabla_\mu{\bar\Phi^{(i)}}_m)$. 
This conserved current is
associated  with  the global invariance of the theory under the action of  a  $U(1)$  group for each field.
We will use this quantity to
classify the remnant of the merger. Note that the Noether charge can also be computed for \textit{each} field.
The integrals for the Komar mass~\eqref{eq:KM} and the number of particles~\eqref{eq:nump_tot} are performed in the entire numerical grid on each time step.

We   monitor  
the energy density of matter as $\rho=n^\mu n^\nu T_{\mu\nu}$
 during the evolution as a measure of the energy left after the collision, where $n^\mu$ is the unitary normal vector to $\Sigma$.
 
Furthermore, in order to determine the deviations from spherical symmetry of the post merger configuration, we compute the moments of inertia $I_{xx}$, $I_{yy}$ and $I_{zz}$ defined by
\begin{equation}\label{eq:inertia}
I_{x^i x^i} = \int_\Sigma \sqrt{\gamma} d^3 x\, \alpha \, \rho(r^2-{x^i}^2) \ .
\end{equation}

\subsection{Gravitational Waves}

Gravitational radiation is extracted from the numerical simulations by computing the Newman-Penrose  scalar  $\Psi_4 = C_{\alpha\beta\gamma\delta}k^{\alpha}m^{\beta}k^{\delta}m^{\gamma}$, where $C_{\alpha\beta\gamma\delta}$ is the Weyl tensor and $k$ and $m$ are two vectors of the null Kinnersley tetrad \cite{Beetle:2004wu,Campanelli:2005ia}. 
Far from the source $\Psi_4$ represents an outward propagating wave and has been used as a measure of the gravitational radiation emitted during the merger of compact objects.
In order to analyze the structure of the radiated waves it is convenient to decompose the signal in -2 spin weighted spherical harmonics as
\begin{eqnarray}
\Psi_4(t,r,\vartheta,\varphi) = \sum_{l, k}\Psi_4^{l,k}(t,r) {}_{-2}Y_{l,  k}(\vartheta,\varphi) \ .
\end{eqnarray}
According to the \emph{peeling theorem}
the leading order decay of $\Psi_4$ is $1/r$ \cite{Wald84}. We use this fact to check the accuracy in the computation of the gravitational waveforms. 
Our description will focus directly on the strongly dominant component $l=2$, $k=2$.

\section{Results}
\label{sec:results}

We have investigated five different cases of the $\ell$-boson star datasets corresponding to different values of compactness. All the initial configurations correspond to two stars of the same type, localized  
on the stable branch, and with total mass (sum of the two stars' mass) that is \textit{larger} than the maximal mass of the model. The stars have  radii (as defined before) ranging from $\mu R_{99}=56.5$ (model N), to $\mu R_{99}=13.5$ (model D), corresponding to a compactness ranging from $\mathcal{C} = 0.0074$ to $\mathcal{C}=0.0838$, respectively.
In this section, we extensively discuss the dynamics for the various cases.

The physical attributes of the initial data and some properties of the end state of the collisions are summarized in Table \ref{tab:R_coherent} (for coherent states) and Table \ref{tab:R_incoherent} (for incoherent states). Note that the properties of the initial data coincide in both tables, but not the end state. The models CHl1N and INl1N that appear in both tables will appear only on the analyzes related to the study of the final state. Concerning the initial data the tables show the typical size $R_{99}$, the frequency, the Komar mass of each star $M_0$, the number of particles in one of the fields of each star ($i.e.$ 1/3 of the total number of particles of each star, as described in Appendix \ref{app:number}) and the compactness. 

We define the time of collision $t_c$, as the time at which the spheres given by   
$R_{99}$ of each star intersect. Furthermore, 
in order to give a simple estimation of the object's size after the merger, we still use the aforementioned definition of $R_{99}$ as the radius of a sphere containing 99\% of the total mass, even when the object is not spherically symmetric. 
The center of such sphere is set at the center of mass, $x=y=z=0$.
\begin{table}
\begin{tabular}{|c|c|c|c|c|c||c|c|c|}
\hline
\textit{Coh.}&\multicolumn{8}{ c |}{$\ell=1$}{}\\
\hline
Model &$\mu R_{99}$ &$\omega/\mu$&$\mu M_0$&$\mu^2 N_0$&$\mathcal{C}$&Rem.&$\mu R\sim$&$\mu t_c$\\
\hline
CHl1N&56.5 &0.990&0.418&0.419&0.0074&BS&
-&0\\
CHl1A&31.5 &0.970&0.697&0.703&0.0221&BS&
24&0\\
CHl1B&27.5&0.962&0.775&0.784&0.0282 &BS
&22&70\\
CHl1C&24.7&0.954&0.837&0.849&0.0391&BH&$\mu r_{\rm AH}=2.1$&150\\
CHl1D&13.5  &0.883&1.13&1.17&0.0838&BH&$\mu r_{\rm AH}=3.4$ &240\\
\hline
\end{tabular}
\caption{Coherent cases for $\ell=1$. $R_{99}$ is the radius that contains 99\% of the mass of the star. $\omega$ is the frequency, $M_0$ is the mass of each star, $N_0$ is the number of particles of each star and $\mathcal{C}$ is the compactness. 
The end state of the simulation can be a localized boson configuration (BS) or a black hole (BH).
$R$ is the radius that encloses 99\% of the mass for scalar field remnant at $\mu t=2500$
while for black holes it labels the radius of the apparent horizon.
}
\label{tab:R_coherent}
\end{table}
\begin{table}
\begin{tabular}{|c|c|c|c|c|c||c|c|c|}
\hline
\textit{Incoh.}&\multicolumn{8}{ c |}{$\ell=1$}{}\\
\hline
Model &$\mu R_{99}$ &$\omega /\mu$&$\mu M_0$&$\mu^2 N_0$&$\mathcal{C}$&Rem.&$\mu R\sim$&$\mu t_c$\\
\hline
INl1N&56.5 &0.990&0.418&0.419&0.0074&BS&
-&0\\
INl1A&31.5&0.970&0.697&0.703&0.0221&BS
&30&0\\
INl1B&27.5&0.962&0.775&0.784&0.0282&BS
&23&0\\
INl1C&24.7&0.954&0.837&0.849&0.0391&BH& $\mu r_{\rm AH}={2.2}$&0\\
INl1D&13.5&0.883&1.13&1.17&0.0838&BH& $\mu r_{\rm AH}=3.6$ &230\\
\hline
\end{tabular}
\caption{Same as Table \ref{tab:R_coherent} for  incoherent states.}
\label{tab:R_incoherent}
\end{table}

\subsection{Aligned stars}\label{sec:aligned}

\begin{figure}[ht!]
\begin{center}
\begin{center}
\includegraphics[scale=0.64, angle=0]{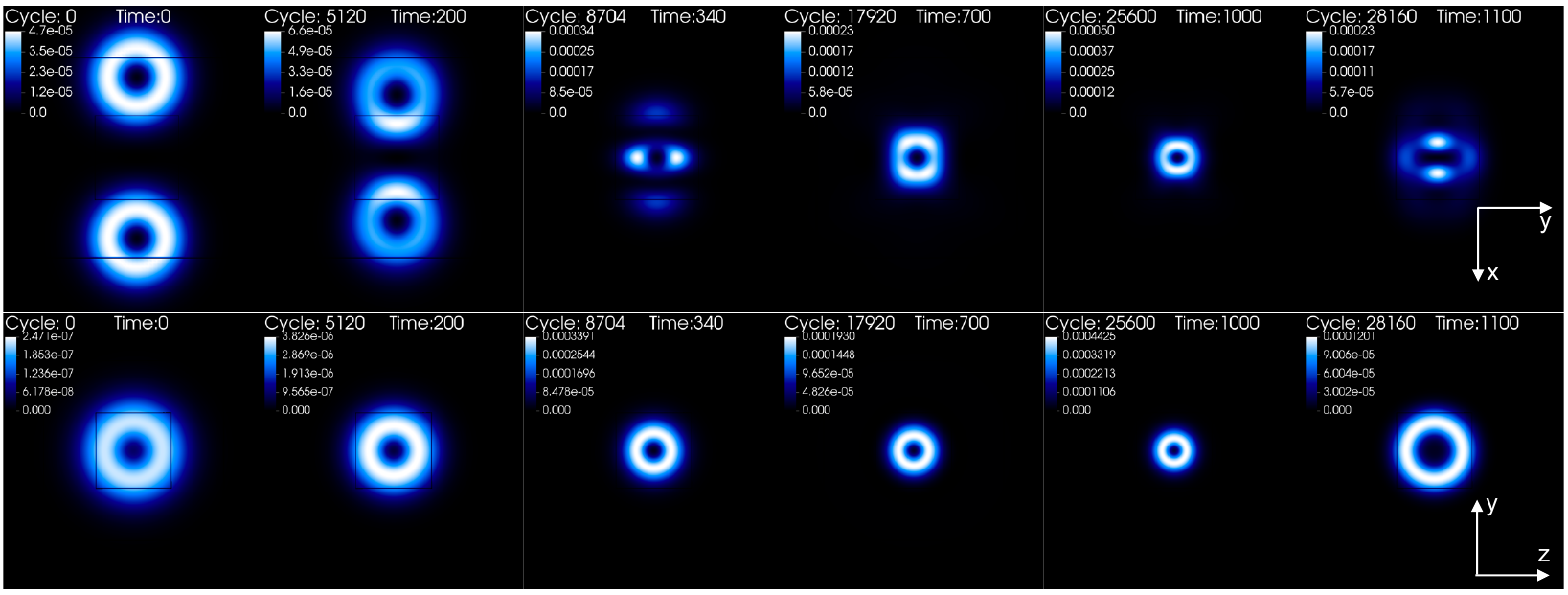}
\end{center}
\caption{Aligned evolutions of a coherent state. Snapshots 
of the scalar field energy density, $\rho/\mu^2$, for model CHl1B on the $z=0$ (top) and $x=0$ (bottom) planes. 
The maximum density of the final configuration for these models is  one order of magnitude larger than the progenitors. Time is given in units of the scalar field mass, $\mu$. Cycle refers to the iteration number.
}
\label{fig:rho_comp_c}
\end{center}
\end{figure}

\begin{figure}[ht!]
\begin{center}
\begin{center}
\includegraphics[scale=0.64, angle=0]{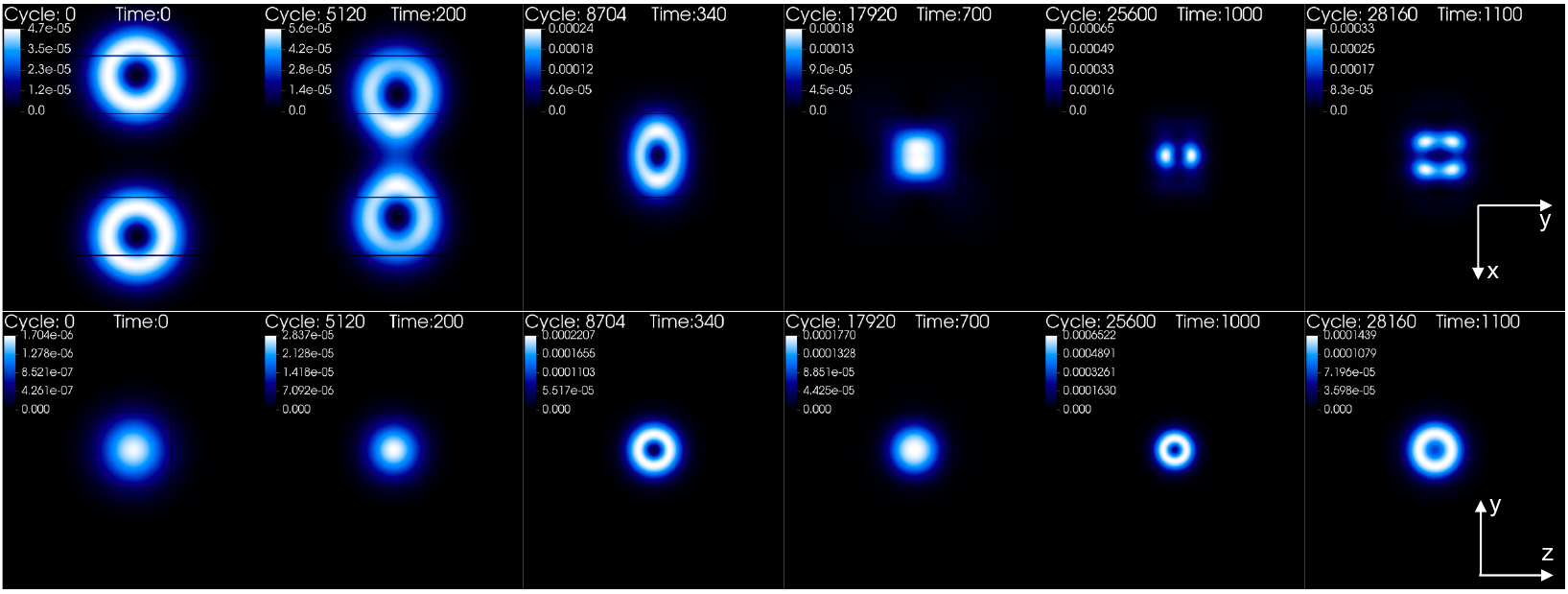}
\end{center}
\caption{Aligned evolutions of an incoherent state. Snapshots of the scalar field energy density, $\rho/\mu^2$, for model INl1B on the $z=0$ (top) and $x=0$ (bottom) planes. 
The maximum density of the final configuration for these models is  one order of magnitude larger than the progenitors.
}
\label{fig:rho_comp_i}
\end{center}
\end{figure}

As discussed above, $\ell$-boson stars are spherically symmetric at the level of the total energy-momentum tensor but have an internal 
frame of reference with respect to the different modes. The phrase ``aligned stars'' refers to both stars having the same orientation. The properties of the remnants are presented in the last columns of Table~\ref{tab:R_coherent} for coherent states and the last columns of Table \ref{tab:R_incoherent} for incoherent states. In particular, the mass of the merger remnant is computed using the Komar integral~\eqref{eq:KM}. Also, some snapshots of the scalar field energy density $\rho$, during the coalescence are displayed in Fig.~\ref{fig:rho_comp_c} and \ref{fig:rho_comp_i}.
Fig.~\ref{fig:rho_comp_c} exhibits the evolution of a coherent state (model CHl1B) whereas Fig.~\ref{fig:rho_comp_i} exhibits the evolution of an incoherent state (model INl1B).
\begin{figure}
\begin{center}
\includegraphics[width=0.4\textwidth]{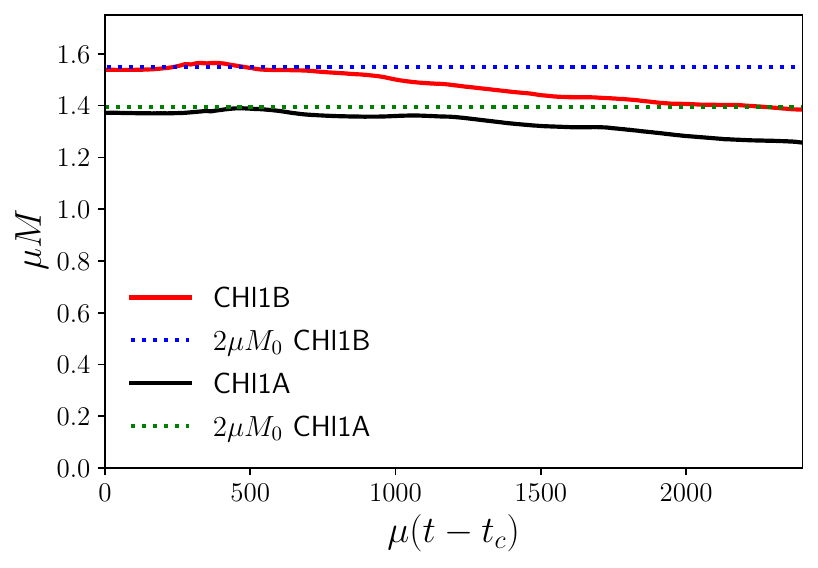}
\includegraphics[width=0.4\textwidth]{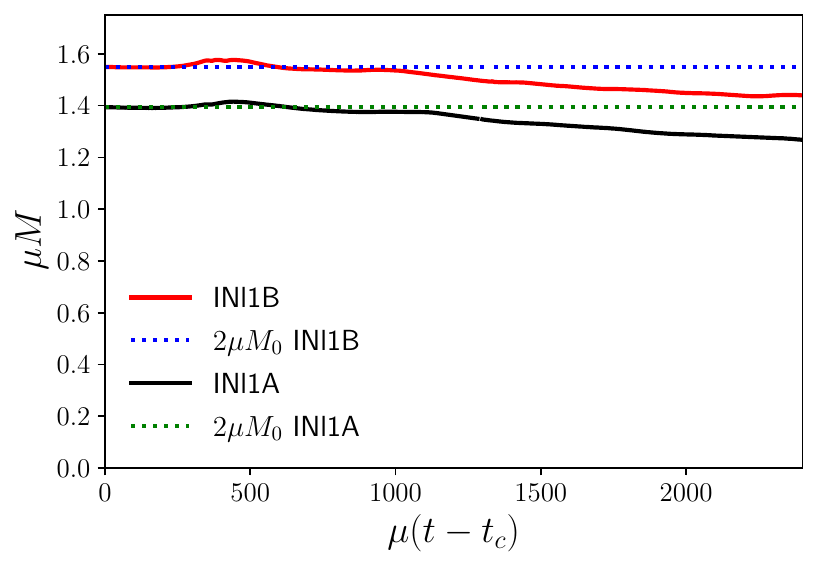}
\end{center}
\caption{Aligned evolutions. Collision remnant mass, for coherent states, models CH1A and CH1B (left) and for incoherent states, models IN1A and IN1B (right). 
} 
\label{fig:mass_bindinge}
\end{figure}

Despite the fact that in all cases the initial mass of the system is above the maximal mass of the model, we observe two qualitatively distinct behaviours. For the most massive and compact models (C and D) a black hole is formed. But for the less compact models (A and B) neither a black hole forms during the simulation time, nor the field disperses away after the collision. In such cases, a bound scalar field configuration remains after the merger, albeit the system is still evolving at the end of the simulation. 
Whereas one cannot state with certainty the final outcome of the system,  the results suggest that the system does tend to an equilibrium lump of scalar fields. This is an asymptotic process, in which the gravitational cooling mechanism plays a key role to allow relaxation, by slowly decreasing the mass of the system. This slow mass ejection can be observed in Fig.~\ref{fig:mass_bindinge}, where we plot  the total mass for models CHl1A, CHl1B, and INl1A, INl1B as a function of time after the merger. 
The horizontal dotted lines indicate the value of twice the initial mass of each star.
From Figs.~\ref{fig:rho_comp_c} and \ref{fig:rho_comp_i} one can observe that the early phase of the encounter is qualitatively similar for both states; but once the objects make contact, the dynamics becomes complex and model dependent. Nonetheless, we observe the same separation between black hole formation and scalar remnant formation, regardless of the states being coherent or incoherent.
The natural question is: what is the system evolving towards, when no black hole forms? In particular, since $\ell$-boson stars involve \textit{precisely} equal amplitudes of the scalar fields involved, these collisions test how \textit{fine-tuned} these configurations are, and if such a destructive process readily creates an asymmetry between the different fields.

\begin{figure}
\includegraphics[height=6.5cm]{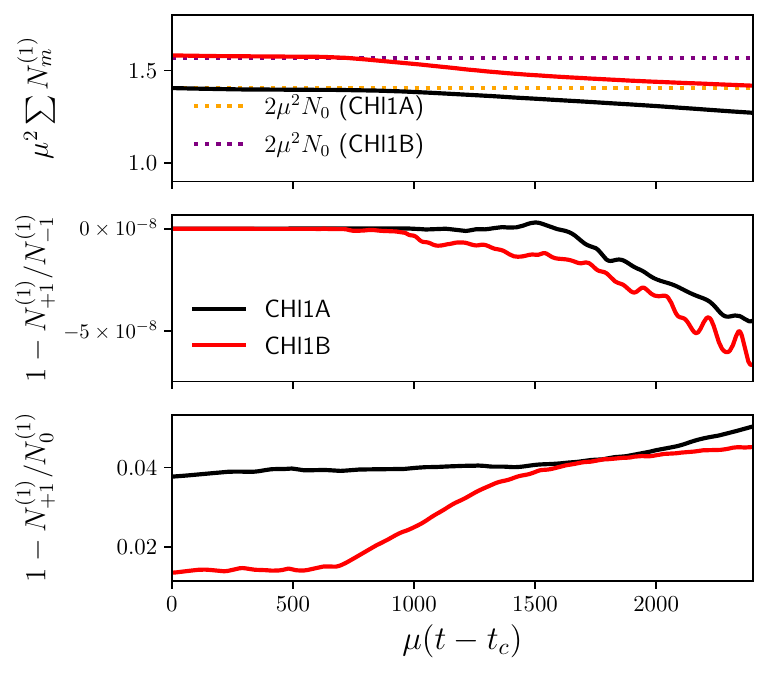}
\caption{Aligned evolutions.
(Top) Total number of particles of the remnant of the merger of coherent states CH1A, CH1B. The dotted lines represent twice the initial number of particles present in the binary. 
(Middle) Ratio between the number of particles associated to the
fields $\Phi_m^{(1)}$ with $m=+1$ and $m=-1$. It remains constant at unity. (Bottom)  Ratio between the number of particles associated to the
fields $\Phi_m^{(1)}$ with $m=+1$ and $m=0$.
At $t=0$ this quantity is not exactly unity, this is effect of the superposition and appears only in this (coherent) case. During  the coalescence a small percentage of their initial value is lost.
There is no leaking or exchange of particles of individual fields during the evolution. 
}
\label{fig:num_particles_co}
\end{figure}

\begin{figure}
\begin{center}
\includegraphics[height=6.5cm]{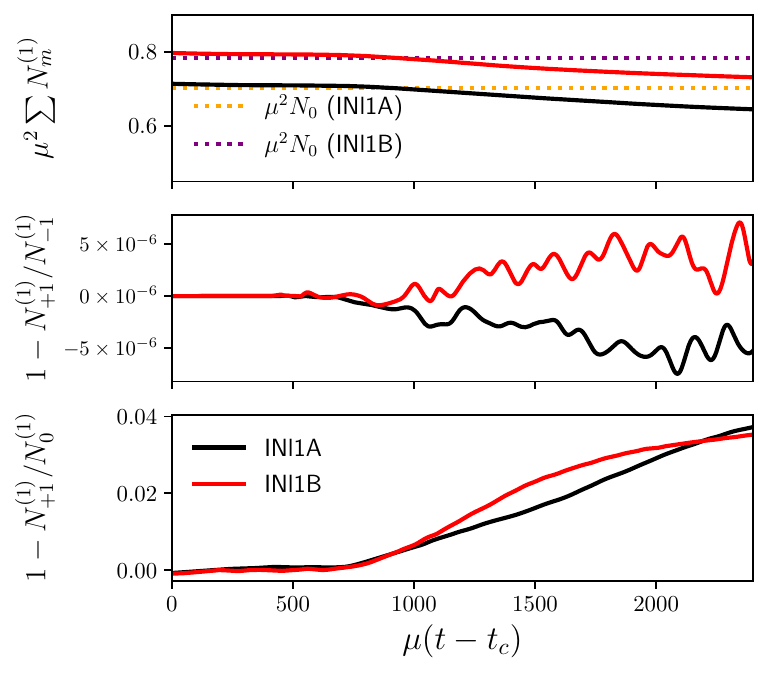}\quad
\includegraphics[height=6.5cm]{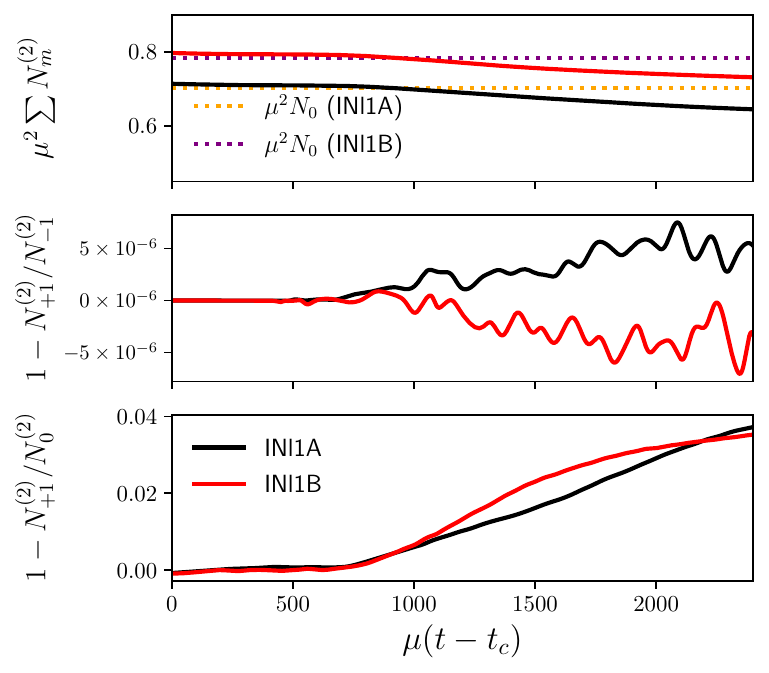}
\end{center}\caption{
Same as Fig.~\ref{fig:num_particles_co}, but now for incoherent states INI1A and INI1B.
}
\label{fig:num_particles_in}
\end{figure}

To tackle this question one may look at the Noether charge, or particle number, in each field. First, we observe that (as discussed in Appendix \ref{app:number}) for a single $\ell$-boson star the total number of particles, $N$, is equally distributed amongst each field. Denoting the Noether charge of each field as $N_m$, then $N_m=\frac{N}{2\ell+1}$. Thus, we can investigate if such equipartition of the Noether charge remains after the collision. 
To this end we have computed the time evolution of the number of particles associated to each scalar field. This is displayed for the coherent models CHl1A, CHl1B in Fig. \ref{fig:num_particles_co} and for the incoherent models  INl1A, INl1B in  Fig. \ref{fig:num_particles_in}.
One observes that, independently of the state, after the merger there is a monotonic loss of Noether charge, consistent with the gravitational cooling process. Moreover, this scalar field leaking is rather democratic; the equipartition between the $m\pm 1$ modes is kept to high accuracy, and the relative difference between the $m=0$ and $m=\pm1$ modes is kept to a few percent. 
Thus, none of the individual Noether charges (corresponding to the individual fields) suffers dramatic preferential losses, even though there is a slight suppression of the $m=\pm 1$ modes with respect to the $m=0$ mode.

\begin{figure}[ht!]
\begin{center}
\begin{center}
\includegraphics[scale=0.65, angle=0]{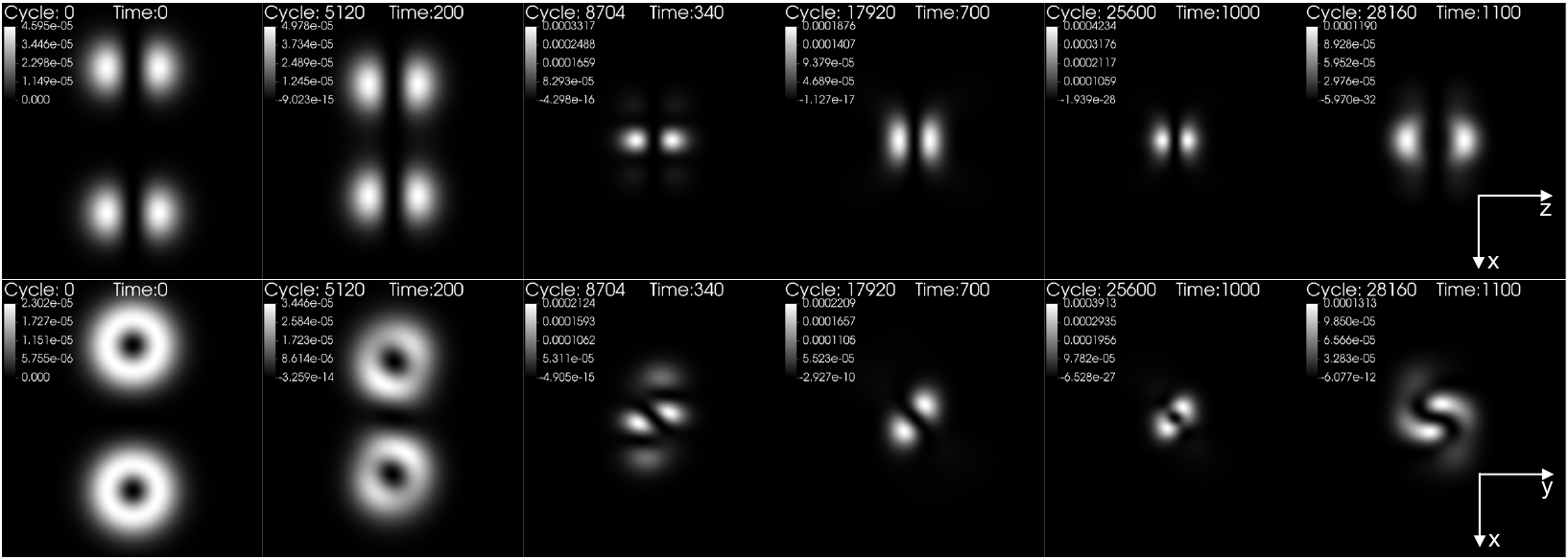}
\end{center}
\caption{Aligned evolutions. Individual currents $j_m /\mu$ for model CHl1B. Top panel: $m=0$ in the $y=0$ plane. Bottom panel: $m=+1$ in the $z=0$ plane.
}
\label{fig:J_comp_ci}
\end{center}
\end{figure}

\begin{figure}[ht!]
\begin{center}
\includegraphics[scale=0.45]{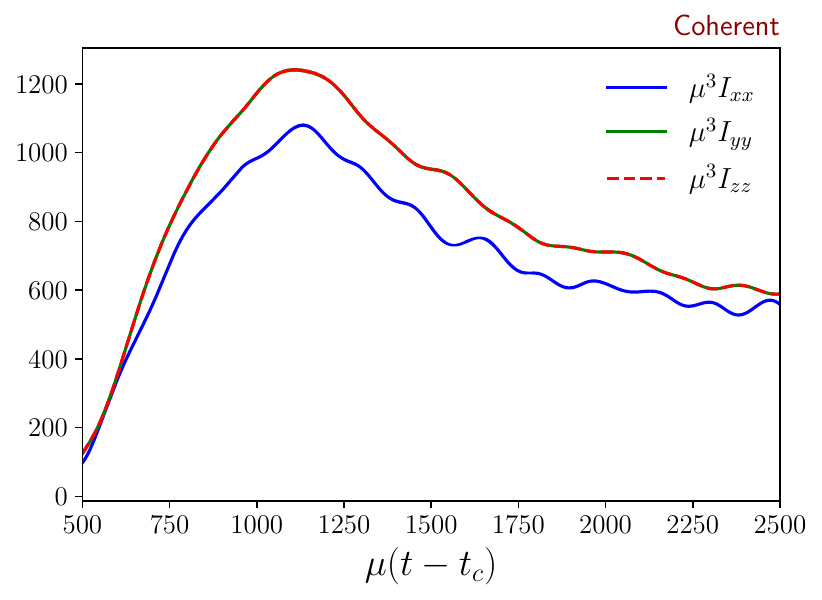}
\includegraphics[scale=0.45]{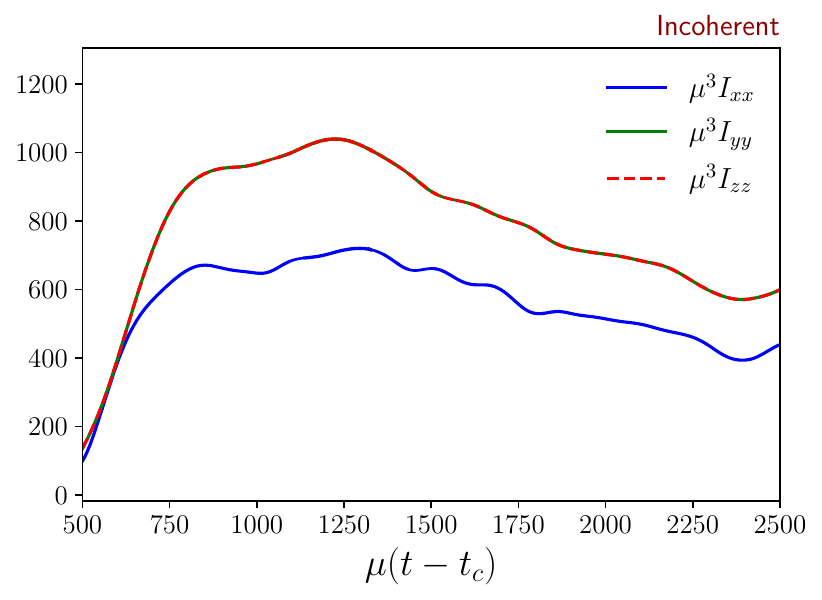}
\caption{Aligned evolutions. Moments of inertia of the post merger configuration as defined in Eq. \eqref{eq:inertia} for the models CHl1B (left panel) and INl1B (right panel). Note that the curves of $I_{yy}$ and $I_{zz}$ coincide perfectly in both cases. 
In the incoherent case, the configuration does not seem to tend to a spherically symmetric distribution of scalar field. Despite the fact that the moments of inertia used in Eq. 4.3 are gauge dependent they represent a good indicator of the presence of nonlinear stability, as has been described in \cite{Saijo:2000qt}.
}
\label{fig:intertia_a}
\end{center}
\end{figure}

To gain further insight into the post-merger dynamics, in Fig.~\ref{fig:J_comp_ci} we have looked at the spatial distribution of the Noether charge during the evolution for model CHl1B. One observes that the $m=0$ field keeps its morphology throughout the evolution. The dynamics arises from the $m=\pm 1$ modes, which yield a non-trivial dynamics as the two lumps become superimposed, in particular altering their morphology. This  can be corroborated in the left panel of Fig.~\ref{fig:intertia_a}, where a difference can be observed between the moments of inertia $I_{xx}$ and $I_{yy}\approx I_{zz}$ after the collision. In the incoherent case, the asymmetry is also evident, as can be seen in the right panel of Fig.~\ref{fig:intertia_a}. 
In both cases a difference between $I_{xx}$ and $I_{yy},I_{zz}$ remains after $\mu t=2500$.
The above analysis is consistent with the remnants being undergoing a relaxation process towards an equilibrium configuration which is a multi-field, but not necessarily spherically symmetric, boson star~\cite{Jaramillo:2020rsv,Sanchis-Gual:2021edp}. In particular, for the coherent state, it could be close to an $\ell$-boson star,  by virtue of the evolution of the moment of inertia, Fig. \ref{fig:intertia_a} (left panel). See also~\cite{Bernal:2009zy,bezares2018gravitational} for the discussion of multistate boson stars in the context of the standard $\ell=0$ boson stars.

\begin{figure}[ht!]
\begin{center}
\includegraphics[width=0.6\textwidth]{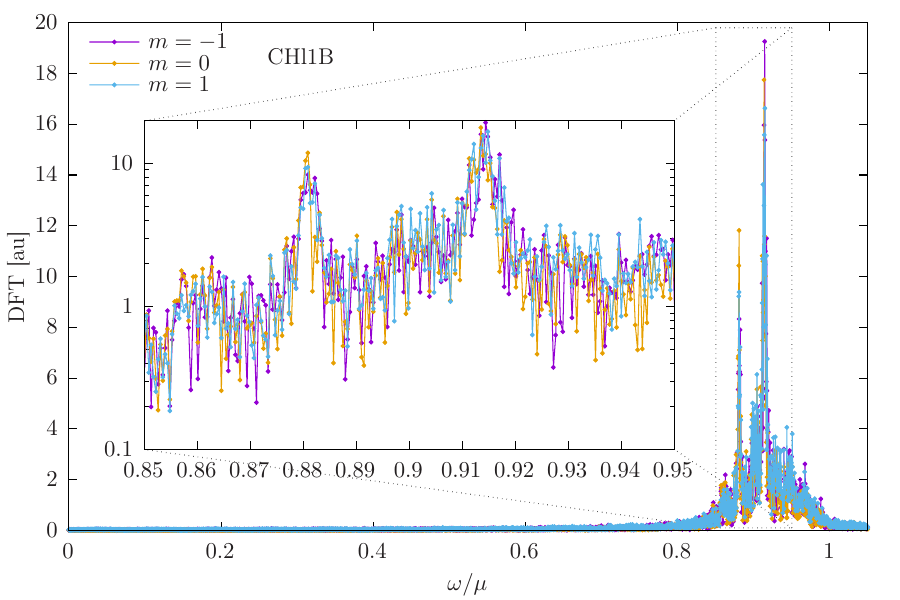}
\caption{DFT in arbitrary units (au) $vs.$ frequency. The inner figure shows a zoom-in of he most relevant region in logarithmic scale.}
\label{fig:dftl}
\end{center}
\end{figure}

We also look at the evolution of the oscillation frequency of the scalar field(s).
During the merger process,  this frequency changes, which can be investigated by performing a spectral analysis. To this end we have evaluated the discrete Fourier transform (DFT) in time of each scalar field component, $m=-1$, $0$ and $1$. For more accuracy, we have actually evaluated the DFT at five different points and then calculated the average.
Figure~\ref{fig:dftl} shows the DFT for the case CHl1B. 
We notice that more than one frequency peak arises, correspond to several bound modes with  $\omega<\mu$. Moreover, at least up to the resolution we could achieve, the spectra of the three $m$ components coincide, not only in the peaks locations, but also in their amplitude. From this viewpoint, therefore, the different modes remain in synchrony, as for the spectrum of the equilibrium states, the $\ell$-boson stars. The key difference is, however, that the $\ell$-boson stars spectra would only show a single peak.

To better resolve the excited modes, one would have to increase the frequency resolution. This resolution, however, 
is proportional to the inverse of the evolution time, which implies a high computational cost  to improve the frequency resolution. In fact, the case shown here was run up to $\mu t=14231$, a much longer  evolution than for every other model in this work. For this reason we have not considered the spectrum of the other models.

\begin{figure}[ht!]
\begin{center}
\includegraphics[scale=0.4]{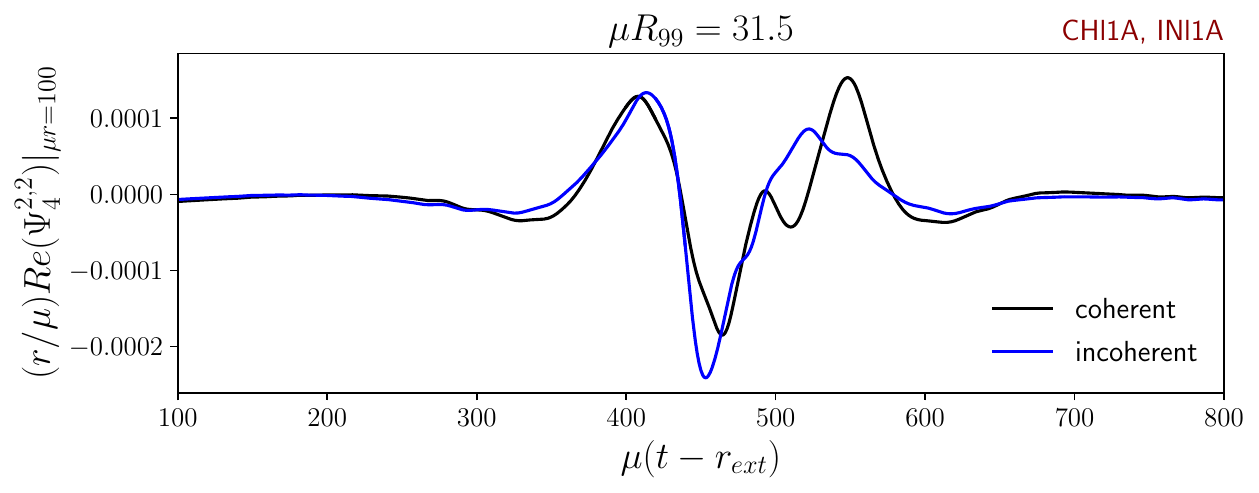}
\includegraphics[scale=0.4]{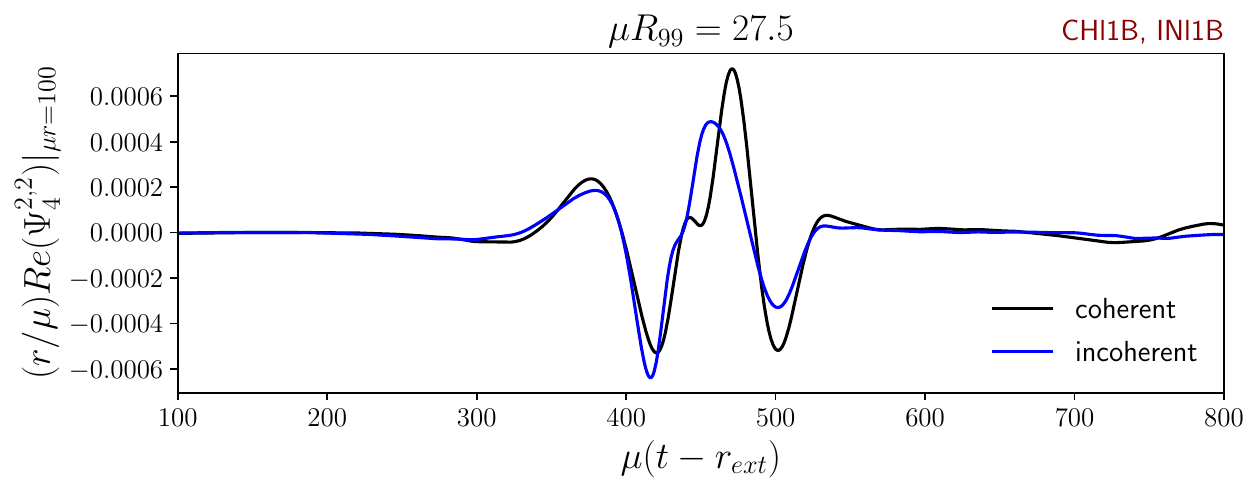}\\
\includegraphics[scale=0.4]{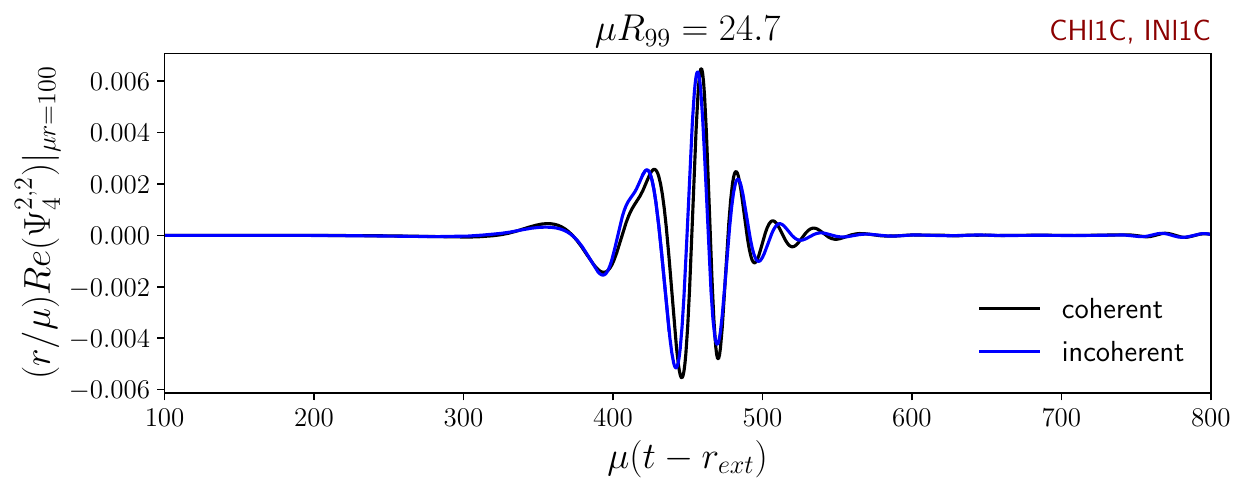}
\includegraphics[scale=0.4]{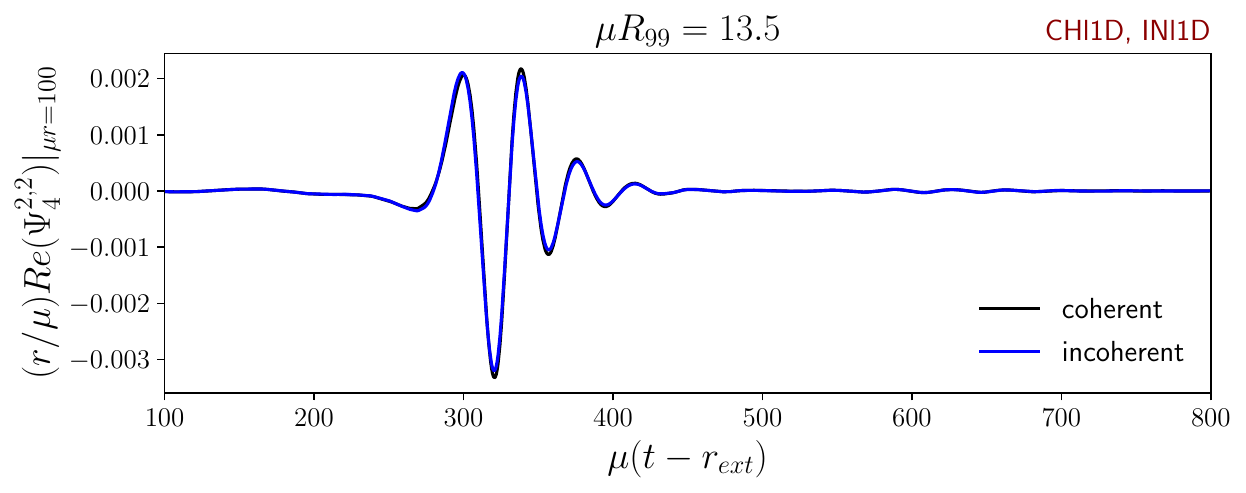}
\caption{Waveforms for $\ell=1$ binaries described in Table \ref{tab:R_coherent} and Table \ref{tab:R_incoherent}. The extraction radius is $\mu r_{\rm ext}=100$.}
\label{fig:GWl1}
\end{center}
\end{figure}

Let us now turn our attention to the GWs emission. In Fig.~\ref{fig:GWl1} we show the real part of the dominant quadrupolar ($l=2$, $k=2$) mode  of  the 
Newman-Penrose scalar $\Psi_4$ as a function of time for both coherent states listed in Table \ref{tab:R_coherent}
and for incoherent sates listed in Table \ref{tab:R_incoherent}.
The signal is extracted at $\mu r_{\rm ext}= 100$ 
and
we have scaled the amplitude with a factor $r$ to better capture the asymptotic behaviour. The horizontal axis has been shifted 
as $t\rightarrow t-r_{\text{ext}}$.
For models CHl1A, INl1A with relatively low compactness ($\mathcal{C}=0.0221$) the GWs signal display some differences between the two. This differences are still visible for models with larger compactness CHl1B and INl1B ($\mathcal{C}=0.0282$).
But for models CHl1C and INl1C the final remnant collapses and the differences in the signal are almost negligible. The waveform indeed resembles the signal of the collision of two black holes.
Finally for the most compact binary CHl1D and INl1D, the final product of the merger forms a black hole and the waveforms are identical for both states. 

\begin{figure}[ht!]
\begin{center}
\includegraphics[width=0.4\textwidth]{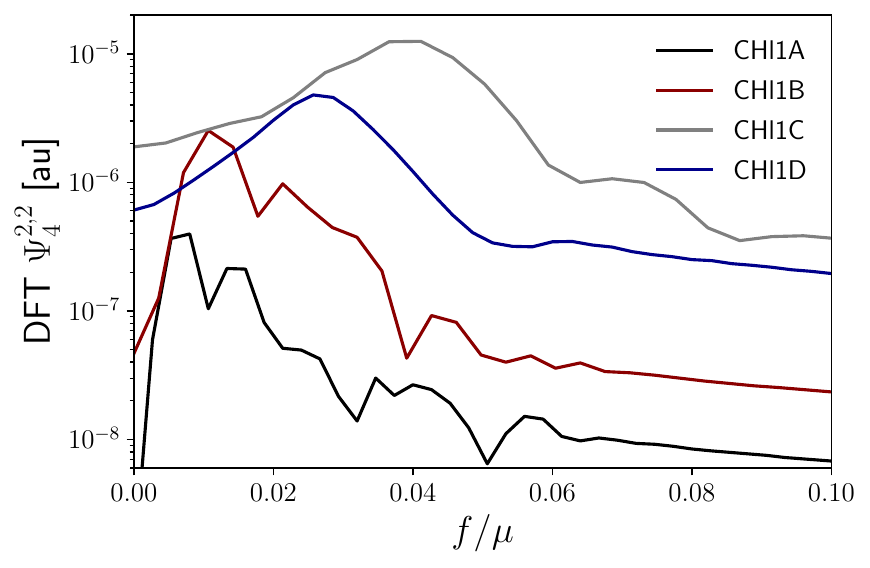}
\includegraphics[width=0.4\textwidth]{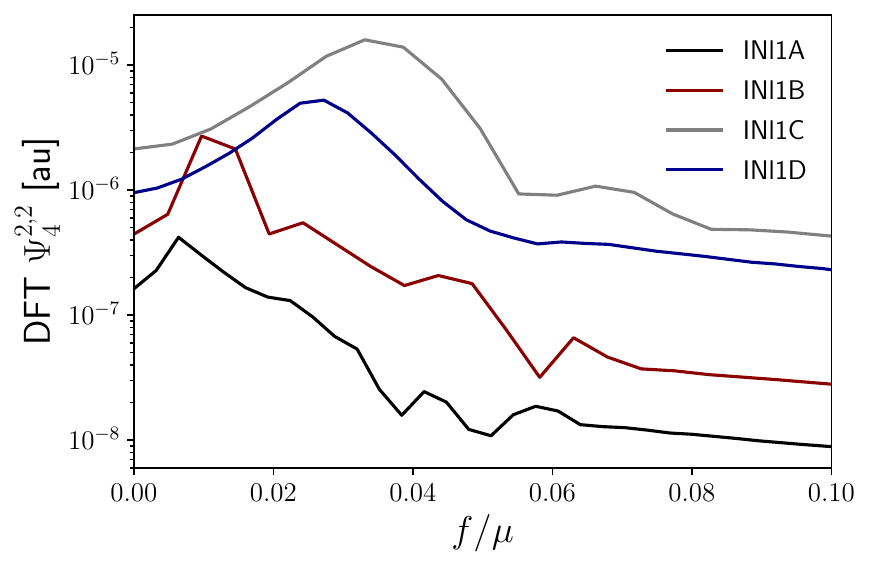}
\caption{Fourier transform of the mode $l=2$, $k=2$ of $\Psi_4$ for coherent (left) and incoherent (right) models.}
\label{fig:fGW}
\end{center}
\end{figure}

\begin{figure}[ht!]
\begin{center}
\includegraphics[scale=0.4]{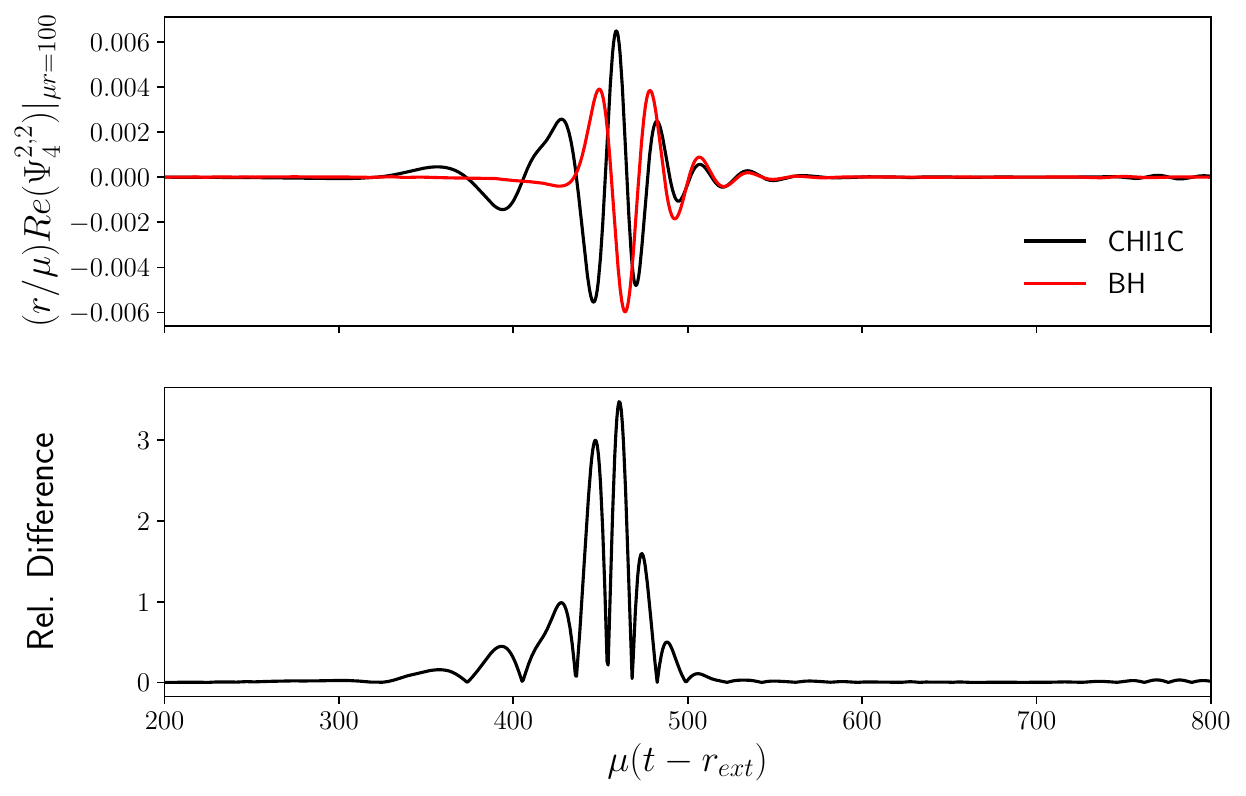}
\includegraphics[scale=0.4]{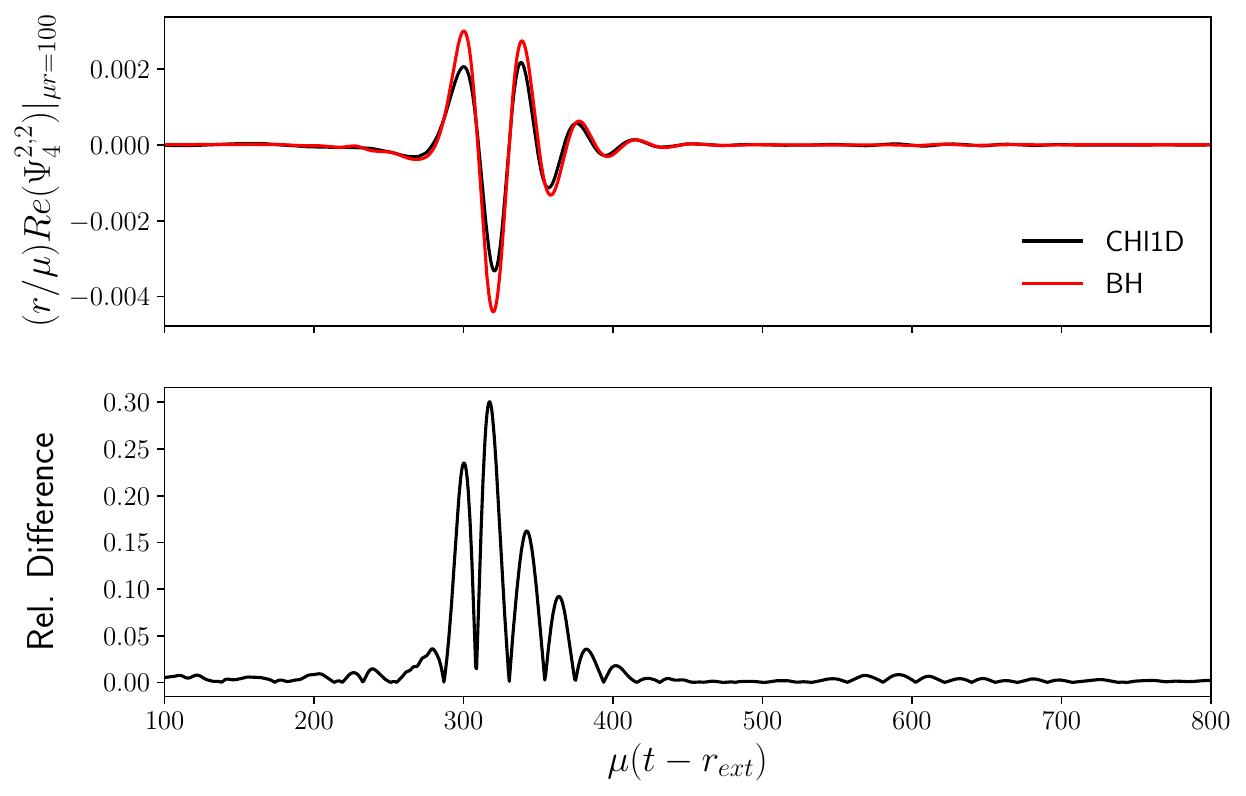}
\caption{Difference between the black hole end state evolutions (CHl1C, CHl1D, INl1C and INl1D) and the corresponding same mass black hole head-on collision. Rel. Difference $=|\Psi^{2,2}_4-\Psi^{2,2}_{4\mathrm{, BH}}|/\mathrm{max}(\Psi^{2,2}_{4\mathrm{, BH}})$.}
\label{fig:GWBH}
\end{center}
\end{figure}

To quantify the differences between waveforms as well as the similarities with the head-on collision of black holes, mentioned in previous statements, we perform the following pair of analyses. First, we calculate the Fourier transform of $\Psi_4$ to obtain the frequencies of the gravitational signal for the different models. The results are shown in Fig.~\ref{fig:fGW}, from which it can be confirmed that the main differences in the signal occur in relation to the compactness of the stars; the peak frequency increases by a factor of 4 between model A and C, whether the superposition is coherent or incoherent. Second, we compute the difference between the collapsing models, CHl1C, CHl1D, INl1C and INl1D. To this end, we compare in each case with the gravitational signal from a head-on collision of equal mass black holes hole with the same total mass as the boson star system. In Fig.~\ref{fig:GWBH} these gravitational waveforms are shown. The more compact model, CHl1D overlaps better with the black hole signal, in particular after the collapse, at the ringdown phase. For the model CHl1C the GW signal differentiates from the black hole signal qualitatively and quantitatively at the first stage, however it matches roughly the frequency and phase of the black hole collision at ringdown. The model CHl1C is close to the BH/BS remnant limit (see Table~\ref{tab:R_incoherent}), it involves more complex dynamics before collapse, which is imprinted in the GW at early times.

The more compact the binary, the larger the amplitude of the gravitational waveform; for the most compact objects the amplitude is almost one order of magnitude larger than for their less compact counterparts. This result applies for both coherent and incoherent states.
This phenomenon is related with the size and geometry of 
the binary.
For $\ell$-boson stars the maximum energy density is not located at the geometrical center of the star; the shape of these objects is more like a spherical shell. As the value of $\ell$ increases, the maximum of the density tends to the external boundary leaving a region with almost zero density at the geometrical center. For very compact binaries the radius $R_{99}$ is considerably smaller than for the less compact ones and the encounter is more violent producing a stronger gravitational signal.

\begin{figure}[ht!]
\begin{center}
\includegraphics[scale=0.4]{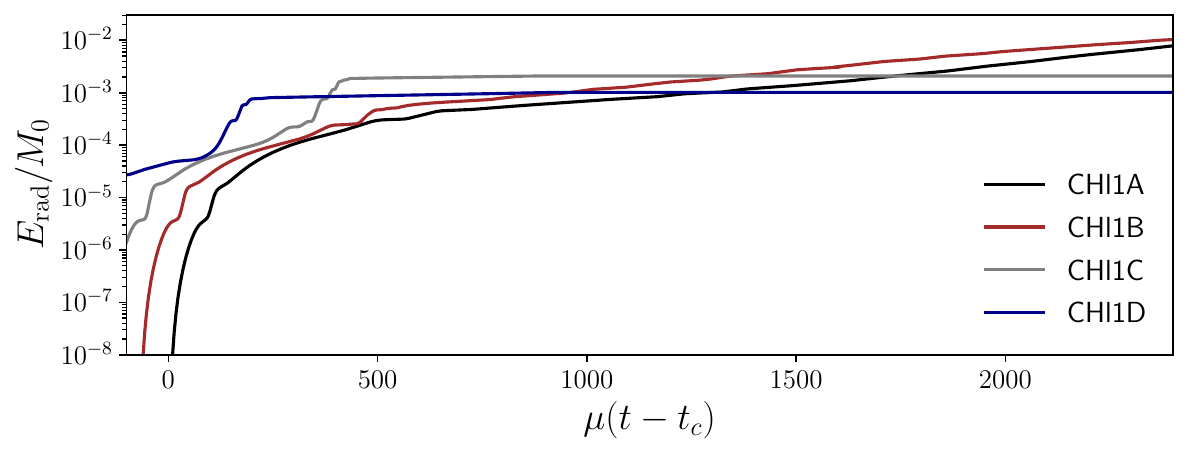}
\includegraphics[scale=0.4]{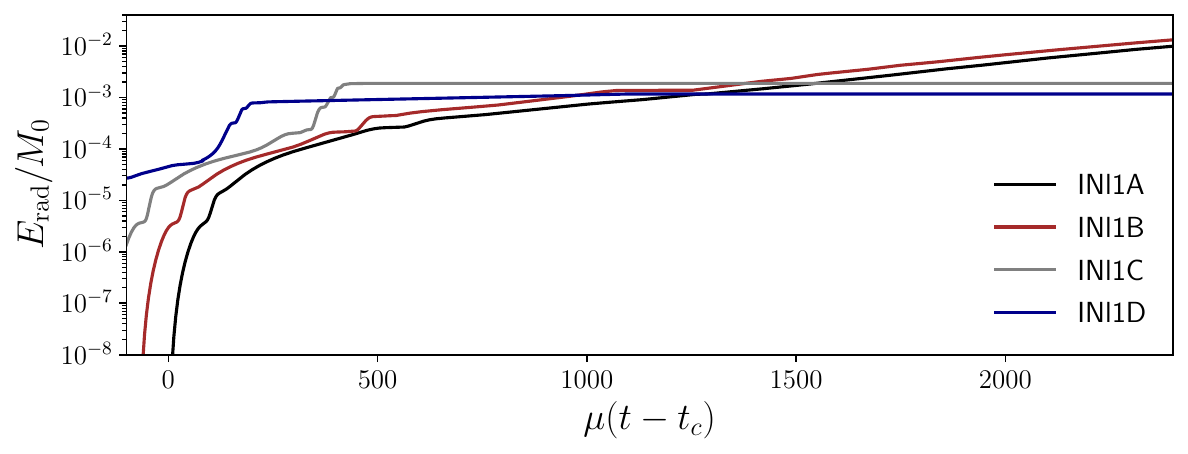}
\caption{Total energy radiated by the gravitational wave.}
\label{fig:E_GW}
\end{center}
\end{figure}

\begin{figure}
\begin{center}
\includegraphics[width=0.4\textwidth]{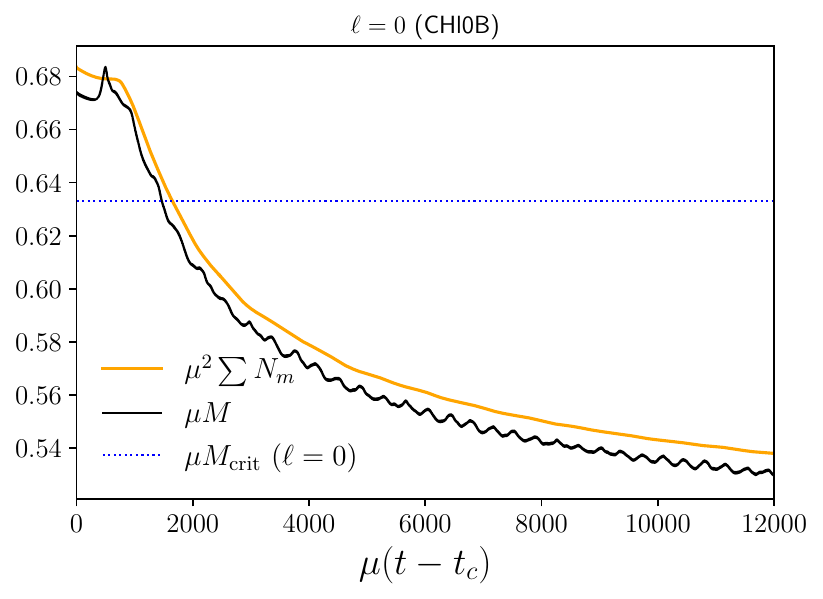}
\includegraphics[width=0.377\textwidth]{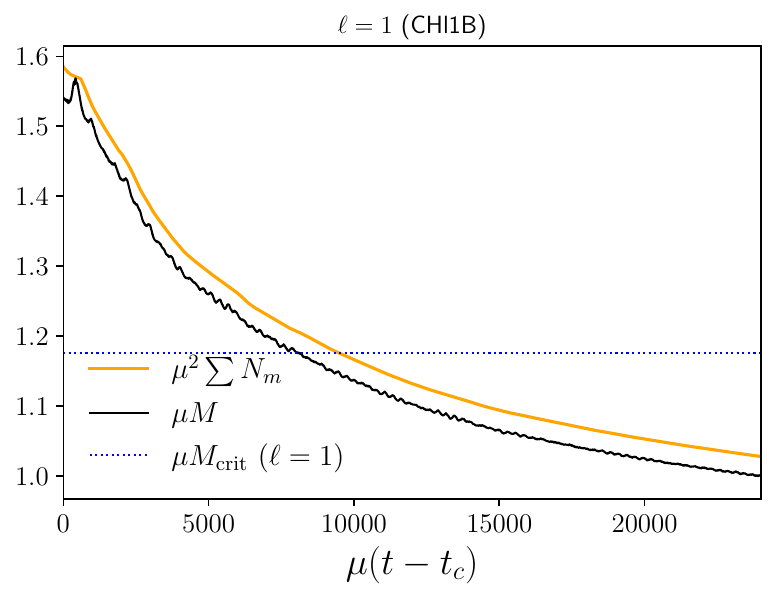}
\end{center}
\caption{Mass and total number of particles for long simulations of the mergers of the $\ell=0$ CHl0B (Left) and $\ell=1$ CHl1B (Right). 
} 
\label{fig:mass_l0l1}
\end{figure}

To further investigate the post-merger behavior, let us compare it for the $\ell=1$ model B and for the  merger of two standard $\ell=0$ boson stars, with the same radius $R_{99}$ as the $\ell=1$ model. The $\ell=0$ corresponding collisions were already discussed in \cite{Palenzuela:2006wp}, where it was shown that
the remnant approaches another $\ell=0$ boson star. In Fig. \ref{fig:mass_l0l1} (left) we can see that the mass decreases in the collision process, reaching below the maximal mass allowed for $\ell=0$ boson stars.
The mass loss can be roughly divided into two phases. There is an initial slower decrease, partly due to GW emission, followed by a larger decrease rate, due to scalar field emission, $i.e.$ gravitational cooling, that remains throughout the evolution.  
The remnant oscillates around a spherical distribution; in fact an equilibrium boson star with $\ell=0$, with the oscillation amplitude decreasing with time.
We stress the main relaxation process to attain the equilibrium state is gravitational cooling. This can be confirmed in Fig.~\ref{fig:E_GW}, where the energy radiated by GW is shown; when compared to the total mass loss, in Fig.~\ref{fig:mass_bindinge}, it is seen that the emitted gravitational radiation is at least one order of magnitude smaller than the the total mass loss, in all the cases.

The right panel of Fig. \ref{fig:mass_l0l1} exhibits the analogous behavior for the collision of the  stars with $\ell=1$. Similarities with the previous case are observable: there is a mass loss (together with a decrease in the number of particles), with a GW emission component occurring initially and an expelling of scalar field  throughout the entire evolution. In this case, however, it is not clear that the final state oscillates around a spherically symmetric configuration, say an $\ell$-boson star. It can be mentioned that the system, as in the previous case, keeps oscillating, but the oscillation amplitudes decrease more slowly in this case. Also binary configurations consisting of boson stars with masses smaller than the CHl1A model have been explored, of particular interest are those cases where the total mass of the system is smaller than the maximum mass of the $\ell=1$ family of solutions, $M_0^{\mathrm{max}}=1.176$, the results for these cases are very similar to those shown in Fig.~\ref{fig:mass_l0l1}, that is, highly dynamic behavior at late times and a continuous and very slow loss of mass compared to the corresponding $\ell=0$ models with the same value of $R_{99}$. For instance, low compactness models CHl1N and INl1N, which can be found in Tables \ref{tab:R_coherent} and \ref{tab:R_incoherent}, are an example of this behaviour since the initial total mass is smaller by half compared to $M_0^{\mathrm{max}}$. In Appendix \ref{app:comparisson} we present the wave forms of  boson stars mergers with $\ell=0$ and $\ell=1$ in order to stress the role of the $\ell$ parameter in the GW emission. 

To establish how spherical the after merger is, one may investigate the evolution of the components of the inertia tensor, described above. For the $\ell=0$ boson stars merger the difference between the diagonal components averages to zero, confirming the tendency to sphericity. For the $\ell$-boson stars with $\ell=1$ case, on the other hand, this is not so. This supports the conclusion that the end state in the merger of the $\ell=0$ boson stars tends towards a new $\ell=0$ boson star, whereas for the  $\ell$-boson stars with $\ell=1$ merger, it does not seem to tend towards a new  $\ell$-boson star with $\ell=1$, albeit still remaining a bound state of the scalar field. It is possible that the asymptotic end state is a localized configuration with less symmetries; our simulations, however, can only raise this possibility, not establish it. 
%
\subsection{Non-aligned stars}

As discussed above, a binary system of $\ell$-boson stars may be given a relative orientation, despite the sphericity of the individual stars. 
We shall now discuss the impact of this feature on the head-on collisions. Let us take the orientation of the left (centered at $\mu x=-25$) $\ell$-boson star fixed and rotate the right star (centered at $\mu x=+25$) using the $R_i$ matrices presented at the end of section \ref{sec:initial_data}.
It turns out that this relative orientation has very noticeable consequences in the outcome of the merger. Specifically we will make rotations around the $y$ and $z$ axes (recall that the collision is along the $x$ axis) for the coherent model B, identifying the cases by the transformation performed, $i.e.$, the $R_y(\delta)$ and $R_z(\delta)$ respectively, where we have chosen the angles of rotation as $\delta=\pi,\frac{\pi}{2}$ and $\frac{\pi}{4}$.

\begin{figure}
\includegraphics[height=6.5cm]{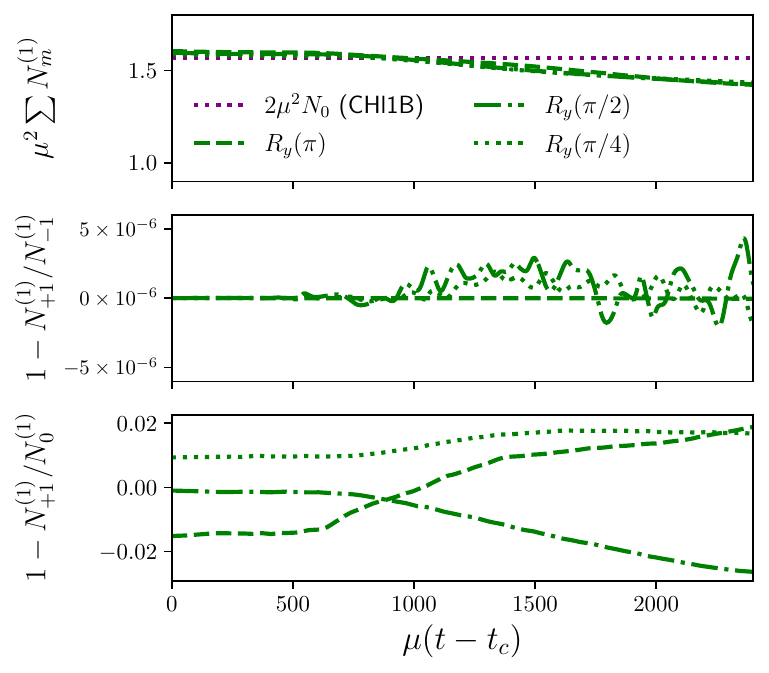}\quad
\includegraphics[height=6.5cm]{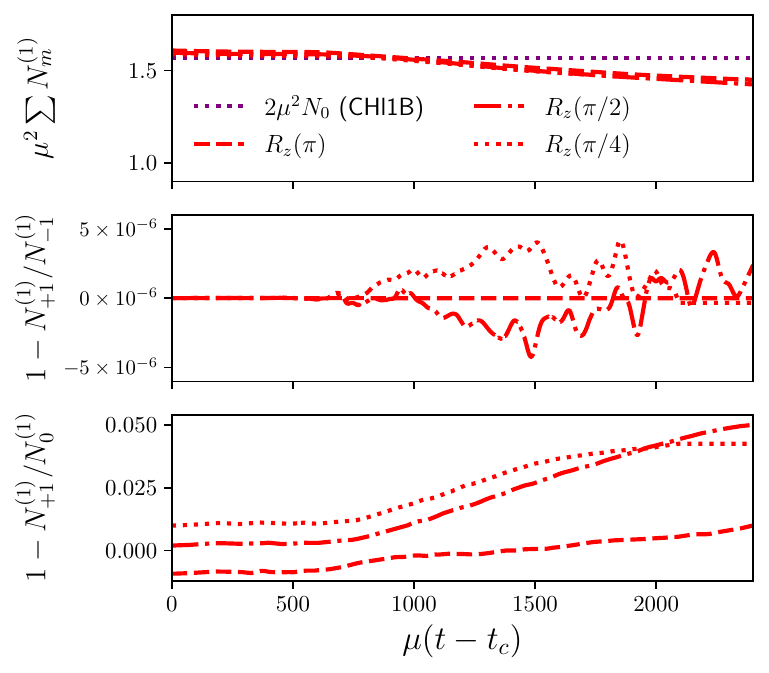}
\caption{
Non aligned stars. 
Total and individual (associated with each field) number of particles. (Left) Rotations around $y$-axis. (Right) Rotations around $z$-axis. As in Fig.~\ref{fig:num_particles_co}, $1-N^{(1)}_{+1}/N_0^{(1)}$ shows expected deviations from unity at $t=0$ due to coherent superposition.
}
\label{fig:num_particles_R}
\end{figure}

As a general observation before describing specific cases, we do not obtain appreciable differences in the evolution of the mass or of the total number of particles with respect to the aligned case, see the upper panels of Fig. \ref{fig:num_particles_R}, despite some particularities that we will address later.  Another similarity with the aligned coherent case is that the number of particles in the $m=+1$ mode is equal to that of the $m=-1$ mode throughout the evolution, as indicated by the middle panels of Fig. \ref{fig:num_particles_R}. 

On the other hand,  the relative behaviour of the number of particles in the modes $m=+1$ and $m=0$ can be different in the non aligned case, although the relative difference is still small,  never exceeding 10\%. 
A difference found with respect to the aligned configuration, is that that there are cases where
the three momenta of inertia become different.

\begin{figure}[ht!]
\subfloat[Scalar field energy density, $\rho/\mu^2$, for model CHl1B. Top panel: $z=0$ plane. Bottom panel: $x=0$ plane.\label{fig:rho_y180}]{\includegraphics[width=\textwidth, angle=0]{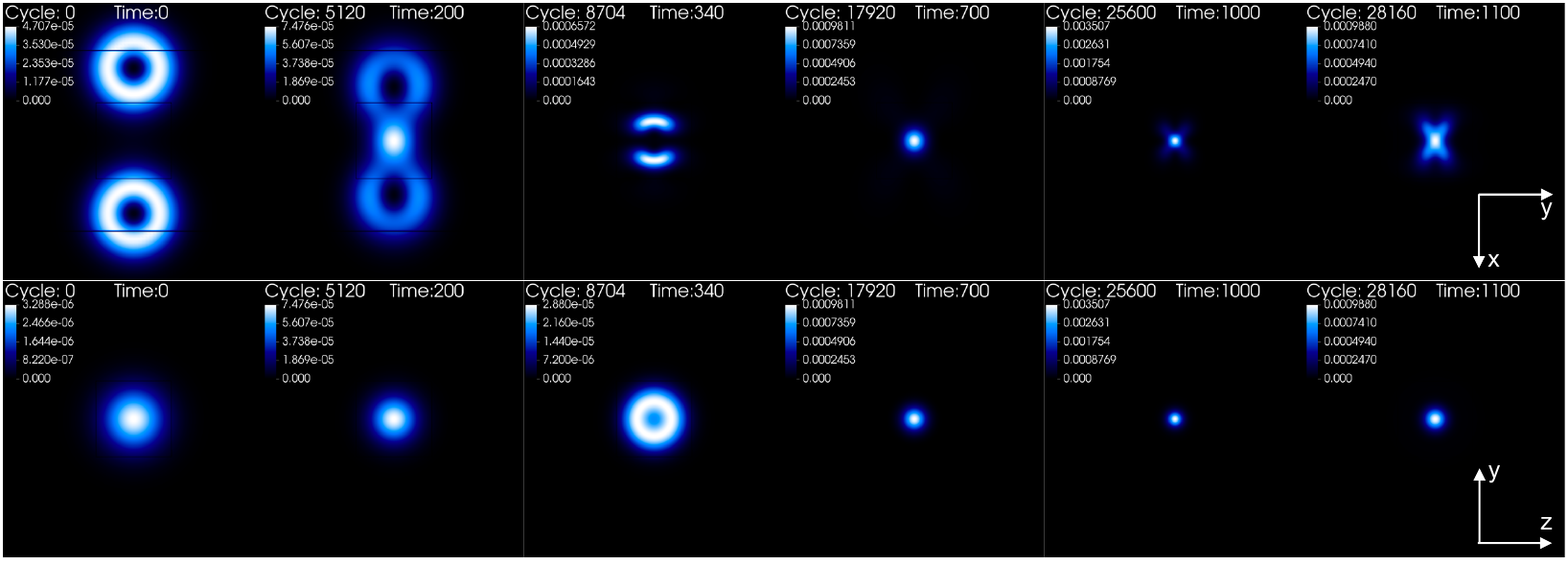}}\\
\subfloat[Individual currents $j_m /\mu$ for model CHl1B. Top panel: $m=0$ in the $y=0$ plane. Bottom panel: $m=+1$ in the $z=0$ plane.\label{fig:J_y180}]{\includegraphics[width=\textwidth, angle=0]{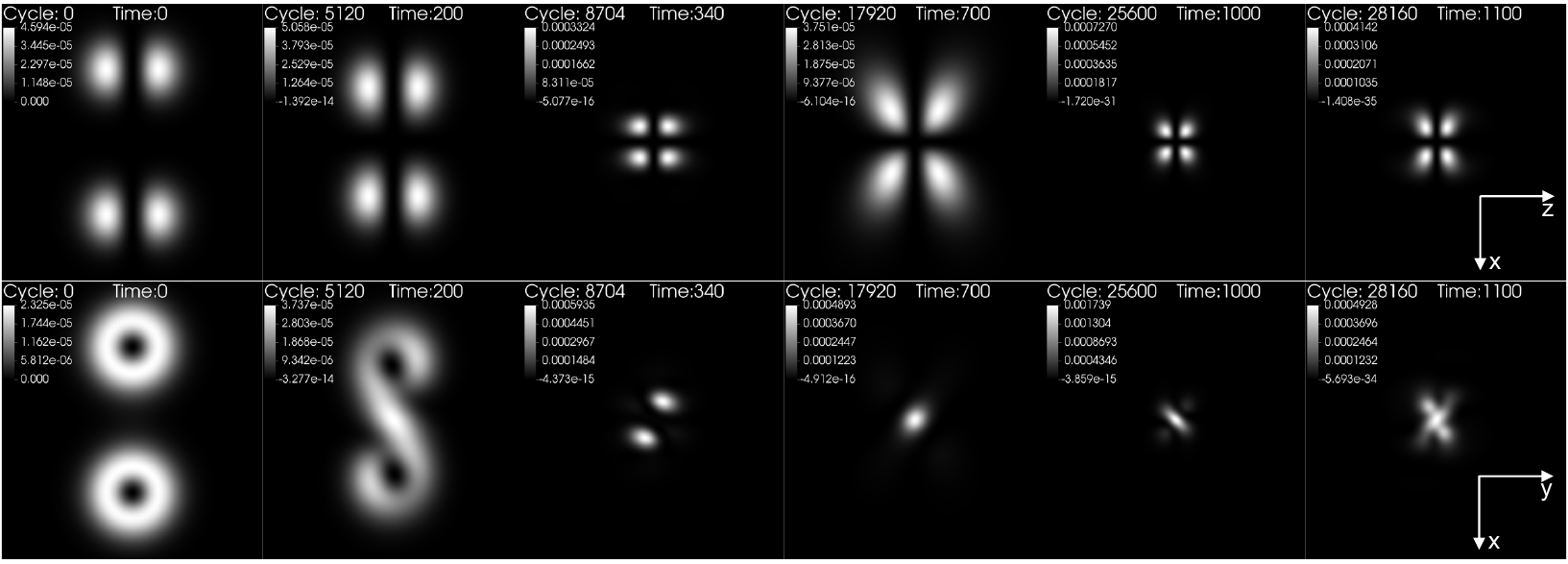}}
\caption{Non aligned stars $R_y(\pi)$.}
\end{figure}

Let us now discuss the specific cases  with $R_y(\pi)$,  $R_y(\pi/2)$ and $R_y(\pi/4)$;
the last two cases, however, are qualitatively similar.
An $\ell=1$  boson star is a composite of an $m=0$ and $m=\pm 1$ modes. The $m=0$ mode has a dipole-like energy distribution, whereas the $m=\pm 1$ modes are toroidal.
In Fig. \ref{fig:rho_y180} we show snapshots of the energy density for the $z=0$ (top panels) and $x=0$ (bottom panels) planes. One observes that the maximum of the density can reach the origin. This behavior is characteristic of this type of merger (rotation $R_y(\pi)$). The bottom panels confirm that 
the symmetry along the (collision) $x$-axis is preserved during the merger.
Fig. \ref{fig:J_y180} shows snapshots for the individual currents for the same collision.
The top panels correspond to the mode $m=0$; one observes that the quadrupolar shape is maintained all along the evolution, presenting a clearly repulsive effect (attributed to the phase difference of $\pi$ in this mode after the rotation) thus preventing this mode from concentrating at the origin.
The bottom panels show the current for the mode $m=+1$ where there is not a well defined shape after the merger.

\begin{figure}[ht!]
\subfloat[Scalar field energy density, $\rho/\mu^2$, for model CHl1B. Top panel: $z=0$ plane. Bottom panel: $x=0$ plane.\label{fig:rho_y90}]
{\includegraphics[width=\textwidth, angle=0]{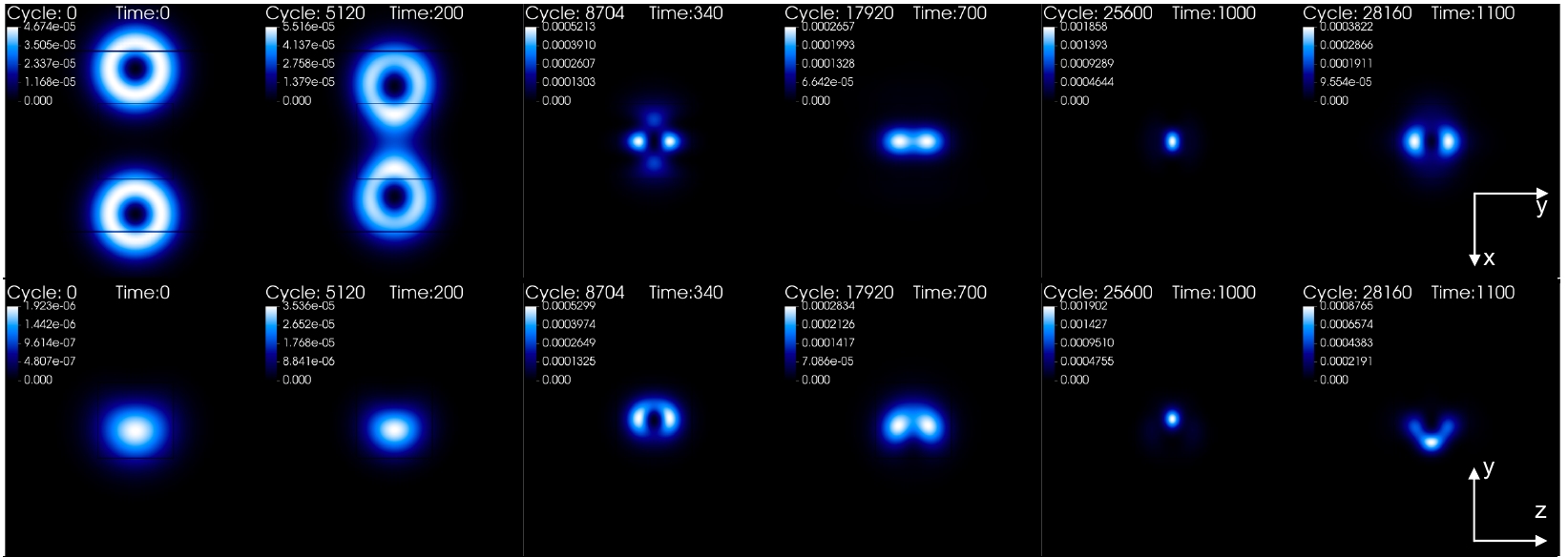}}\\
\subfloat[Individual currents $j_m /\mu$ for model CHl1B. Top panel: $m=0$ in the $y=0$ plane. Bottom panel: $m=+1$ in the $z=0$ plane.
\label{fig:J_y90}]{\includegraphics[width=\textwidth, angle=0]{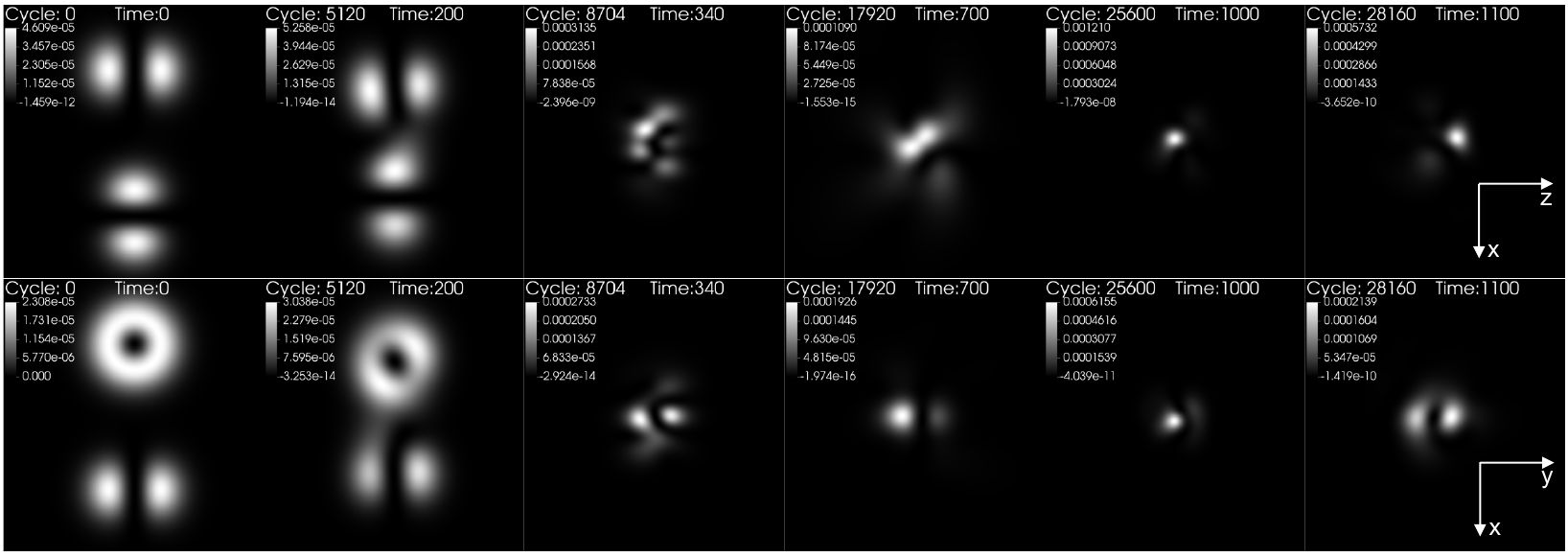}}
\caption{Non aligned stars $R_y(\pi/2)$.}
\end{figure}

\begin{figure}[ht!]
\begin{center}
\includegraphics[scale=0.4]{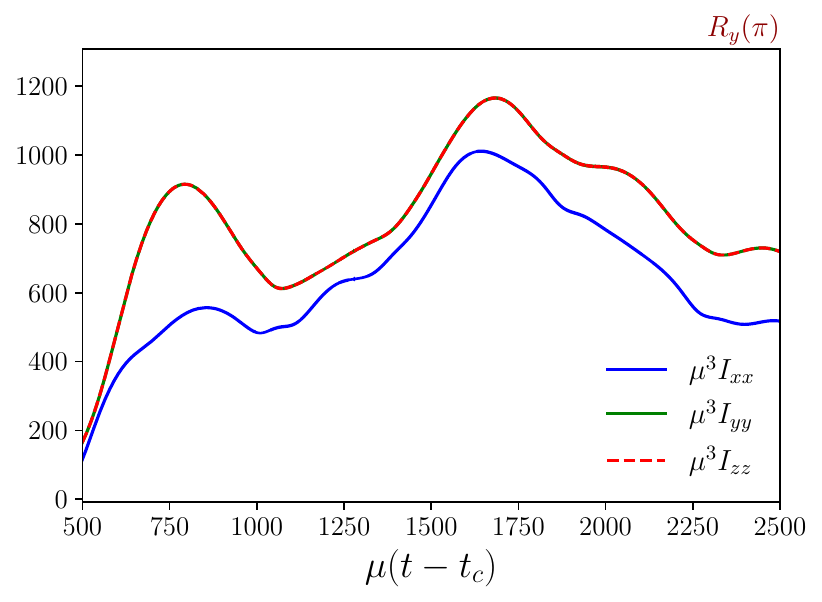}
\includegraphics[scale=0.4]{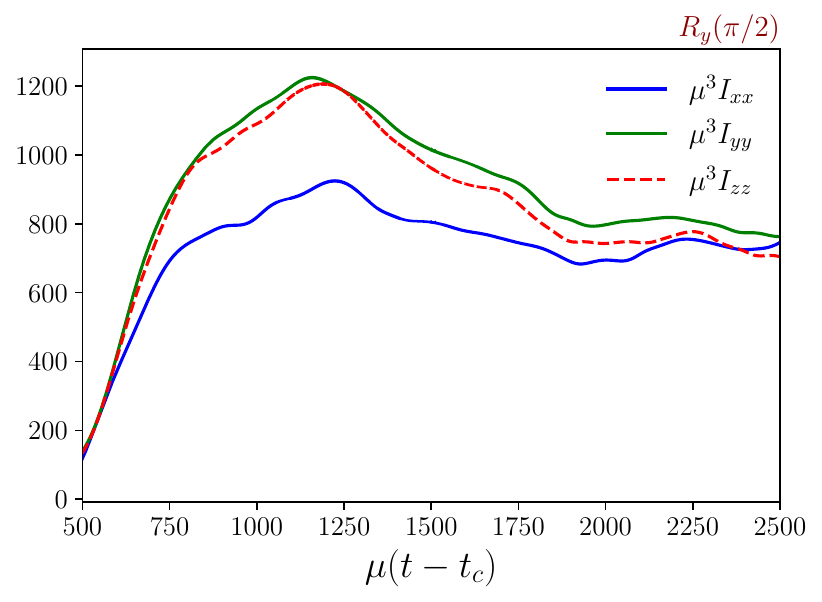}
\caption{Moments of inertia for the non aligned models CHl1B, with $R_y(\pi)$ (left) and $R_y(\pi/2)$ (right).
.}
\label{fig:intertia_Ry}
\end{center}
\end{figure}

Regarding the $R_y(\pi/2)$ collision, there is an even more intricate interaction amongst the modes of each star. 
In Fig. \ref{fig:rho_y90} we present snapshots of the energy distribution in the $z=0$-plane (top panels) and
in the $x=0$-plane (bottom panels).
The noticeable feature is that the  symmetry of the configuration with respect to the collision axis is  lost. Another characteristic is that the maximum density can reach the origin after the merger, despite the hollow shape of each star.
Fig. \ref{fig:J_y90} shows the individual currents for the $m=0$ (top panels) and the $m= + 1$ (bottom panels) modes.
These examples show that the spherical symmetry of the end product of the merger is lost. One can further support this statement by looking at the moments of inertia of the final configuration.
Fig. \ref{fig:intertia_Ry} shows the moments of inertia as a function of time for the rotated case $R_y(\pi)$ and $R_y(\pi/2)$.
For the case with CHl1B-$R_y(\pi)$ one of the three moments of inertia is different.
For the case with CHl1B-$R_y(\pi/2)$ the three moments of inertia remain all different.
We have also studied the evolution of a system where the non alignment is due to a rotation around the $z$ axis. This case shows several effects similar to the previous case, for example, the attraction/repulsion of the modes and a loss of sphericity.

\begin{figure}[ht!]
\subfloat[Scalar field energy density, $\rho/\mu^2$, for model CHl1B. Top panel: $z=0$ plane. Bottom panel: $x=0$ plane.\label{fig:rho_z180}]{\includegraphics[width=\textwidth, angle=0]{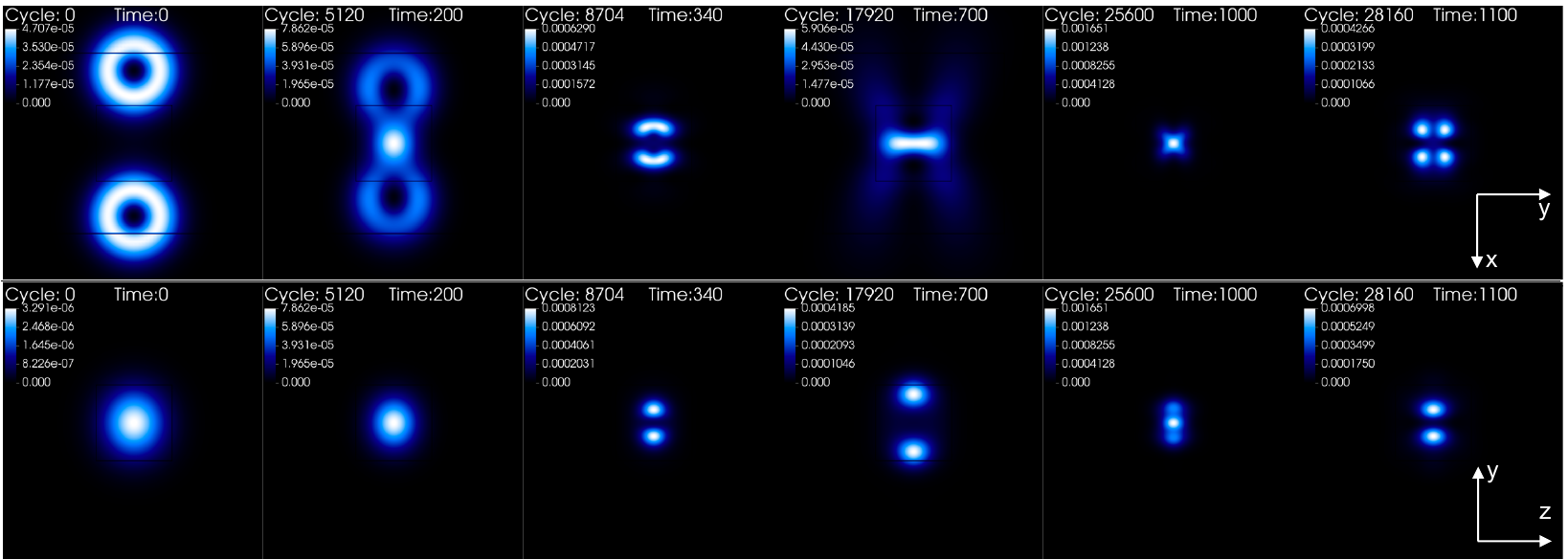}}\\
\subfloat[Individual currents $j_m/\mu$ for model CHl1B. Top panel: $m=0$ in the $y=0$ plane. Bottom panel: $m=+1$ in the $z=0$ plane.\label{fig:J_z180}]{\includegraphics[width=\textwidth, angle=0]{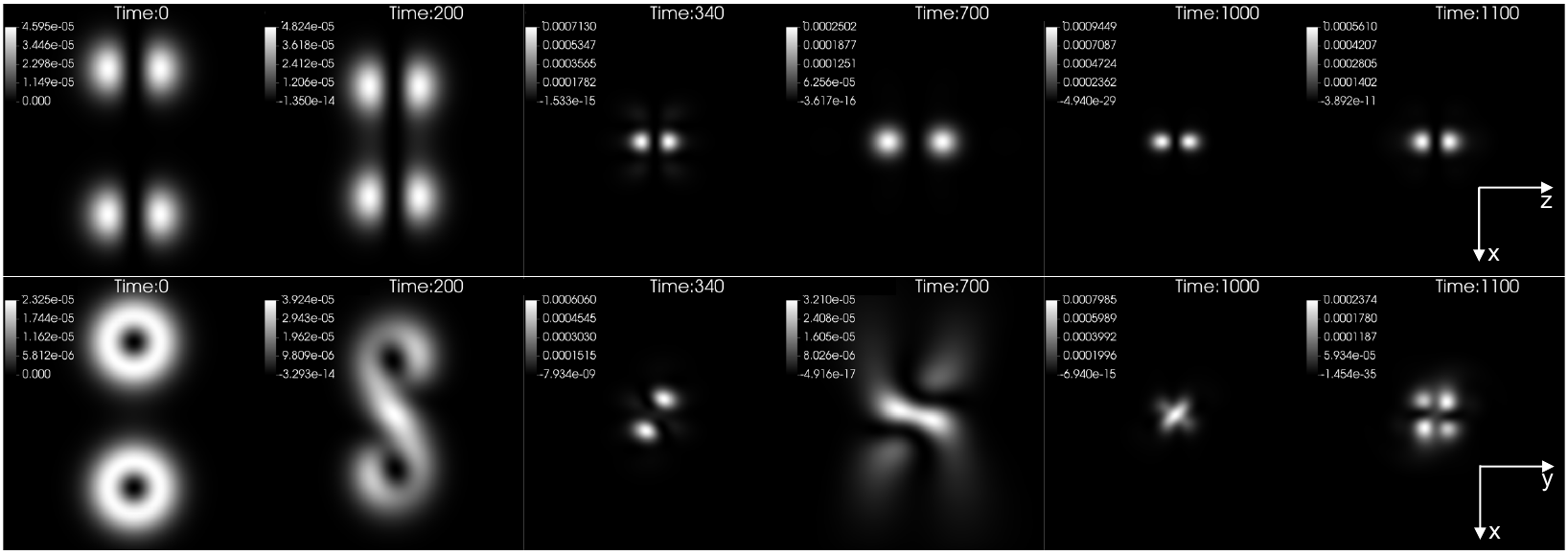}}
\caption{Non aligned stars $R_z(\pi)$.}
\end{figure}

\begin{figure}[ht!]
\subfloat[Scalar field energy density, $\rho/\mu^2$, for model CHl1B. Top panel: $z=0$ plane. Bottom panel: $x=0$ plane.\label{fig:rho_z90}]{\includegraphics[width=\textwidth, angle=0]{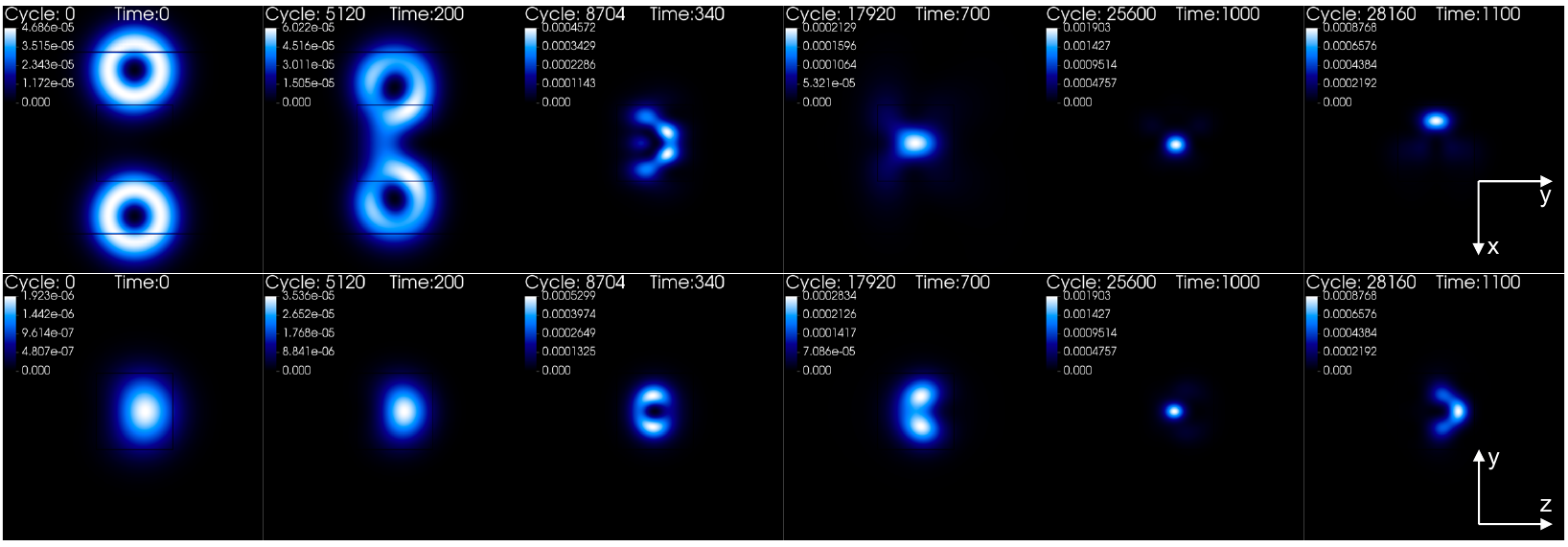}}\\
\subfloat[Individual currents $j_m/\mu$ for model CHl1B. Top panel: $m=0$ in the $y=0$ plane. Bottom panel: $m=+1$ in the $z=0$ plane.\label{fig:J_z90}]{\includegraphics[width=\textwidth, angle=0]{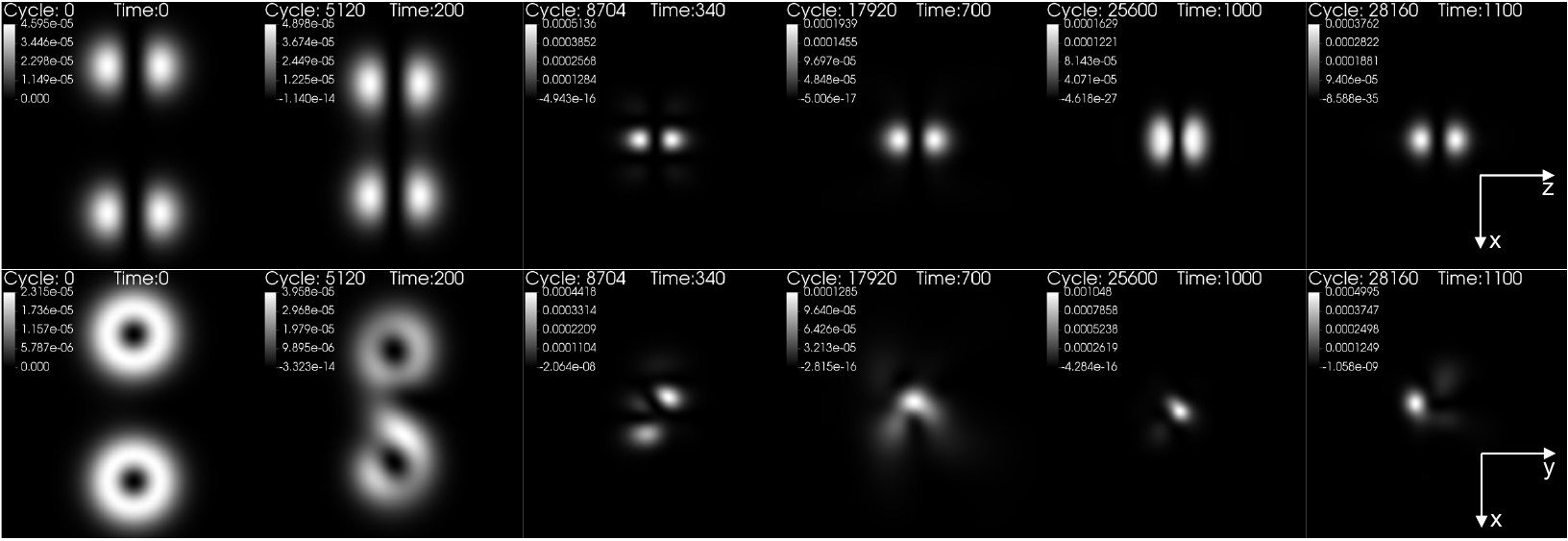}}
\caption{Non aligned stars $R_z(\pi/2)$.}
\end{figure}

In Fig.~\ref{fig:rho_z180}, we present the evolution of the energy density for the $R_z(\pi)$ rotation, and in Fig. \ref{fig:J_z180} we present the corresponding evolution of the modes. 
The energy density diagnosis shows, as in the rest of the cases analyzed, a loss of the original symmetry of each star. As in the aligned case, this kind of rotations keeps unchanged the initial $m=0$ mode, see Fig.~\ref{fig:harmonic} (recall that the spherical harmonic $\ell=1$, $m=0$ does not depend on the angle $\varphi$). During the evolution it can be seen that individually 
the $m=0$ mode maintains its morphology, while 
the $m=\pm 1$ have a drastic change in morphology, yielding the observed change in the energy density.
The $R_z(\pi/2)$ rotation case is shown in Fig.~\ref{fig:rho_z90}. We see a twisting effect in the evolution of the energy density distribution in the $z=0-$plane.  
Concerning the individual modes, the $m=0$ mode tends to keep the original morphology, while the $m\pm 1$ modes loose it, a behavior similar to the previous case.

To conclude, we have observed  that the rotations around the $y-$axis, $R_y$, produce more significant changes in the final morphology than the rotations along the $z-$axis, which also modify the final morphology, but keep the shape of the remnant $m=0$ mode.

\begin{figure}[ht!]
\begin{center}
\includegraphics[scale=0.4]{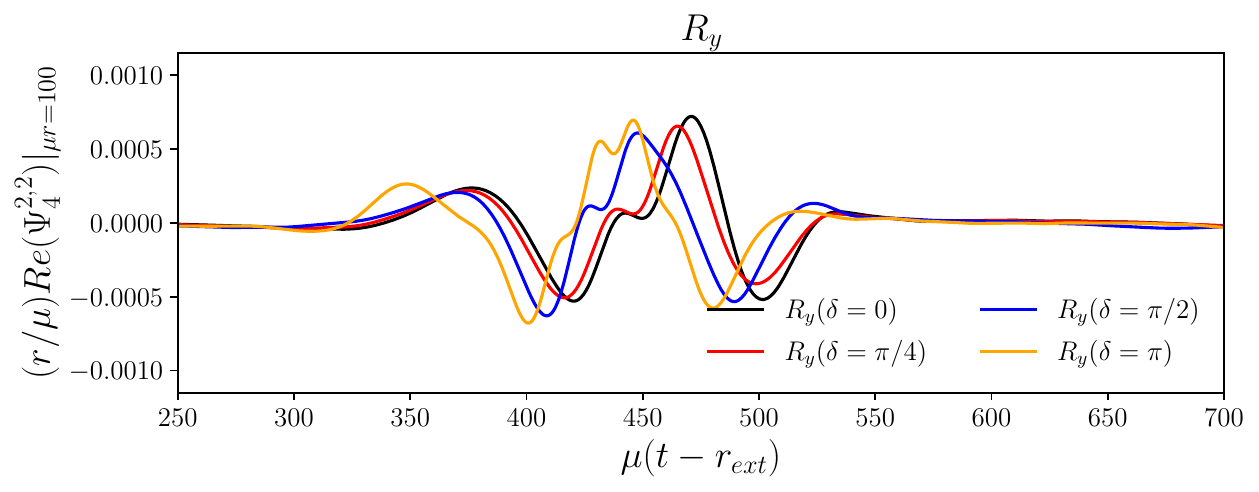}
\includegraphics[scale=0.4]{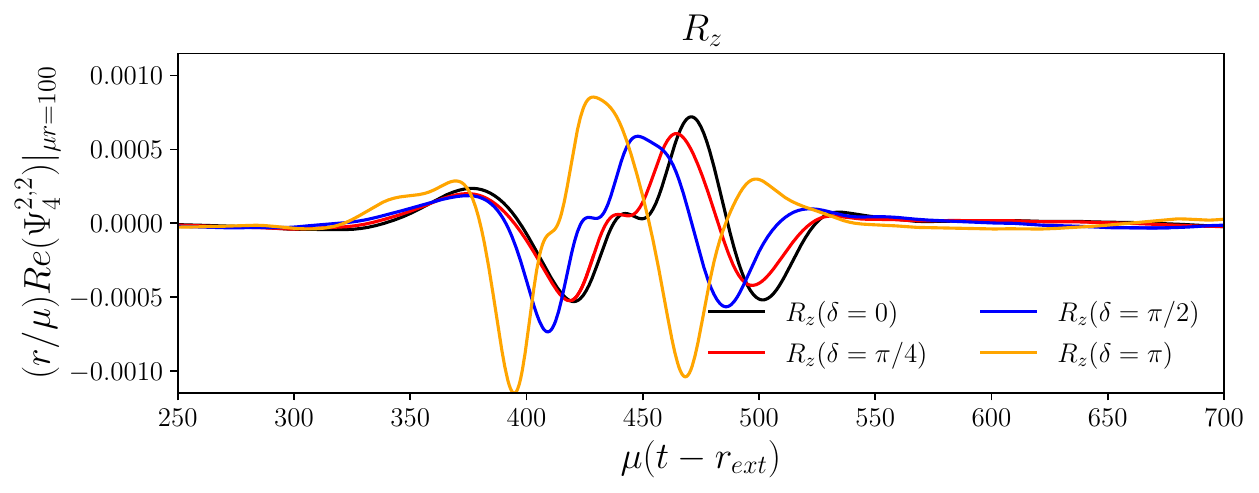}\\
\includegraphics[scale=0.4]{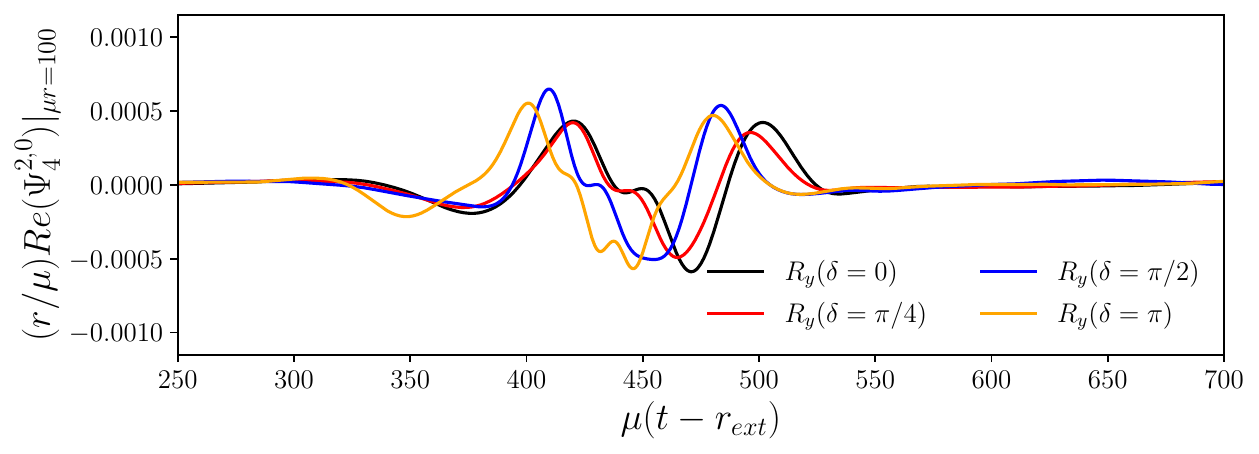}
\includegraphics[scale=0.4]{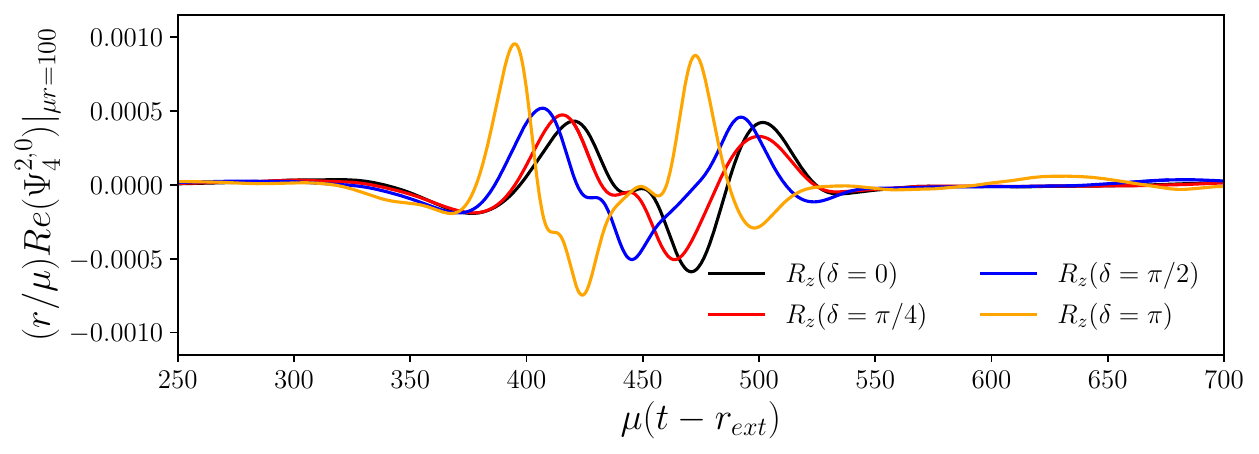}
\caption{Waveforms for non aligned $\ell=1$ binary. CHl1B model. Top panels show the real part of the $l=2$, $k=2$ mode. Bottom panels show the real part of the $l=2$, $k=0$ mode. The extraction radius is $\mu r_{\rm ext}=100$.}
\label{fig:GWl1_na}
\end{center}
\end{figure}

Regarding the GW profiles, we looked for any significant difference that could indicate a waveform dependence (particularly in its amplitude) on the angle and direction of rotation. As shown in Fig. \ref{fig:GWl1_na}, however, in none of the alignments studied for the model B we have observed significant differences; the amplitude of the seven cases is of the same order, although the first peak appears earlier in the non-aligned cases, with a maximum difference in time with respect to the aligned case of $\mu\Delta t \sim 50$ for the $R_y(\pi)$ case. This is true for all $\Psi_4^{l,k}$ but in particular for the dominant $l=2$ $k=0,2$ modes shown in Fig. \ref{fig:GWl1_na}.

Rotations $R_y$ of the initial data cause slightly larger differences in the waveforms than the $R_z$ cases. This is consistent with the fact, already pointed out, that unexpected dynamics on the $m=0$ scalar field may occur, essentially because $R_y$, unlike $R_z$, modifies the distribution of all the individual fields of the rotated star and not only the $m=\pm 1$ modes.

The GW signals just discussed can be compared with the ones previously presented for coherent (aligned) superposition and the incoherent one. For example, comparing the upper right panel of Fig. \ref{fig:GWl1} and \ref{fig:GWl1_na}, one observes that at the level of the GWs. We remark that in hypothetical astrophysical scenarios  an exact alignment should be accidental, and the generic case should be of misaligned stars. Concerning the end state, however, it is unclear if the coherence could have a greater importance than the relative alignment of the stars.

\section{Final remarks}
\label{sec:conclusions}
In this paper we have studied head-on collisions of $\ell$-boson stars, starting from rest. These are \textit{composite} self-gravitating solitons, made up of $2\ell+1$ complex, massive scalar fields~\cite{Alcubierre:2018ahf}. In such scalar lumps, the different scalar fields have precisely the same amplitude, which raises the concern of possible self-tuning. Yet, it has been shown that these solutions are dynamically robust in regions of the parameter space~\cite{Alcubierre:2019qnh,Alcubierre:2021mvs}, at least against small perturbations. Here, we test these solutions against more violent processes: head-on collisions of two such equal stars.

Our simulations consider a variety of cases. As the  two main scenarios we consider that the two colliding $\ell$-boson stars are made up of the same, or of different, scalar fields. These two cases are dubbed coherent and incoherent, respectively. Additionally, we consider different possible orientations of the colliding stars. It may sound strange that such spherical stars have an "orientation"; yet they do. The point is that the composing fields of each star have a multipolar structure - $cf.$ Fig.~\ref{fig:harmonic} -, defined with respect to a pre-established Cartesian reference frame. We can thus choose that these reference frames coincide, or not, for the two colliding stars. These two cases are dubbed aligned and non aligned, respectively.

Independently of the specific characteristics of each model that we have studied, a first generic conclusion one can put forward is the following. The collision of sufficiently massive (and therefore compact stars) stars forms a black hole. This is what happens, $e.g.$, for models C and D in Fig.~\ref{fig:mass_vs_frequency}, corresponding to the collisions CH11C and CH11D in Table~\ref{tab:R_coherent} and IN11C and IN11D in Table~\ref{tab:R_incoherent}. But below a certain mass threshold, the collisions do not lead to horizon formation; a bound state of the different scalar fields remains, that neither disperses nor collapses. This is seen, for instance, for models A and B in Fig.~\ref{fig:mass_vs_frequency}, corresponding to the collisions CH11A and CH11B in Table~\ref{tab:R_coherent} and IN11A and IN11B in Table~\ref{tab:R_incoherent}.

Concerning the end state of collisions that do not form a black hole, the simulations reveal different specificities, depending on the chosen characteristics of the stars and of the collision. Still one sees two generic features. First, the democracy between the different composing fields of each $\ell$-boson star is lost, albeit not dramatically. This feature can be diagnosed from the Noether charge in each field - see $e.g.$ Fig.~\ref{fig:num_particles_co} (for aligned coherent states), Fig.~\ref{fig:num_particles_in} (aligned incoherent states) and Fig.~\ref{fig:num_particles_R} (for non aligned states). Observe, nonetheless, that the balance of particles in the $m=\pm 1$ modes is kept to high accuracy, and the slight imbalance with the $m=0$ mode does not exceed a few percent.
Second, the collision aftermath deviates from spherical symmetry. This can be quantified by looking at the moments of inertia, see Fig.~\ref{fig:intertia_a} (for aligned collisions) and Fig.~\ref{fig:intertia_Ry} (for non aligned collisions). Observe that $I_{xx}=I_{yy}$ in all cases, except the non aligned collisions under a rotation $R_y(\pi/2)$; also note that in some cases the three momenta of inertia, $I_{xx},I_{yy},I_{zz}$ seem to converge after some time, which may be interpreted as a glimpse of a tendency towards sphericity.

Overall, the above description, albeit not entirely conclusive, allows us to answer the question we have set out to investigate: how dynamically robust are $\ell$-boson stars? The answer seems to be two-fold. \textit{Exact}, equilibrium $\ell$-boson stars, with precise equipartition of the number of particles amongst the $2\ell+1$ fields are indeed fine tuned, and will not withstand generic perturbations. This is no surprise, and it was already anticipated by considering non-spherical perturbations of equilibrium $\ell$-boson stars~\cite{Jaramillo:2020rsv}. Yet, $\ell$-boson stars as a particular symmetry enhanced point of a larger family of multi-field bosonic stars, as reported in~\cite{Sanchis-Gual:2021edp}, can be considered to be dynamically robust and long lived. In fact, the violent collisions we have considered, could only produce an imbalance in the Noether charges, of the order of a few  percent.  Of course, the end states of our simulations remain oscillating and have not yet reached a stationary state. We cannot rule out that gravitational cooling and GW emission will work towards a 
spherical distribution, getting rid of the non-spherical modes. However, our simulations do not exhibit strong, generic evidence for this possibility. The very slow convergence towards a final state is not a consequence of using models with initial total mass larger than the maximum mass of the final expected state, an $\ell=1$ boson star, as can be concluded from the CHl1N and INl1N cases presented.

Finally, let us comment that our collisions have generated waveforms which can be quite different from those of BHs. As emphasized in the Introduction, it will be interesting to continue this effort, in order to generate large libraries of alternative waveform templates, to compare with observational data. This effort, of course, will require going towards orbiting binaries.


\acknowledgments
This work was supported in part by the CONACYT Network Projects No. 376127 ``Sombras, lentes y ondas gravitatorias generadas por objetos compactos astrof\'isicos'', No. 304001 ``Estudio de campos escalares con aplicaciones en cosmolog\'ia y astrof\'isica" and No. 140630
``Explorando los confines de las teor\'ias relativistas de la gravitaci\'on y sus consecuencias'', as well as by DGAPA-UNAM through grants 
IN110218, IA103616, IN105920. We also thank the support by the Center for Research and Development in Mathematics and Applications (CIDMA) through the Portuguese Foundation for Science and Technology (FCT - Funda\c{c}\~ao para a Ci\^encia e a Tecnologia), references UIDB/04106/2020, UIDP/04106/2020 and the projects PTDC/FIS-OUT/28407/2017, CERN/FIS-PAR/0027/2019,  PTDC/FIS-AST/3041/2020 and CERN/FIS-PAR/0024/2021.  We further acknowledge support from the European Union's Horizon 2020 research and innovation (RISE)
program H2020-MSCA-RISE-2017 Grant
No. FunFiCO-777740. VJ acknowledges support from CONACYT graduate  grants  program. NSG was also supported by the Spanish Ministerio de Universidades, reference UP2021-044, within the European Union-Next Generation EU.

\appendix
\section{Code validation}
\label{app:codevalid}

\begin{figure}[ht!]
\begin{center}
\begin{center}
\includegraphics[scale=0.4]{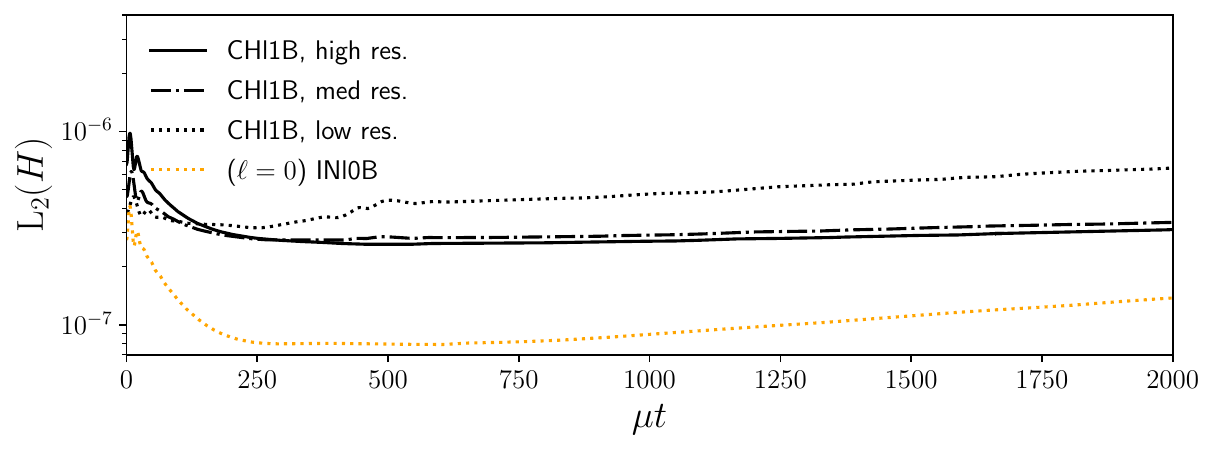}
\includegraphics[scale=0.4]{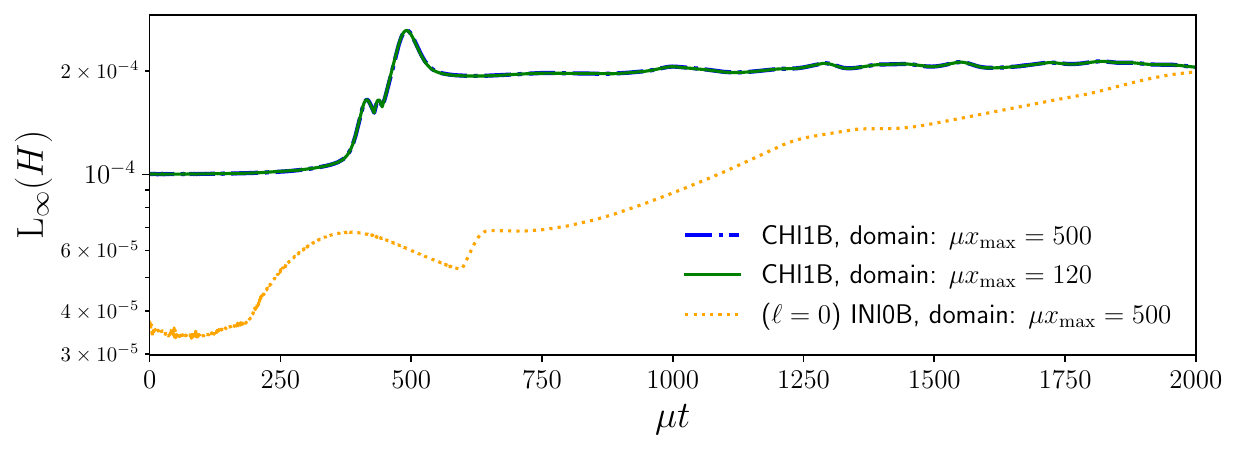}\\
\includegraphics[scale=0.4]{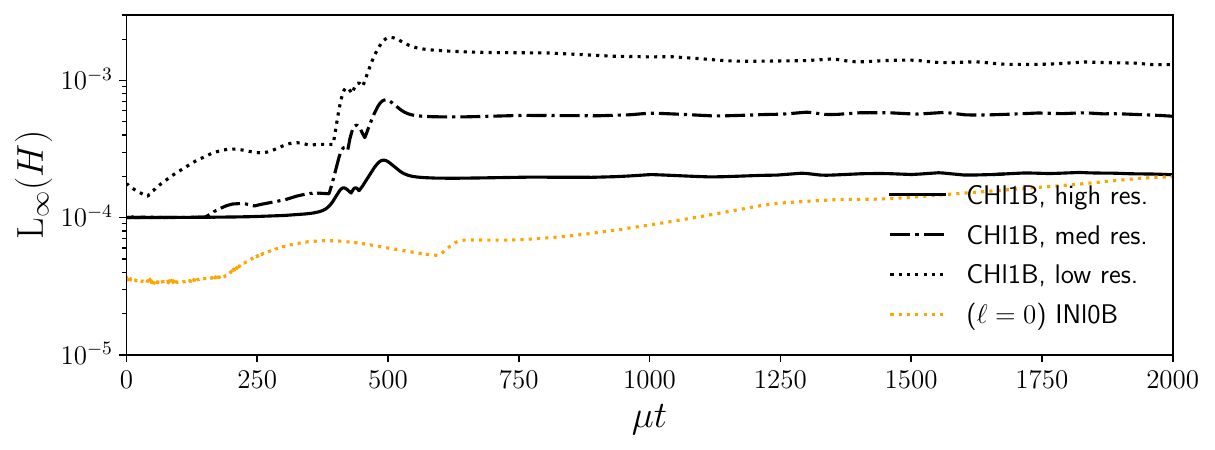}
\end{center}
\caption{$L_2$ and $L_\infty$ norms of the Hamiltonian constraint for the model CHl1B. In this plot, as in the results presented throughout this article the stars have been separated 50 coordinate units.}
\label{fig:ham_l2}
\end{center}
\end{figure}

In all binary boson star configurations considered in this paper we have chosen to separate the stars 50 coordinate units ($\mu x_c=25$). This responds to the fact that this is a distance for which, at least in the solutions studied, the absolute maximum of the violation of the Hamiltonian constraint $H$ at $t=0$ is approximately of the same order as the floor error of the numerical implementation, estimated from the single isolated star simulations. We verified that the $L_\infty$ norm $i.e.$, the maximum of the absolute value of the Hamiltonian constraint, decreases as the separation of the stars is increased. For example, the value obtained if the binary CHl1B is separated by a distance of 70, 50 and 30 coordinate units, is $L_\infty(H)(t=0)=9\times10^{-5},10^{-4},2\times10^{-3}$ respectively. Thus, for the chosen separation of 50, the violation induced for the constraint is of the same order of the interpolation error. $L_\infty(H)$ and the $L_2$ norm of $H$, defined as $L_2(H)=\sqrt{\sum H_i^2/N_g}$, where $N_g$ is the number of points of the grid, are shown in Fig.~\ref{fig:ham_l2} as functions of $t$ in the case where the separation of the stars is 50 units. The left panel shows that part of the initial data numerical error dissipates until $\mu t=500$. The right panel indicates that the maximum of $\abs{H}$ remains contained in the same order of magnitude throughout the simulation, in this plot, we show the results using two different computational domains; in the first one, using a blue line we show the case used in the simulations throughout the article, a box of size $\mu x_\mathrm{max}=500$, in the second case a box of size $\mu x_\mathrm{max}=120$. The maximum of $|H|$ is located (and remains) near the center of the grid.

\begin{figure}[ht!]
\begin{center}
\begin{center}
\includegraphics[scale=0.4]{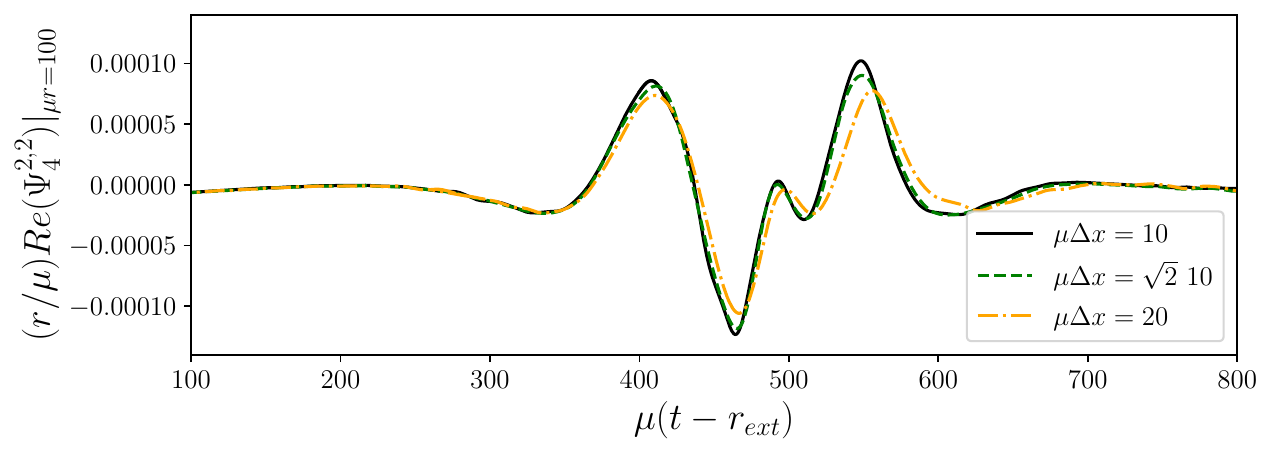}
\includegraphics[scale=0.4]{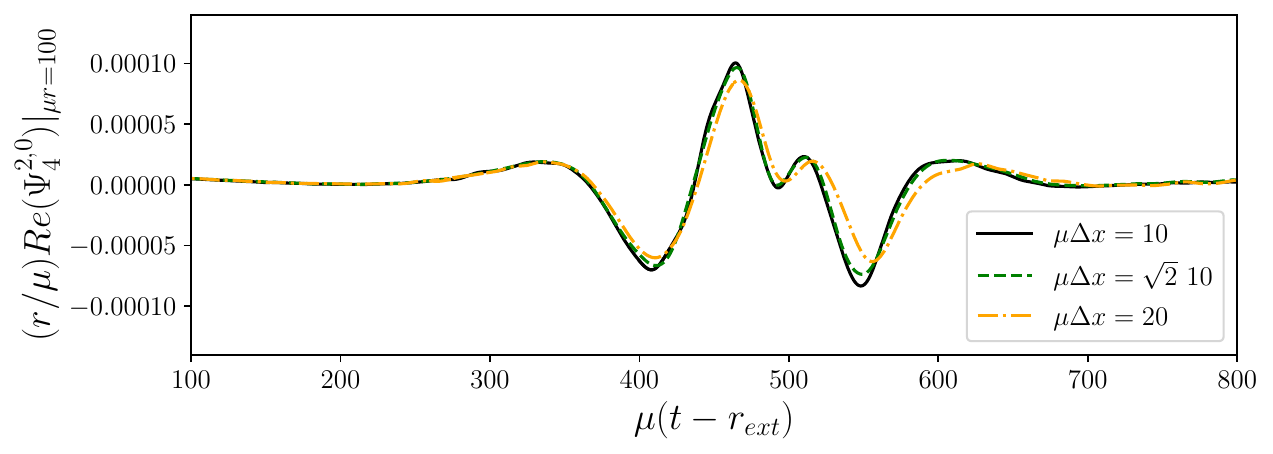}\\
\quad\includegraphics[scale=0.4]{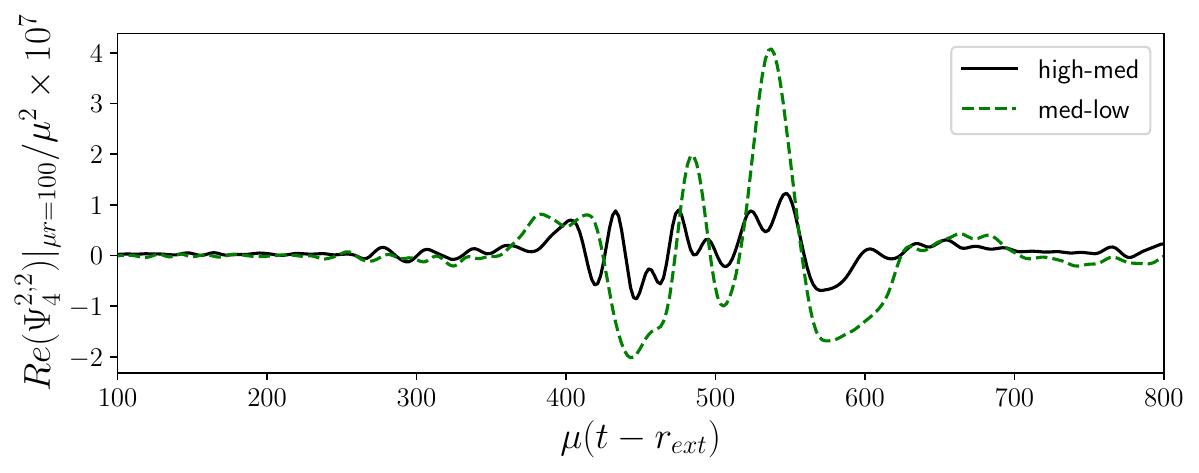}\qquad
\includegraphics[scale=0.4]{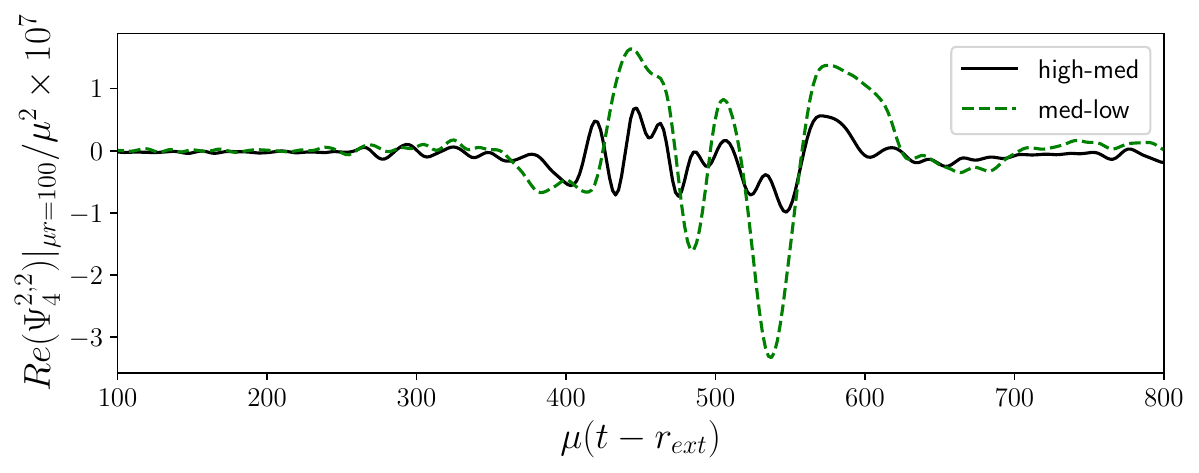}\\
\includegraphics[scale=0.4]{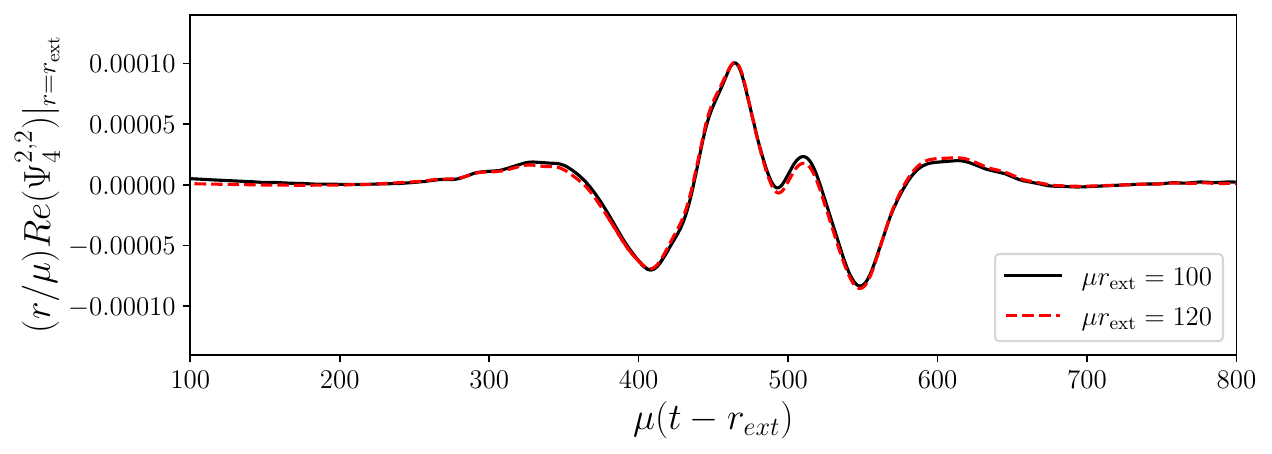}
\includegraphics[scale=0.4]{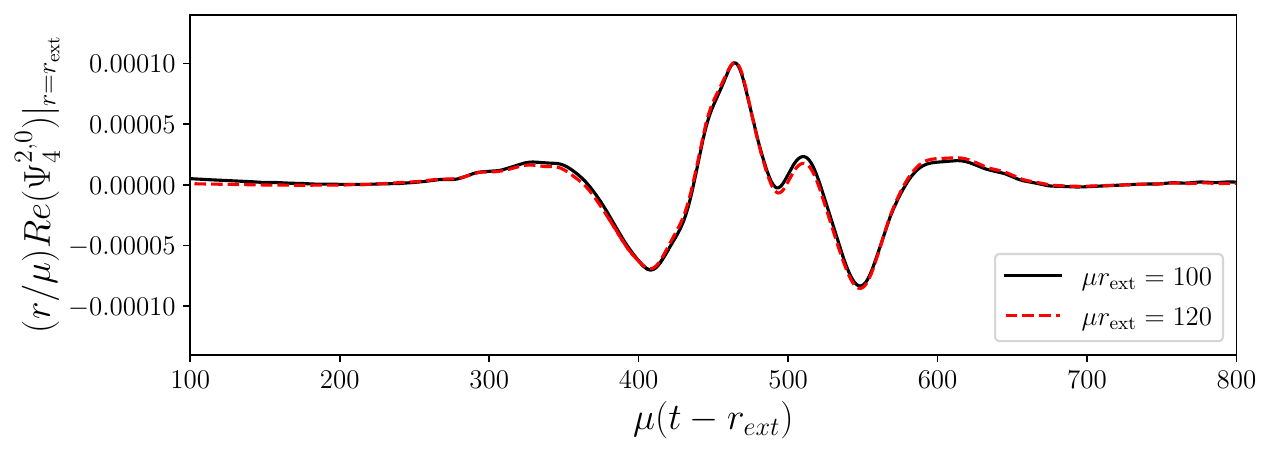}
\end{center}
\caption{Model CHl1A. Top panels show the real part of $r\Psi^{2,2}_4$ and $r\Psi^{2,0}_4$ at $\mu r_\mathrm{ext}=100$ using different resolutions. Their differences are displayed in the middle panel. Bottom panels show overlap of waves extracted at different radii when they are appropriately rescaled.}
\label{fig:GW_convergence}
\end{center}
\end{figure}

To evaluate the evolution of the small constraint violations induced by the initial data superposition of the two boson star solutions and test for convergence, we have analyzed the Hamiltonian and momentum constraint together with some of the analysis quantities and GW outputs using three different resolutions. The coarsest level in the low resolution case is set to $\mu\{\Delta x,\Delta y,\Delta z\}=20$, while for the medium and high resolutions we have $\mu\{\Delta x,\Delta y,\Delta z\}=\sqrt{2}\ 10$, $\mu\{\Delta x,\Delta y,\Delta z\}=10$, respectively. In the medium and low resolutions we have placed six refinement levels with the same spatial distribution as in the high resolution case (which is the one used in this work) described in section \ref{sec:evolution}. In Fig.~\ref{fig:GW_convergence} we have plotted the gravitational waveform of CHl1A, which converges. The left and the bottom panels of Fig.~\ref{fig:ham_l2} show that the Hamiltonian constraint also converges with resolution. The momentum constrains have been corroborated to be consistent with zero at $t=0$, and later in the evolution the three components converge with increasing resolution, as expected. At the middle panel of Fig.~\ref{fig:GW_convergence} the differences between the gravitational signal at different resolutions are given, this helps to establish convergence of the GW and also gives an estimate of $1\times10^{-3}$ for the relative difference when comparing the high and the medium resolutions. Finally, we have generated GW output at different radii in order to check consistency and accuracy of the results; the bottom panels of Fig.~\ref{fig:GW_convergence} show that $\Psi_4^{2,2}$ and $\Psi_4^{2,0}$ overlap when properly rescaled by the factor $1/r$. The extraction surfaces are therefore within the ``wave-zone''.

\section{Number of particles}
\label{app:number}
In this appendix we show that the number of boson particles in a single isolated $\ell$-boson stars, for a fixed value of $\ell$, is equally distributed among the different modes $m$, that compose the star. There is an \textit{equipartition} of the total Noether charge.

The total conserved (Noether) charge associated to all scalar fields of an $\ell$-boson star is given by the zero component of the total current $j^0$, as 
\begin{eqnarray}
N &=& \int   j^0 \alpha\sqrt{\gamma}dx^3 := \int \left( \sum_{m=-\ell}^\ell j_m \right)\alpha\sqrt{\gamma}dx^3 \ ,
\label{eq:Nb}
\end{eqnarray}
where $j^0$ is given by the sum
\begin{equation}
j^0=\sum_{m=-\ell}^\ell\left[\frac{i}{2} g^{0b}({\bar\Phi}_m\nabla_b\Phi_m-\Phi_m\nabla_b{\bar \Phi}_m)\right]=\sum_{m=-\ell}^\ell j_m \ .
\label{eq:j0m}
\end{equation}
For a single $\ell$-boson star
the scalar field  has a time dependence of the form
 $\Phi_m=e^{-i\omega t}\phi_\ell(r)Y^{\ell m}(\vartheta,\varphi)$ and the metric is given by~\eqref{eq:3metric_esph}.
The individual currents $j_m$, can be written, after some simplifications, as
\begin{eqnarray}
j_m
&=&-\frac{1}{\alpha^2(r)}\omega\phi_\ell^2(r)\abs{Y^{\ell m}(\vartheta,\varphi)}^2 \ .
\label{eq:j_m_den}
\end{eqnarray}
Integration of Eq. \eqref{eq:j_m_den} over the 3-element of volume gives the number of particles associated to each field
\begin{align}
N_m= \int \alpha\,j_m\sqrt{\gamma}dx^3  =
-\omega\left[\int_0^\infty dr\frac{a(r) r^2}{\alpha(r)}\phi_\ell^2(r)\right]\left[\int_0^{\pi} \int_0^{2\pi} d\vartheta d\varphi\sin \theta\abs{Y^{\ell m}(\vartheta,\varphi)}^2\right]\ ;
\end{align}
the second integral is equal to 1 due to the normalization of the spherical harmonics. Therefore
\begin{equation}
N_m =  -\omega\int_0^\infty dr\frac{a(r) r^2}{\alpha(r)}\phi_\ell^2(r) \ ,   
\end{equation}
and from Eq.~\eqref{eq:Nb} one obtains
\begin{eqnarray}
N 
  &=& \sum_{m=-\ell}^\ell  \int \alpha\,j_m \, \sqrt{\gamma}dx^3 
  =  ({2\ell+1})\, N_m \ .
\end{eqnarray}
Consequently, the total number of particles of a single $\ell$-boson star is divided equally into the associated number of particles stored in each field
\begin{equation}
N_m=\frac{1}{2\ell+1} N \ .
\end{equation}

\section{Comparison with \texorpdfstring{$\ell=0$}{l=0}  boson stars}
\label{app:comparisson}

The purpose of this appendix is to continue the discussion regarding the comparison 
between boson stars mergers with $\ell=0$ and $\ell=1$, presented at the end of section \ref{sec:aligned}, identifying 
the  effect  of the $\ell$ parameter on the GW signal.
The dynamics and GW signatures of the merger of boson star binaries have been discussed extensively in the past, $e.g.$ \cite{Liebling:2012fv,bezares2017final}.
It has been shown that the gravitational waveform may be very different from that of black holes specially in the early phases.
In our numerical experiments, we have obtained similar results to the ones presented in \cite{bezares2017final,bezares2018gravitational} for the standard $\ell=0$ boson stars merger, in particular that the waveform of the merger of incoherent states is much smaller and occurs after its coherent counterpart. 

As discussed above, for the boson stars with $\ell=1$ mergers, the coherent and incoherent configurations become more similar. Due to this fact, we focus our description of the $\ell=0$ and $\ell=1$ comparison, only for
the corresponding coherent states. In order to perform such comparison of the gravitational waveforms we select stars with the same $R_{99}$.

\begin{figure}[ht!]
\begin{center}
\includegraphics[scale=0.4]{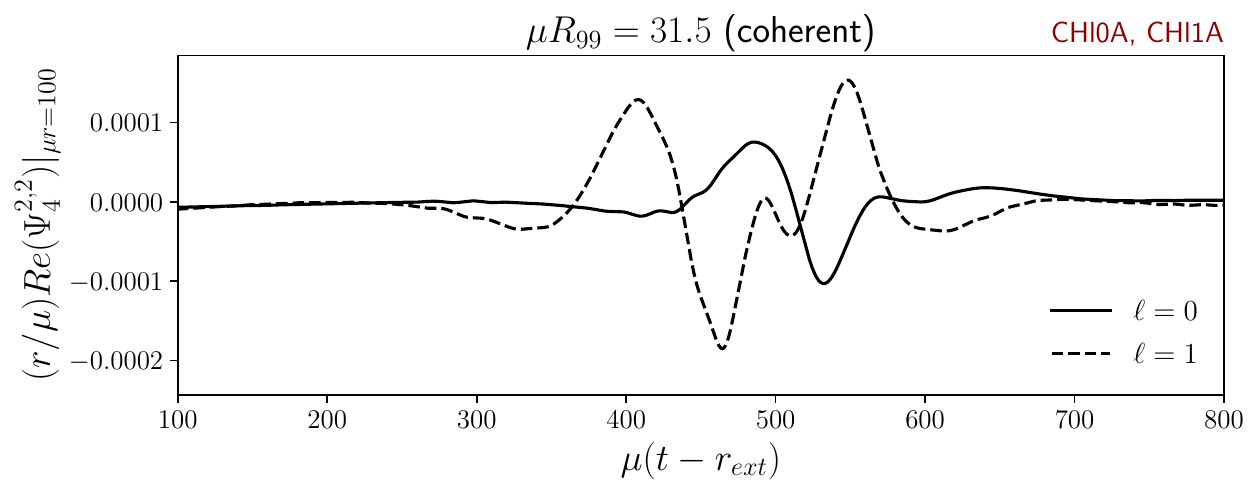}
\includegraphics[scale=0.4]{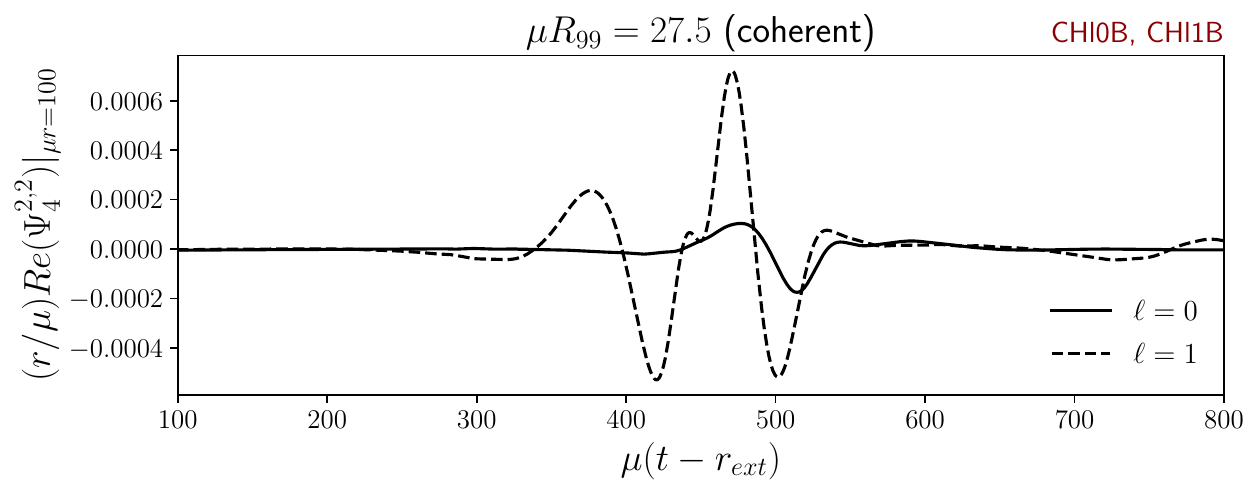}\\
\includegraphics[scale=0.4]{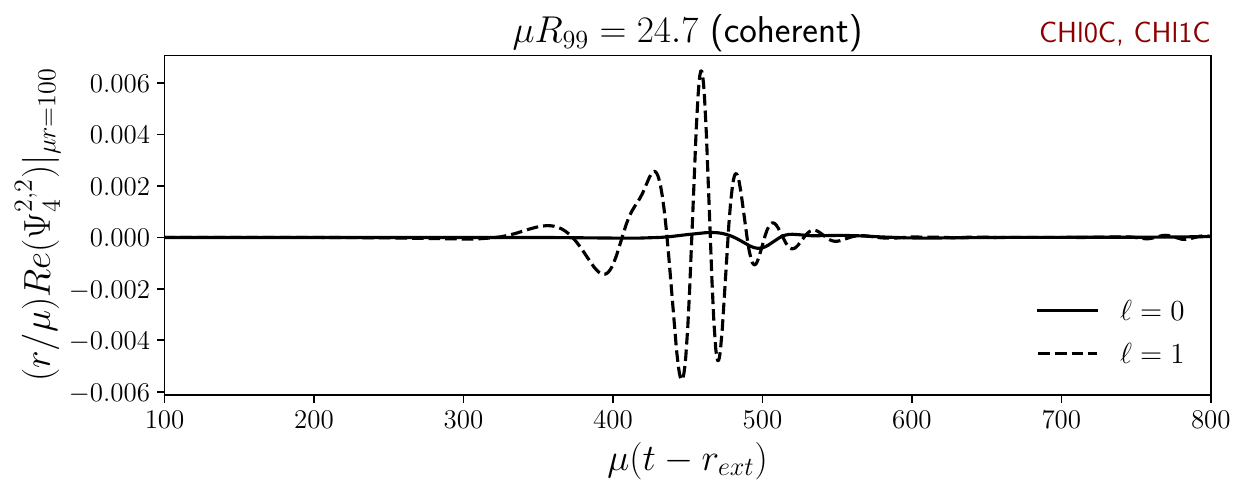}
\includegraphics[scale=0.4]{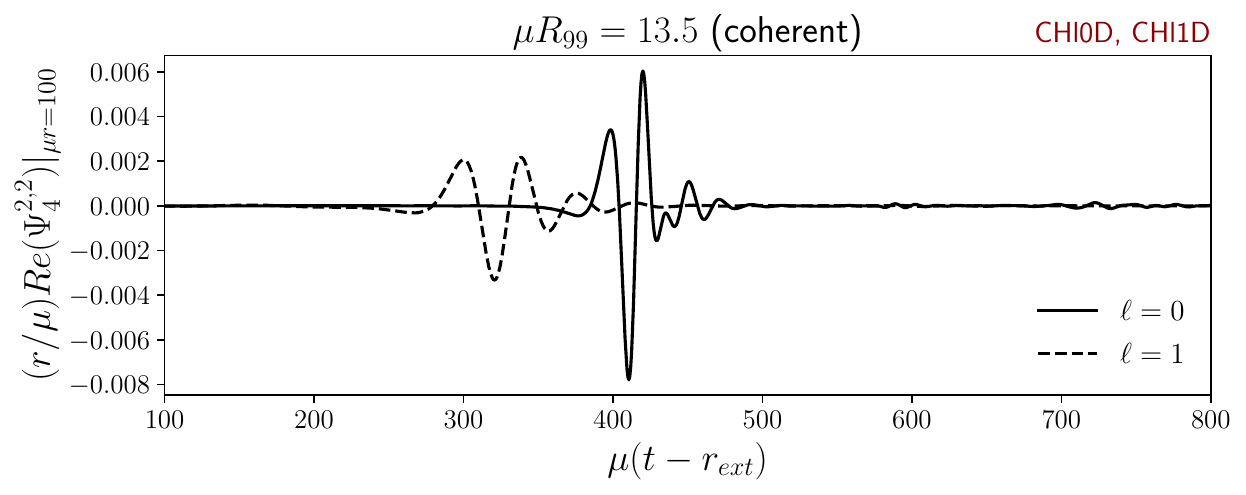}
\end{center}
\caption{GW signal for configurations described in Table \ref{tab:R_coherenta}. The extraction radius is $\mu r=100$. }
\label{fig:GWC}
\end{figure}

In Table \ref{tab:R_coherenta} we shown some properties of the stars. The radii of the stars are the same as models listed in Table \ref{tab:R_coherent}.
\begin{table}
\begin{tabular}{|c|c|c|c|c|c||c|c|c|}
\hline
\textit{Coh.}&\multicolumn{8}{ c |}{$\ell=0$}{}\\
\hline
Model &$\mu R_{99}$ &$\omega/\mu$&$\mu M_0$&$\mu^2 N_0$&$\mathcal{C}$&Rem.&$\mu R\sim$&$\mu t_c$\\
\hline
CHl0A&31.5&0.985&0.296&0.298&0.009 &BS 
&30&0\\
CHl0B&27.5&0.980&0.333&0.335&0.012 &BS 
&30&0\\
CHl0C&24.7&0.976&0.364&0.367&0.015 &BS 
&25&110\\
CHl0D&13.5&0.931&0.548&0.559&0.041 &BH &$\mu r_{AH}=1.25$ &320\\
\hline
\end{tabular}
\caption{Coherent cases for $\ell=0$. $R_{99}$ is the radius that contains 99\% of the mass of the star. $\omega$ is the frequency, $M_0$ is the mass of each star, $N_0$ is the number of particles of each star and $\mathcal{C}$ is the compactness. 
The final remnant can be a localized boson configuration (BS) or a black hole (BH).
$R$ is the radius that encloses 99\% of the mass for scalar field remnant at $\mu t=2500$
while for BHs it labels the radius of the apparent horizon.}
\label{tab:R_coherenta}
\end{table}
In Fig. \ref{fig:GWC} it is shown the waveforms $\psi^{2\, 2}_4$, for models
listed in Tables \ref{tab:R_coherent}
and \ref{tab:R_coherenta}.
For models CHl0A and CHl1A with  $\mu R_{99}=31.5$, the mode $\psi^{2\, 2}_4$ for boson stars with $\ell=0$ has the same order of magnitude despite the fact the compactness is larger for the $\ell=1$ boson stars. 
For models with $\mu R_{99} = 27.5$ (CHl0B and CHl1B) the maximum amplitude of the signal is considerably larger for the $\ell=1$ case.
For models with $\mu R_{99} = 24.7$ (CHl0C and CHl1C) the final product of the collision for $\ell=0$ is a boson star but for $\ell=1$ the remnant is a black hole. Both signals are clearly distinguishable.

Finally, for models with $\mu R_{99} = 13.5$ (CHl0D and CHl1D) the gravitational imprints for $\ell=0$ boson stars and $\ell = 1$ boson stars are quite different from each other despite the fact a black hole forms after the merger in both cases. This happens because in the $\ell =1$ boson star binary, the mass of the BH formed is bigger. 
%
\bibliographystyle{unsrt}
\bibliography{referencias} 

\end{document}